%% file: hybridRD_final.tex
\documentclass[a4paper,twocolumn,11pt,accepted=2021-12-24]{quantumarticle}
\pdfoutput=1
\usepackage[utf8]{inputenc}
\usepackage[english]{babel}
\usepackage[T1]{fontenc}
\usepackage{amsmath}
\usepackage{hyperref}

\usepackage{tikz}
\usepackage{lipsum}

\usepackage{theorem}
\usepackage{amssymb}

\input{preferences}

\newcommand{\beq}{\begin{eqnarray}}
\newcommand{\eeq}{\end{eqnarray}}

\newcommand{\tr}{\mathrm{Tr}}

\begin{document}

\title{One-Shot Hybrid State Redistribution}

\author{Eyuri Wakakuwa}
\email{e.wakakuwa@gmail.com}
\affiliation{Department of Communication Engineering and Informatics, Graduate School of Informatics and Engineering, The University of Electro-Communications, Tokyo 182-8585, Japan}
\affiliation{Department of Computer Science, Graduate School of Information Science and Technology, The University of
Tokyo, Bunkyo-ku, Tokyo 113-8656, Japan
}
\orcid{0000-0002-2445-2701}
\author{Yoshifumi Nakata}
\affiliation{Yukawa Institute for Theoretical Physics, Kyoto university, Kitashirakawa Oiwakecho, Sakyo-ku, Kyoto, 606-8502, Japan}
\affiliation{Photon Science Center, Graduate School of Engineering, The University of
Tokyo, Bunkyo-ku, Tokyo 113-8656, Japan
}
\affiliation{JST, PRESTO, 4-1-8 Honcho, Kawaguchi, Saitama, 332-0012, Japan}
\email{nakata@qi.t.u-tokyo.ac.jp}
\orcid{0000-0003-0290-4698}
\author{Min-Hsiu Hsieh}
\email{min-hsiu.hsieh@foxconn.com}
\affiliation{Centre for Quantum Software \& Information (UTS:QSI), University of Technology Sydney, Sydney NSW, Australia}
\affiliation{Hon Hai (Foxconn) Research Institute, Taipei, Taiwan}
\orcid{0000-0003-1985-4623}
\maketitle

\begin{abstract}
  We consider state redistribution of a ``hybrid'' information source that has both classical and quantum components. The sender transmits classical and quantum information at the same time to the receiver, in the presence of classical and quantum side information both at the sender and at the decoder. The available resources are shared entanglement, and noiseless classical and quantum communication channels. We derive one-shot direct and converse bounds for these three resources, represented in terms of 
the smooth conditional entropies of the source state.
Various coding theorems for two-party source coding problems are systematically obtained by reduction from our results,
including the ones that have not been addressed in previous literatures.
\end{abstract}

\begin{figure*}[t]
  \centering
  \includegraphics[bb={0 10 1200 426}, scale=0.32]{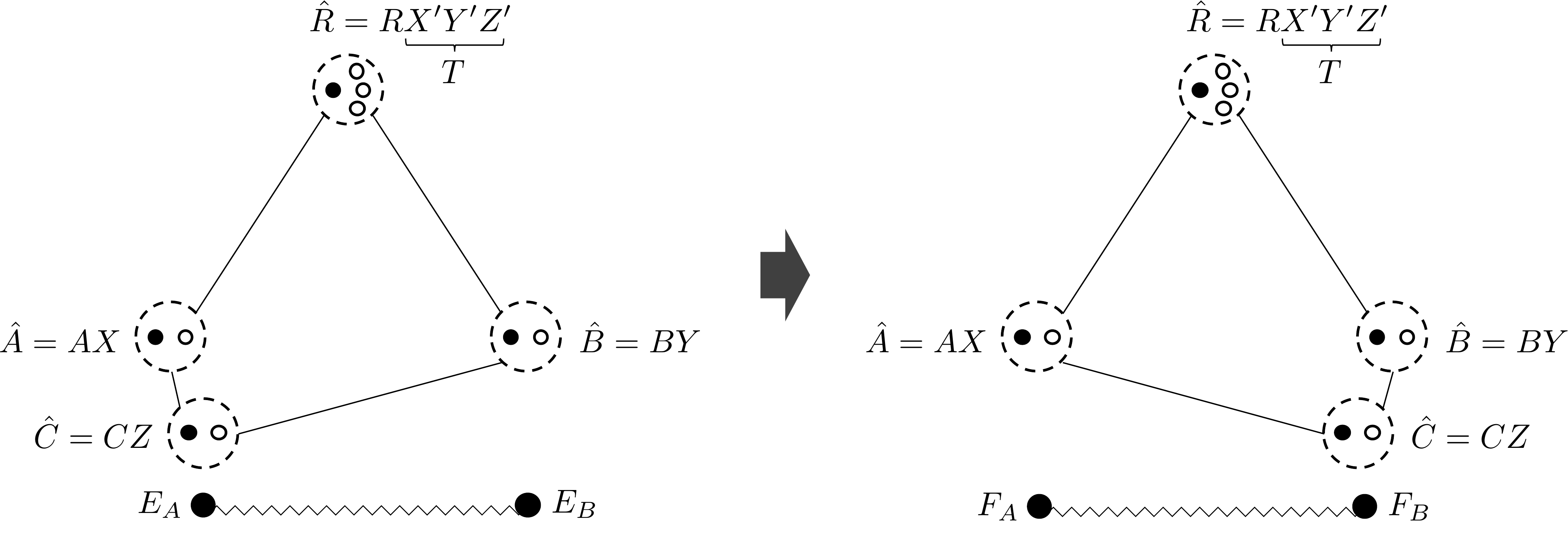}
  \caption{The task of state redistribution for the classical-quantum hybrid source is depicted. 
The black dots and the circles represent classical and quantum parts of the information source, respectively.
The wavy line represents the entanglement resource.
}
  \label{fig:A}
\end{figure*}

\section{Introduction}

Quantum state redistribution is a task in which the sender aims at transmitting quantum states to the receiver, in the presence of quantum side information both at the sender and at the receiver.
The costs of quantum communication and entanglement required for state redistribution have been analyzed in \cite{yard2009optimal,devetak2008exact,ming08} for the asymptotic scenario of infinitely many copies and vanishingly small error, and in \cite{berta2016smooth,1409.4352,anshu2017one} for the one-shot scenario.
Various coding theorems for two-party quantum source coding problems are obtained by reduction from these results as special cases, such as the Schumacher compression \cite{schumacher95}, quantum state merging \cite{horo07} and the fully-quantum Slepian-Wolf \cite{ADHW2009, datta2011apex}.
However, some of the well-known coding theorems cannot be obtained from those results, such as the (fully-classical) Slepian-Wolf (see e.g.~\cite{cover05}) and the classical data compression with quantum side information \cite{devetak2003classical}. 
This is because the results in \cite{yard2009optimal,devetak2008exact,ming08,berta2016smooth} only cover the fully quantum scenario, in which the information to be transmitted and the available resources are both quantum.

In this paper, we generalize the one-shot state redistribution theorem in \cite{berta2016smooth} to a ``hybrid'' situation.
That is, we consider the task of state redistribution in which the information to be transmitted and the side information at the parties have both classical and quantum components.
Not only quantum communication and shared entanglement, but also classical communication is available as a resource.
Our goal is to derive trade-off relations among the costs of the three resources required for achieving the task within a small error.
The main result is that we provide the direct and the converse bounds for the rate triplet to be achievable, in terms of the smooth conditional entropies of the source state and the error tolerance. 
For most of the special cases that have been analyzed in the previous literatures, the two bounds match in the asymptotic limit of infinitely many copies and vanishingly small error,
providing the full characterization of the achievable rate region.
Our result can be viewed as a one-shot generalization of the classically-assisted state redistribution protocol, proposed in \cite{min08}.

Coding theorems for most of the redistribution-type protocols, not only for quantum or classical information source but also for hybrid one, in one-shot scenario are systematically obtained from our result by reduction. 
In this sense, our result completes the one-shot capacity theorems of the redistribution-type protocols in a standard setting. 
As examples, we show that the coding theorems for the fully quantum state redistribution, the fully quantum Slepian-Wolf, 
quantum state splitting, quantum state merging, 
classical data compression with quantum side information, quantum data compression with classical side information and the fully classical Slepian-Wolf and quantum state redistribution with classical side information only at the decoder \cite{anshu2018noisy} can be recovered.
The last one would further lead to the family of quantum protocols in the presence of classical side information only at the decoder, along the same line as the one without classical side information \cite{ADHW2009,devetak2004family}.
In addition, our result also covers some redistribution-type protocols that have not been addressed in the previous literatures.

We note that the cost of resources in the hybrid redistribution-type protocols cannot be fully analyzed by simply plugging the hybrid source and the hybrid channel into the fully quantum setting.
This is because interconversion of classical and quantum communication channels requires the use of entanglement resource, which is not allowed e.g. in the fully classical scenario.

This paper is organized as follows.
In \rSec{prelimi}, we introduce notations and definitions that will be used throughout this paper. 
In \rSec{mainresults}, we provide the formulation of the problem and present the main results.
The results are applied in \rSec{specialcases} to special cases, and compared with the results in the previous literatures.
The proofs of the direct part and the converse part are provided in \rSec{direct} and \rsec{converse}, respectively.
Conclusions are given in \rSec{conclusion}.
The properties of the smooth entropies  used in the proofs are summarized in \rApp{propSmEn}.

\section{Preliminaries}
\lsec{prelimi}

We summarize notations and definitions that will be used throughout this paper. 

\subsection{Notations}

We denote the set of linear operators on a Hilbert space $\ca{H}$ by $\ca{L}(\ca{H})$.
For normalized density operators and sub-normalized density operators, we use the following notations, respectively:
\begin{align}
&
\ca{S}_=(\ca{H}) = \{\rho \in \ca{L}(\ca{H}) : \rho \geq 0, \tr [\rho]=1 \},
\\
&
\ca{S}_{\leq}(\ca{H}) = \{\rho \in \ca{L}(\ca{H}) : \rho \geq 0, \tr [\rho] \leq 1 \}.
\end{align}
A Hilbert space associated with a quantum system $A$ is denoted by ${\mathcal H}^A$, and its dimension is denoted by $d_A$. A system composed of two subsystems $A$ and $B$ is denoted by $AB$. When $M$  and $N$ are linear operators on ${\mathcal H}^A$ and ${\mathcal H}^B$, respectively, we denote $M\otimes N$ as $M^A\otimes N^B$ for clarity.  
In the case of pure states, we abbreviate $|\psi\rangle^A\otimes|\phi\rangle^B$ as $|\psi\rangle^A|\phi\rangle^B$. 
We denote $|\psi\rangle\!\langle\psi|$ simply by $\psi$.

For $\rho^{AB} \in \ca{L}(\ca{H}^{AB})$, $\rho^{A}$ represents ${\rm Tr}_B[\rho^{AB}]$.  
The identity operator is denoted by $I$. 
We denote $(M^A\otimes I^B)\ket{\psi}^{AB}$ as $M^A\ket{\psi}^{AB}$ and $(M^A\otimes I^B)\rho^{AB}(M^A\otimes I^B)^{\dagger}$ as $M^A\rho^{AB}M^{A\dagger}$. 
When ${\mathcal E}$ is a supermap from $\ca{L}(\ca{H}^{A})$ to $\ca{L}(\ca{H}^{B})$, we denote it by $\ca{E}^{A \rightarrow B}$. When $A = B$, we use $\ca{E}^{A}$ for short.
We also denote $({\mathcal E}^{A \rightarrow B} \otimes{\rm id}^C)(\rho^{AC})$ by ${\mathcal E}^{A \rightarrow B} (\rho^{AC})$.  
When a supermap is given by a conjugation of a unitary $U^A$ or a linear operator $W^{A \rightarrow B}$, we especially denote it by its calligraphic font such as 
$
\ca{U}^{A}(X^A):= (U^{A }) X^A (U^{A })^{\dagger}
$
and
$
\ca{W}^{A \rightarrow B}(X^A):= (W^{A \rightarrow B}) X^A (W^{A \rightarrow B})^{\dagger}
$.

The maximally entangled state between $A$ and $A'$, where $\ca{H}^{A} \cong \ca{H}^{A'}$, is defined by
\alg{
\ket{\Phi}^{AA'}:=\frac{1}{\sqrt{d_A}}\sum_{\alpha=1}^{d_A}\ket{\alpha}^A\ket{\alpha}^{A'}
}
with respect to a fixed orthonormal basis $\{\ket{\alpha}\}_{\alpha=1}^{d_A}$. The maximally mixed state on $A$ is defined by $\pi^A:=I^A/d_A$.

For any linear CP map $\ca{T}^{A\rightarrow B}$, there exists a finite dimensional quantum system $E$ and a linear operator $W_{\ca{T}}^{A\rightarrow BE}$ such that $\ca{T}^{A\rightarrow B}(\cdot)={\rm Tr}_E[W_{\ca{T}}(\cdot)W_{\ca{T}}^\dagger]$.
The operator $W_{\ca{T}}$ is called a Stinespring dilation of $\ca{T}^{A\rightarrow B}$ \cite{stinespring1955positive},
and the linear CP map defined by ${\rm Tr}_B[W_{\ca{T}}(\cdot)W_{\ca{T}}^\dagger]$ is called a {\it complementary map} of $\ca{T}^{A\rightarrow B}$.
With a slight abuse of notation, we denote the complementary map by $\ca{T}^{A\rightarrow E}$.

\subsection{Norms and Distances}

For a linear operator $X$, the trace norm is defined as $|\! | X |\! |_1 = \tr[ \sqrt{X^{\dagger}X}]$. 
For subnormalized states $\rho,\sigma\in\ca{S}_\leq(\ca{H})$,
the trace distance is defined by $\|\rho-\sigma\|_1$. 
The generalized fidelity and the purified distance are defined by
\alg{
\bar{F}(\rho,\sigma)
:=
\|\sqrt{\rho}\sqrt{\sigma}\|_1
+
\sqrt{(1-{\rm Tr}[\rho])(1-{\rm Tr}[\sigma])}
}
and
\alg{
P(\rho,\sigma)
:=
\sqrt{1-\bar{F}(\rho,\sigma)^2},
\laeq{dfnPD}
}
respectively (see Lemma 3 in \cite{tomamichel2010duality}).
The trace distance and the purified distance are related as
 \alg{
\frac{1}{2}\|\rho-\sigma\|_1
\leq
P(\rho,\varsigma)
\leq
\sqrt{2\|\rho-\sigma\|_1}
\laeq{relTDPD}
}
for any $\rho,\sigma\in\ca{S}_\leq(\ca{H})$.
The epsilon ball of a subnormalized state $\rho\in\ca{S}_\leq(\ca{H})$ is defined by
\begin{align}
\ca{B}^\epsilon(\rho):=\{\tau\in\ca{S}_\leq(\ca{H})|\:P(\rho,\tau)\leq\epsilon\}.
\label{eq:epsilon}
\end{align}

\begin{figure*}[t]
\begin{center}
\includegraphics[bb={0 20 1010 472}, scale=0.32]{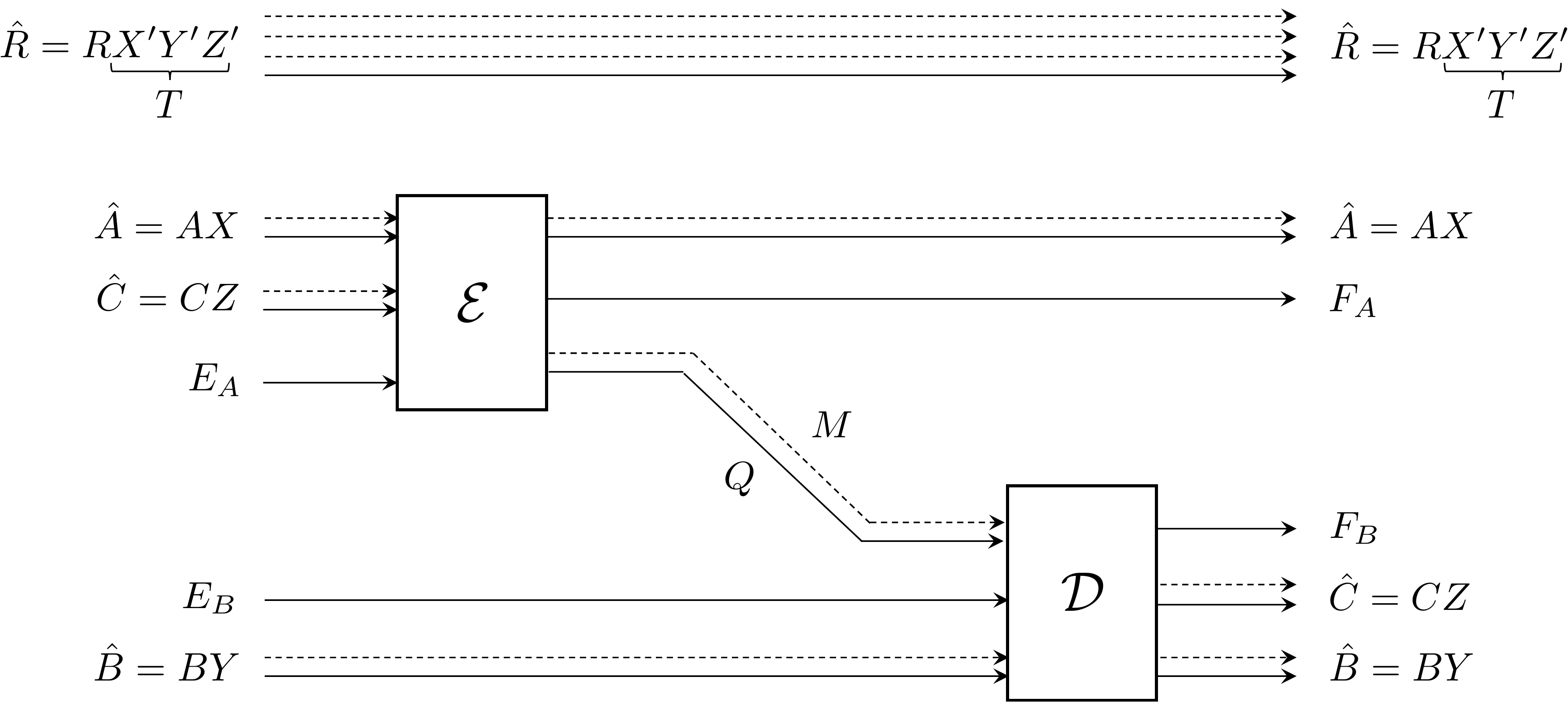}
\end{center}
\caption{
The task of state redistribution for the classical-quantum hybrid source is depicted in the diagram. 
The black lines and the dashed lines represent classical and quantum systems, respectively.
}
\label{fig:B}
\end{figure*}

\subsection{One-Shot Entropies}

For any subnormalized state $\rho\in\ca{S}_\leq(\ca{H}^{AB})$ and normalized state $\varsigma\in\ca{S}_=(\ca{H}^{B})$, define
\alg{
H_{\rm min}(A|B)_{\rho|\varsigma} 
:= \sup \{ \lambda \in \mathbb{R}| 2^{-\lambda} I^A \otimes \varsigma^B \geq \rho^{AB} \}
}
and
\alg{
H_{\rm max}(A|B)_{\rho|\varsigma} := \log{\|\sqrt{\rho^{AB}}\sqrt{I^A\otm\varsigma^B}\|_1^2}.
}
 The conditional min- and max- entropies (see e.g.~\cite{T16}) are defined by
\begin{align}
H_{\rm min}(A|B)_{\rho}& := \sup_{\sigma^B \in \ca{S}_=(\ca{H}^B)}H_{\rm min}(A|B)_{\rho|\sigma}, \\
H_{\rm max}(A|B)_{\rho}& := \sup_{\sigma^B \in \ca{S}_=(\ca{H}^B)}H_{\rm max}(A|B)_{\rho|\sigma},
\end{align}
and the smoothed versions thereof are given by
\begin{align}
H_{\rm min}^\epsilon(A|B)_{\rho}& := \sup_{\hat{\rho}^{AB} \in \ca{B}^\epsilon(\rho)}H_{\rm min}(A|B)_{\hat\rho}, 
\laeq{dfnmine}\\
H_{\rm max}^\epsilon(A|B)_{\rho}& := \inf_{\hat{\rho}^{AB} \in \ca{B}^\epsilon(\rho)}H_{\rm max}(A|B)_{\hat\rho}
\laeq{dfnmaxe}
\end{align}
for $\epsilon\geq0$.
In the case where $B$ is a trivial (one-dimensional) system, we simply denote them as $H_{\rm min}^\epsilon(A)_{\rho}$ and $H_{\rm max}^\epsilon(A)_{\rho}$, respectively.
We define
\alg{
&
H_{*}^{(\iota,\kappa)}(A|B)_{\rho}
\nn\\
&\quad\quad
:=
\max\{H_{\rm min}^{\iota}(A|B)_{\rho}, H_{\rm max}^{\kappa}(A|B)_{\rho}\}
\laeq{dfnHstar}
}
and
\alg{
&
\tilde{I}_{\rm min}^{\epsilon}(A:C|B)_{\rho}
\nn\\
&\quad\quad
:=
H_{\rm min}^{\epsilon}(A|B)_{\rho}
-H_{\rm min}^{\epsilon}(A|BC)_{\rho}.
\laeq{dfnmimMI}
}
We will refer to \req{dfnmimMI} as the {\it smooth conditional min mutual information}.
For $\tau\in\ca{S}(\ca{H}^A)$, we also use the ``max entropy'' in the version of \cite{renner2008security} (see Section 3.1.1 therein).
Taking the smoothing into account, it is defined by
\alg{
H_{\rm max'}^\epsilon(A)_\tau
:=
\inf_{\Pi:{\rm Tr}[\Pi\tau]\geq1-\epsilon}
\log{{\rm rank}[\Pi]},
\laeq{dfnHrank}
}
where the infimum is taken over all projections $\Pi$ such that ${\rm Tr}[\Pi\tau]\geq1-\epsilon$.
The von Neumann entropies and the quantum mutual information are defined by
\alg{
H(A)_\rho
&
:=
-{\rm Tr}[\rho^{A}\log{\rho^{A}}],
\\
H(A|B)_\rho
&:=H(AB)_\rho-H(B)_\rho,
\\
I(A:B)_\rho
&:=H(A)_\rho-H(A|B)_\rho.
}
The properties of the smooth conditional entropies used in this paper are summarized in \rApp{propSmEn}.

\section{Formulation and Results}
\lsec{mainresults}

Consider a classical-quantum source state in the form  of
\alg{
&
\Psi_s^{ABCRXYZX'Y'Z'}
:=
\nn\\
&
\quad
\sum_{x,y,z}p_{xyz}
\proj{x}^X\otm\proj{y}^Y\otm\proj{z}^Z
\quad\quad\quad\quad
\nn \\
&\quad\quad
\otm\proj{\psi_{xyz}}^{ABCR}\otm\proj{xyz}^{X'Y'Z'}.
\!\!
\laeq{sourcestate}
}
Here, $\{p_{xyz}\}_{x,y,z}$ is a probability distribution, $\ket{\psi_{xyz}}$ are pure states, and $\{\ket{x}\}_x$, $\{\ket{y}\}_y$, $\{\ket{z}\}_z$, $\{\ket{xyz}\}_{x,y,z}$ are orthonormal bases. 
The systems $X'$, $Y'$ and $Z'$ are assumed to be isomorphic to $X$, $Y$ and $Z$, respectively.
For the simplicity of notations, we denote $AX$, $BY$, $CZ$, $X'Y'Z'$ and $RX'Y'Z'$ by $\hat{A}$, $\hat{B}$, $\hat{C}$, $T$ and $\hat{R}$, respectively.
Accordingly, we also denote the source state by $\Psi_s^{\hat{A}\hat{B}\hat{C}\hat{R}}$.

We consider a task in which the sender transmits $\hat{C}$ to the receiver (see Figure \ref{fig:A} and \ref{fig:B}). 
The sender and the receiver have access to systems $\hat{A}$ and $\hat{B}$, respectively, as side information.
 The system $\hat{R}$ is the reference system that is inaccessible to the sender and the receiver.
The available resources for the task are the one-way noiseless classical and quantum channels from the sender to the receiver, and an entangled state shared in advance between the sender and the receiver.
We describe the communication resources by a quantum system $Q$ with dimension $2^q$ and a ``classical'' system $M$ with dimension $2^c$.
The entanglement resources shared between the sender and the receiver, before and after the protocol, are given by the maximally entangled states $\Phi_{2^{e+e_0}}^{E_AE_B}$ and $\Phi_{2^{e_0}}^{F_AF_B}$ 
 with Schmidt rank $2^{e+e_0}$ and $2^{e_0}$, respectively.

\bdfn{}
 A tuple $(c,q,e,e_0)$ is said to be achievable within an error $\delta$ for $\Psi_s$, if there exists a pair of an encoding CPTP map $\ca{E}^{\hat{A}\hat{C}E_A\rightarrow \hat{A}QMF_A}$ and a decoding CPTP map $\ca{D}^{\hat{B}QME_B\rightarrow \hat{B}\hat{C}F_B}$, such that
\begin{eqnarray}
\!
\left\|
\ca{D}\circ\ca{E}(\Psi_s^{\hat{A}\hat{B}\hat{C}\hat{R}}\!\otm\!\Phi_{2^{e+e_0}}^{E_AE_B})\!-\!\Psi_s^{\hat{A}\hat{B}\hat{C}\hat{R}}\!\otm\!\Phi_{2^{e_0}}^{F_AF_B}
\right\|_1
\nn\\
\leq
\delta.
\quad\quad
\laeq{qeec}
\end{eqnarray}
\edfn

\noindent
Note that, since $M$ is a classical message, the encoding CPTP map $\ca{E}$ must be such that for any input state $\tau$, the output state $\ca{E}^{\hat{A}\hat{C}E_A\rightarrow \hat{A}QMF_A}(\tau)$ is diagonal in $M$ with respect to a fixed orthonormal basis.
Note also that we implicitly assume that $c,q,e_0\geq0$, while the net entanglement cost $e$ can be negative.

Our goal is to obtain necessary and sufficient conditions for a tuple $(c,q,e,e_0)$ to be achievable within the error $\delta$ for a given source state $\Psi_s$. 
The direct and converse bounds are given by the following theorems:

\bthm{direct}{\bf(Direct part.)}
A tuple $(c,q,e,e_0)$ is achievable within an error $4\sqrt{12\epsilon+6\delta}+\sqrt{2}\epsilon$ for $\Psi_s$ if $d_C\geq2$ and it holds that
\alg{
c+2q
&\geq
\max\{\tilde{H}_{I}^{(3\epsilon/2,\epsilon/2)},\tilde{H}_{I\! I}^{(\epsilon/2)}\}
\nn\\
&\quad\quad\quad\quad\quad\quad\quad
-\log{(\delta^4/2)},
\laeq{neon00t}
\\
c+q+e
&\geq
H_{\rm max}^{\epsilon/2}(CZ|BY)_{\Psi_s}-\log{(\delta^2/2)},
\laeq{neon03t}
\\
q+e
&\geq
H_{\rm max}^{\epsilon/2}(C|BXYZ)_{\Psi_s}-\log{\delta^2},
\laeq{neon04t}
\\
e_0
&\geq 
\frac{1}{2}(H_{\rm max'}^{\epsilon^2/8}(C)_{\Psi_s}-H_{\rm max}^{3\epsilon/2}(C|BXYZ)_{\Psi_s})
\nn\\
&\quad\quad\quad\quad\quad\quad\quad\quad
+\log{\delta},
\laeq{neon05t}
}
where
\alg{
\tilde{H}_{I}^{(\iota,\kappa)}
:=
&
H_{*}^{(\iota,\kappa)}(C|AXYZ)_{\Psi_s}
\nn\\
&\quad+H_{\rm max}^{\kappa}(CZ|BY)_{\Psi_s},
\\
\tilde{H}_{I\! I}^{(\iota)}
:=
&
H_{\rm max}^\iota(C|AXZ)_{\Psi_s}
\nn\\
&\quad
+H_{\rm max}^\iota(C|BXYZ)_{\Psi_s}
}
and $H_{*}^{(\iota,\kappa)}$ is defined by \req{dfnHstar}.

In the case where $d_C=1$, a tuple $(c,0,0,0)$ is achievable for $\Psi_s$ within the error $\delta$ if it holds that
\alg{
c\geq
H_{\rm max}^\epsilon(Z|BY)_{\Psi_s}
-\log{\frac{\delta^2}{2}}.
}
\ethm

\bthm{converse}{\bf(Converse part.)}
Suppose that a tuple $(c,q,e,e_0)$ is achievable within the error $\delta$ for $\Psi_s$.
Then, regardless of the value of $e_0$, it holds that
\alg{
c+2q
&\geq
\max\{\tilde{H}_{I}'^{(\epsilon,\delta)},\tilde{H}_{I\! I}'^{(\epsilon,\delta)}\!-\!\Delta^{(\epsilon,\delta)}\}
\!-\!6f(\epsilon),
\laeq{convv00}\\
c+q+e
&
\geq
H_{\rm min}^{\epsilon}(BYCZ)_{\Psi_s}
\nn\\
&
\;\quad-
H_{\rm min}^{12\epsilon+6\sqrt{\delta}}(BY)_{\Psi_s}-f(\epsilon),
\laeq{convv03}\\
q+e
&
\geq
H_{\rm min}^{\epsilon}(BC|XYZ)_{\Psi_s}
\nn\\
&\quad\;
-\!H_{\rm min}^{11\epsilon+8\sqrt{\delta}}(B|XYZ)_{\Psi_s}\!-\!2f(\epsilon)
\!\!\!
\laeq{convv04}
}
for any $\epsilon>0$.
Here,
$
f(x):=-\log{(1-\sqrt{1-x^2})}
$,
\alg{
\tilde{H}_{I}'^{(\epsilon,\delta)}
:=&
H_{\rm min}^{\epsilon}(AC|XYZ)_{\Psi_s}
\nn\\
&\quad
-
H_{\rm max}^{\epsilon}(A|XYZ)_{\Psi_s}
\nn\\
&
\quad\quad
+
H_{\rm min}^{\epsilon}(BYCZ)_{\Psi_s}
\nn\\
&\quad\quad\quad
-
H_{\rm min}^{12\epsilon+6\sqrt{\delta}}(BY)_{\Psi_s},
\\
\tilde{H}_{I\! I}'^{(\epsilon,\delta)}
:=&
H_{\rm min}^{\epsilon}(AXCZ)_{\Psi_s}
\nn\\
&\quad
-
H_{\rm max}^{\epsilon}(AXZ)_{\Psi_s}
\nn\\
&\quad\quad
+
H_{\rm min}^{\epsilon}(BC|XYZ)_{\Psi_s}
\nn\\
&\quad\quad\quad
-H_{\rm min}^{11\epsilon+8\sqrt{\delta}}(B|XYZ)_{\Psi_s}
}
and
\alg{
\Delta^{(\epsilon,\delta)}
:=
\sup_{\ca{F}}\tilde{I}_{\rm min}^{7\epsilon+4\sqrt{\delta}}(G_A:Y'|M_AAX'Z')_{\ca{F}(\Psi_s)}.
\laeq{dfnDeltaed}
}
The supremum in \req{dfnDeltaed} is taken over all CPTP maps $\ca{F}:\hat{A}\hat{C}\rightarrow AG_AM_A$ such that $\ca{F}(\tau)$ is diagonal in $M_A$ with a fixed orthonormal basis for any $\tau\in\ca{S}(\ca{H}^{\hat{A}\hat{C}})$, and 
\begin{eqnarray}
\!\!
\inf_{\{\omega_{xyz}\}}
\!
P
\!
\left(
\!
\ca{F}(\Psi_s^{\hat{A}\hat{C}\hat{R}}),
\sum_{x,y,z}p_{xyz}
\psi_{xyz}^{A\hat{R}}
\otm
\omega_{xyz}^{G_AM_A}
\!
\right)
\nn\\
\leq
2\sqrt{\delta},
\quad\quad
\laeq{conditionaldecoupling}
\end{eqnarray}
where we informally denoted $\psi_{xyz}^{AR}\otm\proj{xyz}^T$ by $\psi_{xyz}^{A\hat{R}}$.
\ethm

\noindent
The proofs of \rThm{direct} and \rThm{converse} will be provided in \rSec{direct} and \rSec{converse}, respectively.

We also consider an asymptotic scenario of infinitely many copies and vanishingly small error.
A rate triplet $(c,q,e)$ is said to be {\it asymptotically achievable} if, for any $\delta>0$ and sufficiently large $n\in\mbb{N}$, there exists $e_0\geq0$ such that the tuple $(nc,nq,ne,ne_0)$ is achievable within the error $\delta$ for the one-shot redistribution of the state $\Psi_s^{\otm n}$.
The achievable rate region is defined as the closure of the set of achievable rate triplets.
The following theorem provides a characterization of the achievable rate region:

\bthm{asymptotic}{\bf(Asymptotic limit.)}
In the asymptotic limit of infinitely many copies and vanishingly small error, 
the inner and outer bounds for the achievable rate region are given by
\alg{
c+2q
&\geq
\max\{\tilde{H}_{I},\tilde{H}_{I\!I}\},
\laeq{neon1-2}
\\
c+q+e
&\geq
H(CZ|BY)_{\Psi_s},
\laeq{neon3-2}
\\
q+e
&\geq
H(C|BXYZ)_{\Psi_s},
\laeq{neon4-2}
\\
e_0
&
\geq
\frac{1}{2}I(C:BXYZ)_{\Psi_s}
\laeq{neon5-2}
}
and
\alg{
c+2q
&\geq
\max\{\tilde{H}_{I},\tilde{H}_{I\!I}-\tilde{\Delta}\},
\laeq{neon1-2o}
\\
c+q+e
&\geq
H(CZ|BY)_{\Psi_s},
\laeq{neon3-2o}
\\
q+e
&\geq
H(C|BXYZ)_{\Psi_s},
\laeq{neon4-2o}
}
respectively.
Here,
\alg{
&
\tilde{H}_{I}
:=H(C|AXYZ)_{\Psi_s}+H(CZ|BY)_{\Psi_s},
\laeq{katatteru1}
\\
&
\tilde{H}_{I\! I}
:=
H(C|AXZ)_{\Psi_s}+H(C|BXYZ)_{\Psi_s}
\laeq{katatteru2}
}
and
\alg{
\tilde{\Delta}
:=
\lim_{\delta\rightarrow0}\lim_{n\rightarrow\infty}\frac{1}{n}\Delta^{(\epsilon,\delta)}(\Psi_s^{\otm n}),
\laeq{dfntildeDelta}
}
where $\Delta^{(\epsilon,\delta)}$ is defined in \rThm{converse}. 
\ethm

\rThm{asymptotic} immediately follows from the one-shot direct and converse bounds (\rThm{direct} and \rThm{converse}).
This is due to the fully-quantum asymptotic equipartition property \cite{tomamichel2009fully},
 which implies that the smooth conditional entropies are equal to the von Neumann conditional entropy in the asymptotic limit of infinitely many copies.
That is, for any $\rho\in\ca{S}_=(\ca{H}^{PQ})$ and $\epsilon>0$, it holds that
\alg{
&
\lim_{n\rightarrow\infty}
\frac{1}{n}H_{\rm min}^\epsilon(P^n|Q^n)_{\rho^{\otm n}}
\nn\\
&=
\lim_{n\rightarrow\infty}
\frac{1}{n}H_{\rm max}^\epsilon(P^n|Q^n)_{\rho^{\otm n}}
\\
&=
H(P|Q)_{\rho}.
\laeq{FQAEP}
}
A simple calculation using this relation and the chain rule of the conditional entropy implies that the R.H.S.s of \req{neon00t}-\req{neon05t} and \req{convv00}-\req{convv04} coincide with those of \req{neon1-2}-\req{neon5-2} and \req{neon1-2o}-\req{neon4-2o}, respectively, in the asymptotic limit of infinitely many copies.

Due to the existence of the term $\tilde{\Delta}$ in Inequality \req{neon1-2o}, the direct and converse bounds in \rThm{asymptotic} do not match in general.
In many cases, however, it holds that $\tilde{\Delta}=0$ and thus the two bounds matches.
This is due to the following lemma about the property of $\Delta^{(\epsilon,\delta)}$:

\blmm{propDelta}
The quantity $\Delta^{(\epsilon,\delta)}$ defined in \rThm{converse} is nonnegative, and is equal to zero if there is no classical side information at the decoder (i.e. ${\rm dim}Y={\rm dim}Y'=1$) or if there is neither quantum message nor quantum side information at the encoder (i.e. ${\rm dim}A={\rm dim}C=1$).
The quantity $\tilde{\Delta}$ satisfies the same property due to the definition \req{dfntildeDelta}.
\elmm

\noindent
A proof of \rLmm{propDelta} will be provided in \rSec{propDelta}.
To clarify the general condition under which $\tilde{\Delta}=0$ is left as an open problem.

\brmk
The results presented in this section are applicable to the case where the sender and the receiver can make use of the resource of classical shared randomness.
To this end, it is only necessary to incorporate the classical shared randomness as a part of classical side information $X$ and $Y$.
\ermk

\begin{figure*}[t]
\begin{center}
\includegraphics[bb={0 20 1256 469}, scale=0.39]{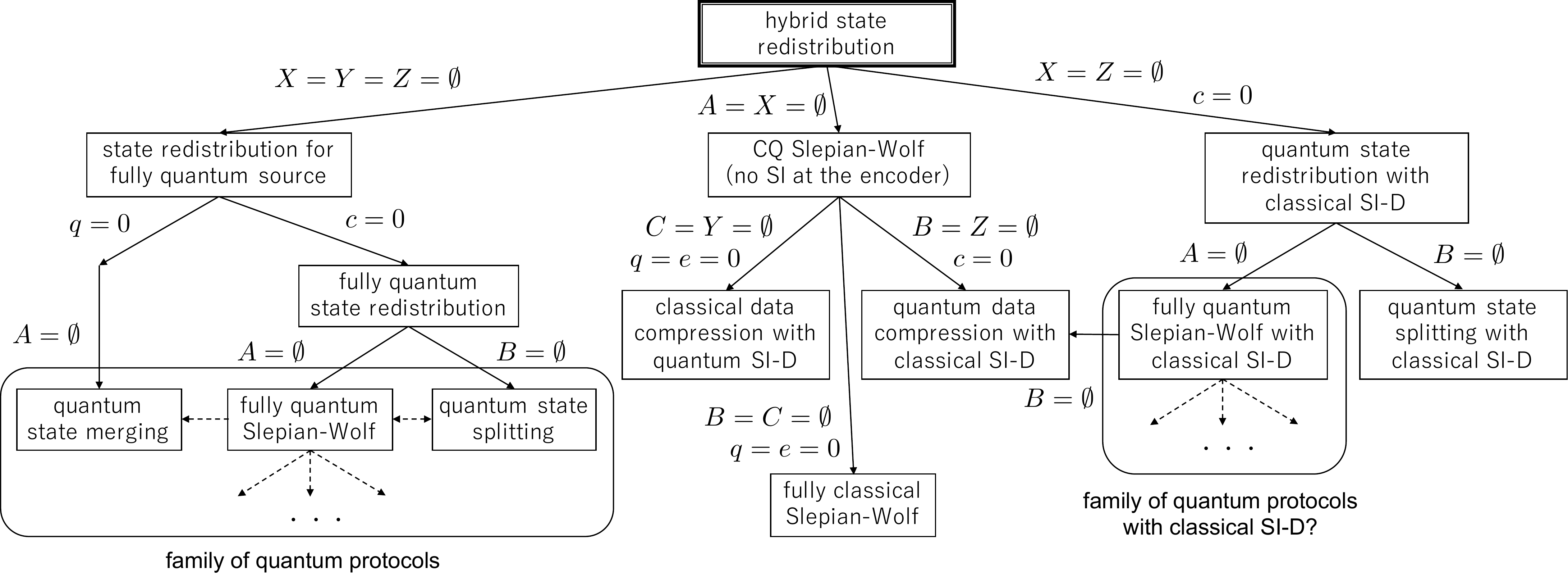}
\end{center}
\caption{
The relation among special cases of communication scenario analyzed in \rSec{specialcases} are depicted. ``SI'' and ``SI-D'' stand for ``side information'' and ``side information at the decoder'', respectively.
See Table \ref{tb:notations} below for the notations.
}
\label{fig:specialcases}
\end{figure*}

\begin{table*}[t]
\renewcommand{\arraystretch}{1.5}
  \begin{center}
    \begin{tabular}{|c|c|c|c|c|c|} \hline
	 & \multicolumn{3}{c}{information source}                  &  \multicolumn{2}{|c|}{available resources}  \\\cline{2-6}
            &   { \begin{tabular}{c} \!\!\!\!\!\!side information\!\!\!\!\!\! \\[-2mm] \!\!\!\!\!\!\!\!at the encoder\!\!\!\!\!\!\!\!  \end{tabular} } 
            &  { \begin{tabular}{c} \!\!\!\!\!\!side information\!\!\!\!\!\! \\[-2mm] \!\!\!\!\!\!\!\!at the decoder\!\!\!\!\!\!\!\!  \end{tabular} }  
            &  { \begin{tabular}{c} \!\!\!\!\!\!\!\!information\!\!\!\!\!\!\!\! \\[-2mm] \!\!\!\!\!\!to be transmitted\!\!\!\!\!\!  \end{tabular} } &  communication  & { \begin{tabular}{c} shared \\[-2mm] correlation \end{tabular} } \\ \hline
    quantum   &    $A$  &      $B$    & $C$   &  $q$   & $e$  \\\hline
   classical &   $X$ & $Y$ & $Z$ & $c$ &- \\\hline
    \end{tabular}
  \end{center}
  \caption{}
  \label{tb:notations}
\end{table*}

\section{Reduction to Special Cases}
\lsec{specialcases}

In this section, we apply the results presented in \rSec{mainresults} to special cases of source coding (see Figure \ref{fig:specialcases} in the next page).
In principle, the results cover all special cases where some of the components $A$, $B$, $C$, $X$, $Y$ or $Z$ are assumed to be one-dimensional, and where $c$, $q$ or $e$ is assumed to be zero.

Among them, we particularly consider the cases with no classical component in the source state and with no side information at the encoder, which have been analyzed in previous literatures.
We also consider quantum state redistribution with classical side information at the decoder, which has not been addressed before.
We investigate both the one-shot and the asymptotic scenarios.
The one-shot direct and converse bounds are obtained from \rThm{direct} and \rThm{converse}, respectively, and the asymptotic rate region is obtained from \rThm{asymptotic}. 
The analysis presented below shows that, for the tasks that have been analyzed in previous literatures, the bounds obtained from our results coincide with the ones obtained in the literatures.
It should be noted, however, that the coincidence in the one-shot scenario is only up to changes of the types of entropies and the values of the smoothing parameters.
All entropies are for the source state $\Psi_s$.
We will use \rLmm{onedimHminmax} in \rApp{propSmEn} for the calculation of entropies.

\subsection{No Classical Component in The Source State}

First, we consider the case where there is no classical component in the source state.
It is described by setting $
X=Y=Z=\emptyset
$.
By imposing several additional assumptions, the scenario reduces to different protocols.

\subsubsection{Fully Quantum State Redistribution}
Our hybrid scenario of state redistribution reduces to the fully quantum scenario, by additionally assuming that
$
c=0
$.
The one-shot direct part is given by
\alg{
2q
&\geq
H_*^{(3\epsilon/2,\epsilon/2)}(C|A)+H_{\rm max}^{\epsilon/2}(C|B)
\nn\\
&\quad\quad\quad\quad\quad\quad\quad\quad\quad\quad
-\log{(\delta^4/2)},
\\
\!\!\!q+e
&
\geq
H_{\rm max}^{\epsilon/2}(C|B)
-\log{(\delta^2/2)},
\\
e_0
&
\geq 
\frac{1}{2}(H_{\rm max'}^{\epsilon^2/8}(C)-H_{\rm max}^{3\epsilon/2}(C|B))+\log{\delta}.\!\!\!
}
An example of the tuple satisfying the above conditions is
\alg{
q
&=
\frac{1}{2}(H_*^{(3\epsilon/2,\epsilon/2)}(C|A)+H_{\rm max}^\epsilon(C|B)
\nn\\
&\quad\quad\quad\quad\quad\quad\quad\quad\quad\quad\quad-\log{(\delta^4/2)}),
\laeq{FQSRa1}\\
e
&=
\frac{1}{2}(-H_{\rm max}^\epsilon(C|A)
+H_{\rm max}^\epsilon(C|B)+1),
\laeq{FQSRa2}
\\
e_0
&
=
\frac{1}{2}(H_{\rm max'}^{\epsilon^2/8}(C)-H_{\rm max}^{3\epsilon/2}(C|B))+\log{\delta}.
}
The achievability of $q$ and $e$ given by \req{FQSRa1} and \req{FQSRa2} coincides with the result of \cite{berta2016smooth} (see also \cite{anshu2017one}).
The one-shot converse bound is represented as
\alg{
2q
&\geq
 H_{\rm min}^\epsilon(AC)- H_{\rm max}^\epsilon(A) +H_{\rm min}^\epsilon(BC)
\nn \\
& \quad\quad\quad\quad\quad
-H_{\rm min}^{12\epsilon+6\sqrt{\delta}}(B)-8f(\epsilon),
\laeq{FQSRconv1}\\
q+e
&\geq
 H_{\rm min}^\epsilon(BC)-H_{\rm min}^{12\epsilon+6\sqrt{\delta}}(B)-f(\epsilon).
}
The condition \req{FQSRconv1} in the above coincides with Inequality (104) in \cite{berta2016smooth}.
The rate region for the asymptotic scenario is obtained from
\rThm{asymptotic}, which yields
\alg{
2q
&\geq
H(C|A)+H(C|B),
\\
q+e
&
\geq
H(C|B).
}
A simple calculation implies that the above rate region is equal to the one obtained in
Ref.~\cite{devetak2008exact,yard2009optimal}.

\subsubsection{Fully Quantum Slepian-Wolf}
The fully-quantum Slepian-Wolf protocol is obtained by setting
$
A=\emptyset
$,
$
c=0
$.
The one-shot direct part obtained from \rThm{direct} reads
\alg{
2q
&\geq
H_*^{(3\epsilon/2,\epsilon/2)}(C)+H_{\rm max}^{\epsilon/2}(C|B)
\nn\\
&
\quad\quad\quad\quad\quad\quad\quad\quad\quad\quad
-\log{(\delta^4/2)},
\\
\!\!q+e
&\geq
H_{\rm max}^{\epsilon/2}(C|B)-\log{(\delta^2/2)},
\\
e_0
&
\geq 
\frac{1}{2}(H_{\rm max'}^{\epsilon^2/8}(C)-H_{\rm max}^{3\epsilon/2}(C|B))+\log{\delta}.\!\!\!
}
An example of the rate triplet $(q,e,e_0)$ satisfying the above inequalities is
\alg{
q
&=
\frac{1}{2}(H_*^{(3\epsilon/2,\epsilon/2)}(C)+H_{\rm max}^{\epsilon/2}(C|B)
\nn\\
&
\quad\quad\quad\quad\quad\quad\quad\quad\quad\quad
-\log{(\delta^4/2)}),
\\
e&=
\frac{1}{2}(-H_*^{(3\epsilon/2,\epsilon/2)}(C)+H_{\rm max}^{\epsilon/2}(C|B)+1),\!\!
\\
e_0
&
=
\frac{1}{2}(H_{\rm max'}^{\epsilon^2/8}(C)-H_{\rm max}^{3\epsilon/2}(C|B))+\log{\delta}.
}
The  result is equivalent to the one given by \cite{datta2011apex} (see Theorem 8 therein), with respect to $q$ and $e$.
Note, however, that our achievability bound requires the use of initial entanglement resource of $e+e_0$ ebits, whereas the one by \cite{datta2011apex} does not.
The one-shot converse bound is obtained from \rThm{converse}, which yields
\alg{
2q
&\geq
 H_{\rm min}^\epsilon(C)+ H_{\rm min}^\epsilon(BC)
 \nn\\
 &\quad\quad\quad\quad
  -H_{\rm min}^{12\epsilon+6\sqrt{\delta}}(B)-6f(\epsilon),
\\
q+e
&\geq
 H_{\rm min}^\epsilon(BC)-H_{\rm min}^{12\epsilon+6\sqrt{\delta}}(B)-f(\epsilon).
}
From \rThm{asymptotic},
the two-dimensional achievable rate region for the asymptotic scenario is given by
\alg{
2q
&\geq
H(C)+H(C|B),
\\
q+e
&\geq
H(C|B),
}
which coincides with the result obtained in \cite{ADHW2009}.
It should be noted that various coding theorems for quantum protocols are obtained from that for the fully quantum Slepian-Wolf protocol, which is referred to as the family of quantum protocols \cite{ADHW2009,devetak2004family}.

\subsubsection{Quantum State Splitting}
The task in which $
B=\emptyset
$,
$
c=0
$ is called quantum state splitting.
The one-shot direct part is represented as
\alg{
2q
&\geq
H_*^{(3\epsilon/2,\epsilon/2)}(C|A)+H_{\rm max}^{\epsilon/2}(C)
\nn\\
&
\quad\quad\quad\quad\quad\quad\quad\quad\quad
-\log{(\delta^4/2)},
\\
q+e
&
\geq
H_{\rm max}^{\epsilon/2}(C)
-\log{(\delta^2/2)},
\\
e_0
&
\geq 
\frac{1}{2}(H_{\rm max'}^{\epsilon^2/8}(C)-H_{\rm max}^{3\epsilon/2}(C))+\log{\delta}.
\laeq{deltae0}
}
Note that if a triplet $(q,e,e_0)$ is achievable, then $(q,e+e_0,0)$ is also achievable. 
Thus, an example of an achievable rate pair $(q,e)$ is
\alg{
q
&=
\frac{1}{2}(H_*^{(3\epsilon/2,\epsilon/2)}(C|A)+H_{\rm max}^{\epsilon/2}(C)-\log{(\delta^4/2)}),
\\
e
&=
\frac{1}{2}(-H_*^{(3\epsilon/2,\epsilon/2)}(C|A)
+H_{\rm max}^{\epsilon/2}(C)+1)
+
\delta e_0,
}
where we have denoted the R.H.S. of \req{deltae0} by $\delta e_0$.
This coincides with Lemma 3.5 in \cite{berta11}, up to an extra term $\delta e_0$.
The one-shot converse bound is given by
\alg{
2q
&\geq
 H_{\rm min}^\epsilon(AC)- H_{\rm max}^\epsilon(A) +H_{\rm min}^\epsilon(C)
 \nn\\
 &\quad\quad
+\log{(1-22\epsilon-16\sqrt{\delta})}-6f(\epsilon),
\\
q+e
&\geq
 H_{\rm min}^\epsilon(C)
 +\log{(1-22\epsilon-16\sqrt{\delta})}-f(\epsilon).
}
The rate region for the asymptotic scenario yields
\alg{
2q
&\geq
H(C|A)+H(C),
\\
q+e
&
\geq
H(C).
}
An example of a rate pair satisfying this condition is
\alg{
q
&=
\frac{1}{2}(H(C)+H(C|A)),
\\
e
&
=
\frac{1}{2}(H(C)-H(C|A)),.
}
This result coincides with Equality (6.1) in \cite{ADHW2009}, under the correspondence $|\Psi_s\rangle^{ACR}=U_\ca{N}^{R'\rightarrow AC}\ket{\varphi}^{R'R}$ with $U_\ca{N}^{R'\rightarrow AC}$ being some isometry.

\subsubsection{Quantum State Merging}
Quantum state merging is a task in which
$
A=\emptyset
$,
$
q=0
$.
The one-shot direct part is given by
\alg{
c
&\geq
H_*^{(3\epsilon/2,\epsilon/2)}(C)+H_{\rm max}^{\epsilon/2}(C|B)-\log{(\delta^4/2)},
\\
e
&\geq
H_{\rm max}^{\epsilon/2}(C|B)-\log{\delta^2},
\laeq{entoneSM}
\\
e_0
&
\geq 
\frac{1}{2}(H_{\rm max'}^{\epsilon^2/8}(C)-H_{\rm max}^{3\epsilon/2}(C|B))+\log{\delta}.
}
The achievability of the entanglement cost \req{entoneSM} is equal to the one given by \cite{DBWR2010} (see Theorem 5.2 therein).
The one-shot converse bound is obtained from \rThm{converse}, which yields
\alg{
c
&\geq
H_{\rm min}^\epsilon(C)+H_{\rm min}^{\epsilon}(BC)
\nn\\
&\quad\quad\quad\quad-H_{\rm min}^{12\epsilon+6\sqrt{\delta}}(B)-6f(\epsilon),\!
\\
e
&\geq
H_{\rm min}^{\epsilon}(BC)-H_{\rm min}^{11\epsilon+8\sqrt{\delta}}(B)-2f(\epsilon).
}
The rate region for the asymptotic setting is obtained from
\rThm{asymptotic} as
\alg{
c
&\geq
H(C)+H(C|B),
\\
e
&\geq
H(C|B).
}
This rate region is equivalent to the results in \cite{horo05,horo07}.
Note, however, that the protocols in \cite{horo05,horo07} are more efficient than ours, in that the catalytic use of entanglement resource is not required.

\subsection{No Side Information At The Encoder}

Next, we consider scenarios in which there is no classical or quantum side information at the encoder.
This corresponds to the case where $A=X=\emptyset$.
We consider three scenarios by imposing several additional assumptions.

\subsubsection{Classical Data Compression with Quantum Side Information at The Decoder}
The task of classical data compression with quantum side information was analyzed in \cite{devetak2003classical}.
This is obtained by additionally setting
$
Y=C=\emptyset
$,
$
q=e=e_0=0
$.
The one-shot direct and converse bounds are
given by
\alg{
c
&\geq
H_{\rm max}^\epsilon(Z|B)
-\log{\frac{\delta^2}{2}},
\\
c
&\geq
H_{\rm min}^\epsilon(BZ)
-
H_{\rm min}^{12\epsilon+6\sqrt{\delta}}(B)
-f(\epsilon),
}
respectively.
This result is equivalent to the one obtained in \cite{renes2012one} (see also \cite{tomamichel2013hierarchy}).
In the asymptotic limit, the achievable rate region is given by
$
c
\geq
H(Z|B)
$,
which coincides with the result by \cite{devetak2003classical}.

\subsubsection{Quantum Data Compression with Classical Side Information at The Decoder}

The task of quantum data compression with classical side information at the decoder was analyzed in \cite{khanian2020distributed}.
This is obtained by imposing additional assumptions
$
Z=B=\emptyset
$,
$
c=0
$.
In the entanglement ``unconsumed'' scenario ($e=0$),
the direct bounds for the one-shot 
case is given by
\alg{
q
&\geq
\frac{1}{2}(
H_*^{(3\epsilon/2,\epsilon/2)}(C)+H_{\rm max}^{\epsilon/2}(C|Y))
-\log{\frac{\delta^4}{2}},
\\
e_0
&
\geq 
\frac{1}{2}(H_{\rm max'}^{\epsilon^2/8}(C)-H_{\rm max}^{3\epsilon/2}(C|Y))+\log{\delta}.
}
Note that the entanglement is used only catalytically.
Thus, in the asymptotic regime, the achievable quantum communication rate in the entanglement unassisted scenario $(e=e_0=0)$ is obtained due to the cancellation lemma (Lemma 4.6 in \cite{deve08}), which reads
\alg{
q
&\geq
\frac{1}{2}(
H(C)+H(C|Y)).
\laeq{QcompCsideDA}
}
In the case where the unlimited amount of entanglement is available, the converse bounds on the quantum communication cost in the one-shot and the asymptotic scenarios read
\alg{
q
&\geq
\frac{1}{2}(
H_{\rm min}^\epsilon(C)+H_{\rm min}^\epsilon(C|Y)-\Delta^{\epsilon,\delta})
-6f(\epsilon),
\\
q
&\geq
\frac{1}{2}(
H(C)+H(C|Y)-\tilde{\Delta})
\nn\\
&\quad\quad\quad
+\frac{1}{2}\log{(1-22\epsilon-16\sqrt{\delta})}.
\laeq{QcompCsideCA}
}
The asymptotic result \req{QcompCsideDA} coincides with Theorem 7 in \cite{khanian2020distributed}, and \req{QcompCsideCA} is similar to Theorem 5 therein.
It is left open, however, whether the quantity $\tilde{\Delta}$ is equal to the function $I_{(n,\delta)}$ that appears in Theorem 5 of \cite{khanian2020distributed} (see Definition 2 in the literature).

\subsubsection{Fully Classical Slepian-Wolf}
In the fully classical scenario, the Slepian-Wolf problem is given by
$
B=C=\emptyset
$ in addition to $X=A=\emptyset$, and
$
q=e=e_0=0
$.
The one-shot achievability is given by
\alg{
c\geq
H_{\rm max}^\epsilon(Z|Y)
-\log{\frac{\delta^2}{2}},
}
and the one-shot converse bound reads
\alg{
c\geq
H_{\rm min}^\epsilon(YZ)
-
H_{\rm min}^{12\epsilon+6\sqrt{\delta}}(Y)
-f(\epsilon),
}
which are equivalent to the result obtained in \cite{renner2005simple}.
It is easy to show that
the well-known achievable rate region
$
c
\geq
H(Z|Y)
$
follows from \rThm{asymptotic}.

\subsection{Quantum State Redistribution with Classical Side Information at The Decoder}

We consider a scenario in which $X=Z=\emptyset$ and $c=0$.
This scenario can be regarded as a generalization of the fully quantum state redistribution, that incorporates classical side information at the decoder \cite{anshu2018noisy}.
The one-shot direct bound is represented by
\alg{
2q
&\geq
\max\{\tilde{H}_{I}^{(3\epsilon/2,\epsilon/2)},\tilde{H}_{I\! I}^{(\epsilon/2)}\}-\log{(\delta^4/2)},
\\
q+e
&\geq
H_{\rm max}^{\epsilon/2}(C|BY)-\log{(\delta^2/2)},
\\
e_0
&
\geq 
\frac{1}{2}(H_{\rm max'}^{\epsilon^2/8}(C)-H_{\rm max}^{3\epsilon/2}(C|BY))+\log{\delta},
}
where
\alg{
&
\tilde{H}_{I}^{(3\epsilon/2,\epsilon/2)}
:=H_{*}^{(3\epsilon/2,\epsilon/2)}(C|AY)+H_{\rm max}^{\epsilon/2}(C|BY),
\\
&
\tilde{H}_{I\! I}^{(\epsilon/2)}
:=
H_{\rm max}^{\epsilon/2}(C|A)+H_{\rm max}^{\epsilon/2}(C|BY).
}
The converse bound is also obtained from \rThm{converse}.
The inner and outer bounds for the achievable rate region in the asymptotic limit is given by
\alg{
2q
&\geq
\tilde{H}_{I\!I},
\\
q+e
&\geq
H(C|BY),
\\
e_0
&\geq
\frac{1}{2}I(C:BY),
}
and
\alg{
2q
&\geq
\max\{\tilde{H}_{I},\tilde{H}_{I\!I}-\tilde{\Delta}\},
\\
q+e
&\geq
H(C|BY),
}
respectively,
where
\alg{
&
\tilde{H}_{I}
:=H(C|AY)+H(C|BY),
\\
&
\tilde{H}_{I\! I}
:=
H(C|A)+H(C|BY).
}
We may also obtain its descendants by further assuming $A=0$ or $B=0$, which are generalizations of the fully quantum Slepian-Wolf and quantum state splitting.

It is expected that various quantum communication protocols with classical side information only at the decoder are obtained by reduction from the above result,
similarly to the family of quantum protocols \cite{ADHW2009,devetak2004family}.
We, however, leave this problem as a future work.

\section{Proof of The Direct Part (\rThm{direct})}
\lsec{direct}

We prove \rThm{direct} based on the following propositions:

\bprp{direct}
A tuple $(c,q,e,e_0)$ is achievable within the error $4\sqrt{12\epsilon+6\delta}$ for $\Psi_s$ if $d_C\geq2$ and it holds that 
\alg{
c+q-e
&\geq
H_{\rm max}^\epsilon(CZ|AX)_{\Psi_s}-\log{\frac{\delta^2}{2}},
\laeq{neon1}
\\
q-e
&\geq
H_{\rm max}^\epsilon(C|AXYZ)_{\Psi_s}-\log{\delta^2},
\laeq{neon2}
\\
c+q+e
&\geq
H_{\rm max}^\epsilon(CZ|BY)_{\Psi_s}-\log{\frac{\delta^2}{2}},
\laeq{neon3}
\\
q+e
&\geq
H_{\rm max}^\epsilon(C|BXYZ)_{\Psi_s}-\log{\delta^2},
\laeq{neon4}
\\
e_0
&=\frac{1}{2}(\log{d_C}-q-e).
\laeq{neon5}
}
In the case where $d_C=1$ and $q=e=e_0=0$, the classical communication rate $c$ is achievable within the error $\delta$ if it holds that
\begin{eqnarray}
c\geq
\max\{H_{\rm max}^\epsilon(Z|AX)_{\Psi_s},H_{\rm max}^\epsilon(Z|BY)_{\Psi_s}\}
\nn\\
-
\log{\frac{\delta^2}{2}}.
\quad\quad
\laeq{matahi}
\end{eqnarray}
\eprp

\bprp{direct2}
A tuple $(c,q,e,e_0)$ is achievable within an error $4\sqrt{12\epsilon+6\delta}$ for $\Psi_s$ if $d_C\geq2$ and it holds that
\alg{
c+2q
&\geq
\max\{\tilde{H}_{I}^{(\epsilon)},\tilde{H}_{I\! I}^{(\epsilon)}\}-\log{(\delta^4/2)},
\laeq{neon00}
\\
\!\!c+q+e
&\geq
H_{\rm max}^\epsilon(CZ|BY)_{\Psi_s}-\log{(\delta^2/2)},\!\!\!
\laeq{neon03}
\\
q+e
&\geq
H_{\rm max}^\epsilon(C|BXYZ)_{\Psi_s}-\log{\delta^2},
\laeq{neon04}
\\
e_0
&\geq 
\frac{1}{2}(\log{d_C}-H_{\rm max}^\epsilon(C|BXYZ)_{\Psi_s})
\nn\\
&\quad\quad\quad\quad\quad\quad\quad\quad\quad
+\log{\delta},
\laeq{neon05}
}
where
\alg{
\tilde{H}_{I}^{(\epsilon)}
:=
&H_{*}^\epsilon(C|AXYZ)_{\Psi_s}
\nn\\
&
+H_{\rm max}^\epsilon(CZ|BY)_{\Psi_s},
\\
\tilde{H}_{I\! I}^{(\epsilon)}
:=
&
H_{\rm max}^\epsilon(C|AXZ)_{\Psi_s}
\nn\\
&
+H_{\rm max}^\epsilon(C|BXYZ)_{\Psi_s}
}
and
\alg{
&
H_{*}^\epsilon(C|AXYZ)_{\rho}
:=
\nn\\
&\quad
\max\{H_{\rm min}^\epsilon(C|AXYZ)_{\rho}, H_{\rm max}^\epsilon(C|AXYZ)_{\rho}\}.
}

In the case where $d_C=1$, a tuple $(c,0,0,0)$ is achievable for $\Psi_s$ within the error $\delta$ if it holds that
\alg{
c\geq
H_{\rm max}^\epsilon(Z|BY)_{\Psi_s}
-\log{\frac{\delta^2}{2}}.
\laeq{mongen}
}
\eprp

Proofs of \rPrp{direct} and \rPrp{direct2} will be given in the following subsections.
In Section \rsec{PBD}, we prove the {\it partial bi-decoupling theorem}, which is a generalization of the bi-decoupling theorem \cite{ming08,berta2016smooth}.
Based on this result, we prove \rPrp{direct} in \rSec{prfPrpdirect}. 
We adopt the idea that a protocol for state redistribution can be constructed from sequentially combining protocols for the (fully quantum) reverse Shannon and the (fully quantum) Slepian-Wolf.
In Section \rsec{prfthmdirect}, we extend the rate region in \rPrp{direct} by incorporating teleportation and dense coding, thereby 
proving \rPrp{direct2}.
Finally, we prove \rThm{direct} from \rPrp{direct2} in \rSec{seru}.

\subsection{Partial Bi-Decoupling}
\lsec{PBD}

The idea of the bi-decoupling theorem was first introduced in \cite{ming08}, and was improved in \cite{berta2016smooth} to fit more into the framework of the one-shot information theory.
The approach in  \cite{berta2016smooth} is based on the decoupling theorem in \cite{DBWR2010}.  
In this subsection, we generalize those results by using the direct part of randomized partial decoupling \cite{wakakuwa2021one} to incorporate the hybrid communication scenario.

\subsubsection{Direct Part of Partial Decoupling}

We first present the direct part of randomized partial decoupling (Theorem 3 in \cite{wakakuwa2021one}).
Let $\Psi^{\hat{C}\hat{S}}$ be a subnormalized state in the form of
\begin{align}
\Psi^{\hat{C}\hat{S}}=\sum_{j,k=1}^J\outpro{j}{k}^{Z}\otimes\psi_{jk}^{CS}\otimes\outpro{j}{k}^{Z'}.\laeq{romanof}
\end{align}
Here, $Z$ and $Z'$ are $J$-dimensional quantum system with a fixed orthonormal basis $\{\ket{j}\}_{j=1}^J$, $\hat{C}\equiv ZC$, $\hat{S}\equiv Z'S$ and $\psi_{jk}\in\ca{L}(\ca{H}^{C}\otimes\ca{H}^{S})$ for each $j$ and $k$.
 Note that the positive-semidefiniteness of $\Psi^{\hat{C}\hat{S}}$ implies $\psi_{jj}\geq0$ for all $j$ and the subnormalization condition implies $\sum_{j=1}^J{\rm Tr}[\psi_{jj}]\leq1$.
Consider a random unitary $U$ on $\hat{C}$ in the form of
\begin{align}
U:=\sum_{j=1}^J\outpro{j}{j}^{Z}  \otimes U_j^{C},
\laeq{RUrpd}
\end{align}
where $U_j \sim {\sf H}_j$ for each $j$, and ${\sf H}_j$ is the Haar measure on the unitary group on $\ca{H}^{C}$. 
The averaged state obtained after the action of the random unitary $U$ is given by
\alg{
\Psi_{\rm av}^{\hat{C}\hat{S}}
&
:=\mbb{E}_{U} [ 
U^{\hat{C}} ( \Psi^{\hat{C}\hat{S}} ) U^{\dagger {\hat{C}}}]
\laeq{SEK}
\\
&=
\sum_{j=1}^Jp_j\proj{j}^{Z}\otimes\pi^C\otm\psi_{j}^{S}\otimes\proj{j}^{Z'},
\laeq{SEK2}
}
where $p_j:={\rm Tr}[\psi_{jj}]$ and $\psi_{j}:=p_j^{-1}\psi_{jj}$.
Consider also the permutation group $\mbb{P}$ on $[1,\cdots,J]$, and  define a unitary $G_\sigma$ for any $\sigma\in\mbb{P}$ by
\alg{
G_\sigma:=\sum_{j=1}^J\outpro{\sigma(j)}{j}^{Z}.
\label{eq:RPrpd}
}
We assume that the permutation $\sigma$ is chosen at random according to the uniform distribution on $\mbb{P}$.

\begin{figure}[t]
\begin{center}
\includegraphics[bb={0 30 780 213}, scale=0.3]{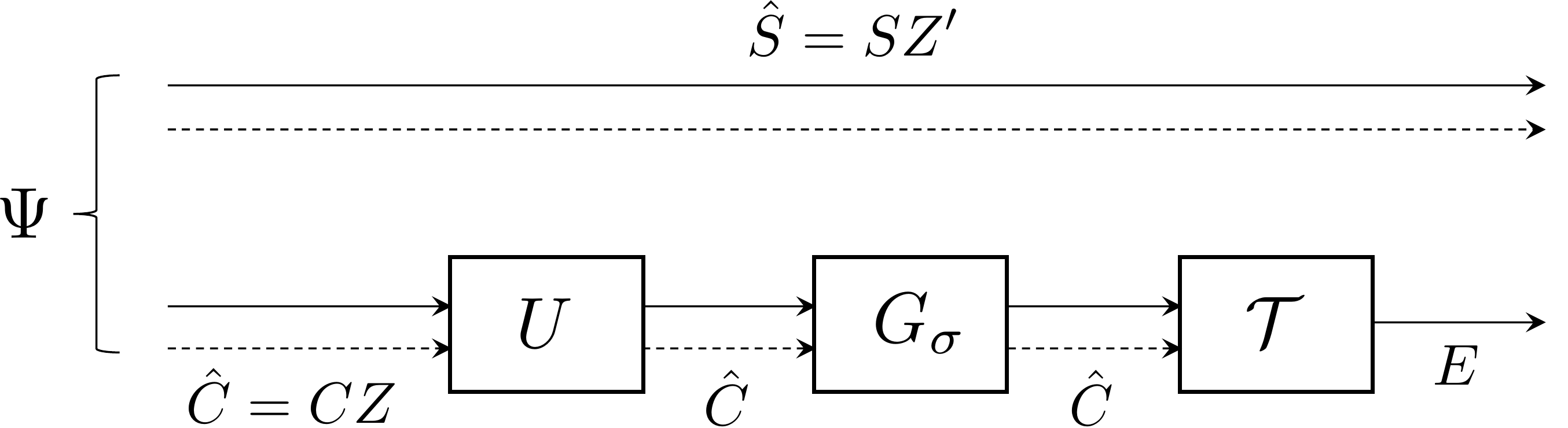}
\end{center}
\caption{
The situation of partial decoupling is depicted.
}
\label{fig:partialdecoupling}
\end{figure}

 Suppose that the state $\Psi^{\hat{C}\hat{S}}$ is transformed by unitaries $U$ and $G_\sigma$, and then is subject to the action of a quantum channel (linear CP map) $\ca{T}^{\hat{C}\rightarrow E}$ (see Figure \ref{fig:partialdecoupling}).
 The final state is represented as
 \alg{
 &
 \ca{T}^{{\hat{C}} \rightarrow E} (  (G_\sigma^Z U^{\hat{C}}) \Psi^{\hat{C}\hat{S}}   (G_\sigma^Z U^{\hat{C}})^\dagger )
 \nn\\
 &\quad\quad
 =
  \ca{T}^{{\hat{C}} \rightarrow E} \circ \ca{G}_\sigma^Z  \circ \ca{U}^{\hat{C}} ( \Psi^{\hat{C}\hat{S}} ).
 }
We consider how close the final state is, on average over all $U$, to the averaged final state $\ca{T}^{{\hat{C}} \rightarrow E} \circ \ca{G}_\sigma^Z ( \Psi_{\rm av}^{\hat{C}\hat{S}} )$, for typical choices of the permutation $\sigma$. 
The following theorem is the direct part of the randomized partial decoupling theorem, which provides an upper bound on the average distance between  $\ca{T}^{\hat{C} \rightarrow E} \circ \ca{G}_\sigma^Z  \circ \ca{U}^{\hat{C}} ( \Psi^{\hat{C}\hat{S}} )$ and $\ca{T}^{\hat{C} \rightarrow E} \circ \ca{G}_\sigma^Z ( \Psi_{\rm av}^{\hat{C}\hat{S}} )$.
Although the original version in \cite{wakakuwa2021one} is applicable to any $J\geq1$, in this paper we assume that $J\geq2$.

\blmm{SmoothExMarkov} {\bf(Corollary of Theorem 3 in \cite{wakakuwa2021one})}
Consider a subnormalized state $\Psi^{\hat{C}\hat{S}}\in\ca{S}_\leq(\ca{H}^{\hat{C}\hat{S}})$ that is decomposed as \req{romanof}. Let $\ca{T}^{\hat{C} \rightarrow E}$ be a linear trace non-increasing CP map with the complementary channel $\ca{T}^{\hat{C}\rightarrow F}$.
Let $U$ and $G_\sigma$ be random unitaries given by (\ref{eq:RUrpd}) and (\ref{eq:RPrpd}), respectively, and fix arbitrary $\epsilon,\mu\geq0$.
It holds that
\begin{align}
&\mbb{E}_{\sigma,U } \left[ \left\|
\ca{T}^{\hat{C} \rightarrow E} \circ \ca{G}_\sigma^Z  \circ \ca{U}^{\hat{C}} ( \Psi^{\hat{C}\hat{S}} )
\right.\right.\nn\\
&\quad\quad\quad\quad\quad\quad
\left.\left.-\ca{T}^{\hat{C} \rightarrow E} \circ \ca{G}_\sigma^Z ( \Psi_{\rm av}^{\hat{C}\hat{S}} )
\right\|_1 \right]\nonumber\\
&\leq  
\begin{cases}
2^{-\frac{1}{2}H_I}
+
2^{-\frac{1}{2}H_{I\!I}}
+4(\epsilon+\mu+\epsilon\mu)
&
\!
(d_C\geq2),
\\
2^{-\frac{1}{2}H_I}
+4(\epsilon+\mu+\epsilon\mu)
&
\!
(d_C=1),
\end{cases}
\!
\laeq{SmExMa}
\end{align}
where
$
\Psi_{\rm av}^{\hat{C}\hat{S}}:=\mbb{E}_{U}[ \ca{U}^{\hat{C}} ( \Psi^{\hat{C}\hat{S}} )]
$.
The exponents $H_I$ and $H_{I\!I}$ are given by
\alg{
&
\!\!
H_I=
\log{(J-1)}+
H_{\rm min}^\epsilon(\hat{C}|\hat{S})_{\Psi}
\nn\\
&
\quad\quad\quad\quad\quad\quad\quad\quad
-H_{\rm max}^\mu(\hat{C}|F)_{\ca{C}(\tau)},
 \\
&
\!\!
H_{I\!I}=
H_{\rm min}^\epsilon(\hat{C}|\hat{S})_{\ca{C}(\Psi)}-H_{\rm max}^\mu(C|FZ)_{\ca{C}(\tau)}.
}
Here, $\ca{C}$ is the completely dephasing operation on $Z$ with respect to the basis $\{|j\rangle\}_{j=1}^J$, and
$\tau$ is the Choi-Jamiolkowski state of $\ca{T}^{\hat{C}\rightarrow F}$ defined by $\tau^{\hat{C}F}:=\ca{T}^{\hat{C}'\rightarrow F}(\Phi^{\hat{C}\hat{C}'})$.
The state $\Phi^{\hat{C}\hat{C}'}$ is the maximally entangled state in the form of
\alg{
|\Phi\rangle^{\hat{C}\hat{C}'}=\frac{1}{\sqrt{J}}\sum_{j=1}^J\ket{jj}^{ZZ'}|\Phi_r\rangle^{CC'}.
\laeq{maxentdfn}
}  
\elmm

\subsubsection{Partial Decoupling under Partial Trace}

\begin{figure}[t]
\begin{center}
\includegraphics[bb={0 40 768 259}, scale=0.3]{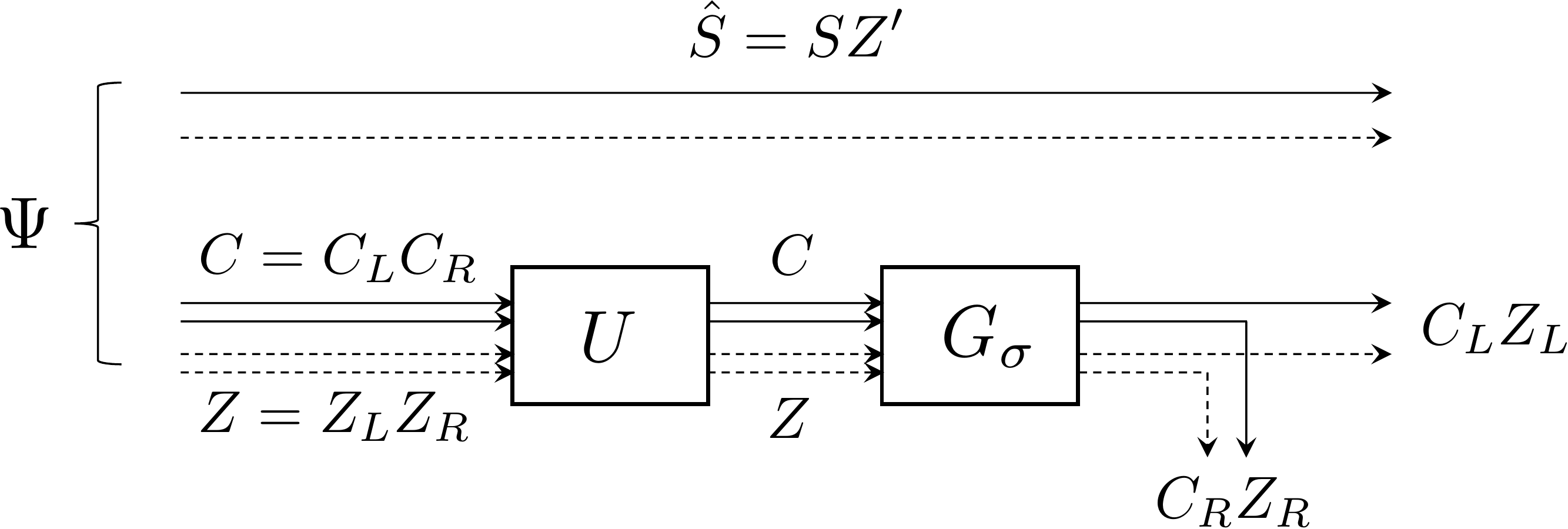}
\end{center}
\caption{
The situation of partial decoupling under partial trace is depicted.
}
\label{fig:partialdecouplingPT}
\end{figure}

We apply \rLmm{SmoothExMarkov} to a particular case where the channel $\ca{T}$ is the partial trace (see Figure \ref{fig:partialdecouplingPT}).

\blmm{tus}
Consider the same setting as in \rLmm{SmoothExMarkov}, and suppose that $Z=Z_LZ_R$, $C=C_LC_R$.
We assume that $Z_L$ and $Z_R$ are equipped with fixed orthonormal bases $\{\ket{z_L}\}_{z_L=1}^{J_L}$ and $\{\ket{z_R}\}_{z_R=1}^{J_R}$, respectively, thus $J=J_LJ_R$ and the orthonormal basis of $Z$ is given by $\{\ket{z_L}\ket{z_R}\}_{z_L,z_R}$.
Fix arbitrary $\epsilon\geq0$.
If $d_C\geq2$ and
\alg{
\log{\frac{d_{C_L}^2}{d_{Z_R}d_{C}}}
&
\leq
H_{\rm min}^\epsilon(\hat{C}|\hat{S})_{\Psi}+\log{\frac{\delta^2}{2}},
\laeq{ppt1}
\\
\log{\frac{d_{C_L}^2}{d_{C}}}
&
\leq
H_{\rm min}^\epsilon(\hat{C}|\hat{S})_{\ca{C}(\Psi)}+\log{\delta^2},
\laeq{ppt2}
}
then it holds that
\alg{
&
\mbb{E}_{\sigma,U }  \left\|
{\rm Tr}_{Z_RC_R} \circ \ca{G}_\sigma^Z  \circ \ca{U}^{\hat{C}} ( \Psi^{\hat{C}\hat{S}} ) \right.
\nn\\
&\quad\quad
\left.
-{\rm Tr}_{Z_RC_R} \circ \ca{G}_\sigma^Z ( \Psi_{\rm av}^{\hat{C}\hat{S}} )
\right\|_1
\leq  
4\epsilon+2\delta,
\laeq{ppt}
}
where
$
\Psi_{\rm av}^{\hat{C}\hat{S}}:=\mbb{E}_{U \sim {\sf H}_{\times}}[ \ca{U}^{\hat{C}} ( \Psi^{\hat{C}\hat{S}} )]
$.
The same statement also holds in the case of $d_C=1$, in which case the condition \req{ppt2} can be removed.
\elmm

\bprf
We apply \rLmm{SmoothExMarkov} by the correspondence $\mu=0$, $E=Z_LC_L$, $F=Z_RC_R$, $J=d_Z$ and $\ca{T}^{\hat{C}\rightarrow Z_LC_L}={\rm id}^{Z_LC_L}\otm{\rm Tr}_{Z_RC_R}$.
It follows that Ineq.~\req{ppt} holds if $d_C\geq2$ and
\alg{
&
\log{(d_Z-1)}+
H_{\rm min}^\epsilon(\hat{C}|\hat{S})_{\Psi}
\nn\\
&\quad\quad
-H_{\rm max}(\hat{C}|Z_R'C_R')_{\ca{C}(\tau)}
+\log{\delta^2}\geq0,
\laeq{PDB-Prf-1} \\
&
H_{\rm min}^\epsilon(\hat{C}|\hat{S})_{\ca{C}(\Psi)}-H_{\rm max}(C|Z_R'C_R'Z)_{\ca{C}(\tau)}
\nn\\
&\quad\quad\quad\quad\quad\quad\quad\quad\quad\quad\quad\quad
+\log{\delta^2}\geq 0.
\laeq{PDB-Prf-2} 
}
Here, $\tau$ is the Choi-Jamiolkowski state of the complementary channel of $\ca{T}^{\hat{C}\rightarrow Z_LC_L}$, and is given by
\alg{
\!\!
\tau^{\hat{C}Z_R'C_R'}=\pi^{Z_L}\otm\pi^{C_L}\otm\Phi^{Z_RZ_R'}\otm\Phi^{C_RC_R'}.
\!\!
}
Using the additivity of the max conditional entropy (\rLmm{addcondmax} in \rApp{propSmEn}),
the entropies are calculated to be
\alg{
&
H_{\rm max}(\hat{C}|Z_R'C_R')_{\ca{C}(\tau)}
\nn\\
&\quad\quad
=\log{d_{Z_L}}+\log{d_{C_L}}-\log{d_{C_R}},
 \\
&
H_{\rm max}(C|Z_R'C_R'Z)_{\ca{C}(\tau)}
\nn\\
&\quad\quad
=\log{d_{C_L}}-\log{d_{C_R}}.
}
Thus, Inequalities \req{PDB-Prf-1} and \req{PDB-Prf-2} are equivalent to
\alg{
&
\log{(d_Z-1)}+
H_{\rm min}^\epsilon(\hat{C}|\hat{S})_{\Psi}
\nn\\
&\quad\quad\quad\quad\quad
-\log{\frac{d_{Z_L}d_{C_L}}{d_{C_R}}}
+\log{\delta^2}\geq0,
\\
&
H_{\rm min}^\epsilon(\hat{C}|\hat{S})_{\ca{C}(\Psi)}-\log{\frac{d_{C_L}}{d_{C_R}}}+\log{\delta^2}\geq 0.
}
Noting that $d_Z=d_{Z_L}d_{Z_R}$, $d_C=d_{C_L}d_{C_R}$ and that $(d_Z-1)/d_Z\geq1/2$,
the above two inequalities follow from \req{ppt1} and \req{ppt2}, respectively.
Thus, the proof in the case of $d_C\geq2$ is done.
The proof for the case of $d_C=1$ proceeds along the same line.
\QED
\eprf

\begin{figure*}[t]
\begin{center}
\includegraphics[bb={0 30 1188 280}, scale=0.4]{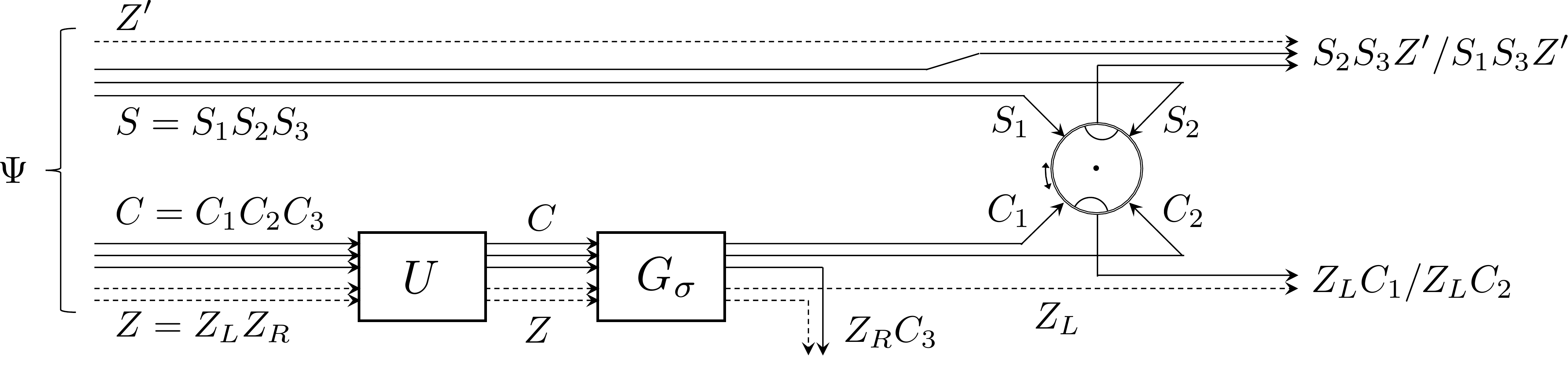}
\end{center}
\caption{
The situation of partial bi-decoupling is depicted.
As represented by the rotary, we consider two cases where $S_1C_2$ or $S_2C_1$ are traced out.
}
\label{fig:partialbidecoupling}
\end{figure*}

\subsubsection{Partial Bi-Decoupling Theorem}

Based on \rLmm{tus}, we introduce a generalization of the ``bi-decoupling theorem''\cite{ming08,berta2016smooth} that played a crucial role in the proof of the direct part of one-shot fully quantum state redistribution.
We consider the case where systems $C$ and $S$ are composed of three subsystems.
The following lemma provides a sufficient condition under which a {\it single} pair of $\sigma$ and $U$ simultaneously achieves partial decoupling of a state, from the viewpoint of two different choices of subsystems (see Figure \ref{fig:partialbidecoupling} in the next page).

\blmm{tus2}
{\bf (Partial bi-decoupling.)}
Consider the same setting as in \rLmm{SmoothExMarkov}, assume $Z=Z_LZ_R$, $C=C_1C_2C_3$, $S=S_1S_2S_3$
and fix arbitrary $\epsilon\geq0$.
If $d_C\geq2$ and
\alg{
\log{\frac{d_{C_1}^2}{d_{Z_R}d_{C}}}
&\leq
H_{\rm min}^\epsilon(\hat{C}|Z'S_2S_3)_{\Psi}+\log{\frac{\delta^2}{2}},
\laeq{PBD1}\\
\log{\frac{d_{C_1}^2}{d_{C}}}
&\leq
H_{\rm min}^\epsilon(\hat{C}|Z'S_2S_3)_{\ca{C}(\Psi)}\!+\!\log{\delta^2}\!,\!
\laeq{PBD2}\\
\log{\frac{d_{C_2}^2}{d_{Z_R}d_{C}}}
&\leq
H_{\rm min}^\epsilon(\hat{C}|Z'S_1S_3)_{\Psi}+\log{\frac{\delta^2}{2}},
\laeq{PBD3}\\
\log{\frac{d_{C_2}^2}{d_{C}}}
&\leq
H_{\rm min}^\epsilon(\hat{C}|Z'S_1S_3)_{\ca{C}(\Psi)}\!+\!\log{\delta^2}\!,\!
\laeq{PBD4}
}
there exist $\sigma$ and $U$ such that
\begin{align}
&
 \left\|
{\rm Tr}_{Z_RC_2C_3} \circ \ca{G}_\sigma^Z  \circ \ca{U}^{\hat{C}} ( \Psi^{\hat{C}S_2S_3Z'} ) 
\right.
\nn \\
&
\quad\quad
\left.
-{\rm Tr}_{Z_RC_2C_3} \circ \ca{G}_\sigma^Z ( \Psi_{\rm av}^{\hat{C}S_2S_3Z'} )
\right\|_1
\leq  
12\epsilon+6\delta,
\laeq{PBD5}
 \\
&
 \left\|
{\rm Tr}_{Z_RC_1C_3} \circ \ca{G}_\sigma^Z  \circ \ca{U}^{\hat{C}} ( \Psi^{\hat{C}S_1S_3Z'} )
\right.
\nn \\
&
\quad\quad
\left.
 -{\rm Tr}_{Z_RC_1C_3} \circ \ca{G}_\sigma^Z ( \Psi_{\rm av}^{\hat{C}S_1S_3Z'} )
\right\|_1
\leq  
12\epsilon+6\delta.
\laeq{PBD6}
\end{align}
The same statement also holds if $d_C=1$, in which case the conditions \req{PBD2} and \req{PBD4} can be removed.
\elmm

\bprf
Suppose that $d_C\geq2$ and the inequalities \req{PBD1}-\req{PBD4} are satisfied.
We apply \rLmm{tus} under the correspondence $C_R=C_\alpha C_3$, $S=S_\alpha S_3$ and $C_L=C_{\bar{\alpha}}$, where $\alpha=1,2$ and $\bar{\alpha}=2,1$ for each.
It follows that 
\alg{
&
\mbb{E}_{\sigma,U}
 \left\|
{\rm Tr}_{Z_RC_\alpha C_3} \circ \ca{G}_\sigma^Z  \circ \ca{U}^{\hat{C}} ( \Psi^{\hat{C}S_\alpha S_3Z'} )
\right.
\nn \\
&
\quad\quad\quad
\left.
 -{\rm Tr}_{Z_RC_\alpha C_3} \circ \ca{G}_\sigma^Z ( \Psi_{\rm av}^{\hat{C}S_\alpha S_3Z'} )
\right\|_1
\leq  
4\epsilon+2\delta.
}
Markov's inequality implies that there exist $\sigma$ and $U$ that satisfy both \req{PBD5} and \req{PBD6}, which completes the proof in the case of $d_C\geq2$.
The proof in the case of $d_C=1$ proceeds along the same line.
\QED
\eprf

\subsection{Proof of \rPrp{direct}}
\lsec{prfPrpdirect}

To prove \rPrp{direct}, we
follow the lines of the proof of the direct part of the fully quantum state redistribution protocol in \cite{ming08}.
The key idea is that a protocol for state redistribution can be constructed from sequentially combining a protocol for the fully quantum reverse Shannon and that for the fully quantum Slepian-Wolf.
We generalize this idea to the ``hybrid'' scenario  (see Figure \ref{fig:stateredistributiondirect} in page \pageref{fig:stateredistributiondirect}).
We only consider the case where $d_C\geq2$.
The proof for the case of $d_C=1$ is obtained along the same line.

\subsubsection{Application of The Partial Bi-Decoupling Theorem}
\lsec{consED}

Consider the ``purified'' source state
\alg{
&
\ket{\Psi}^{ABCRXYZT}:=
 \nn\\
&
\quad\quad
\sum_{x,y,z}\sqrt{p_{xyz}}
\ket{x}^X\ket{y}^Y\ket{z}^Z
\ket{\psi_{xyz}}^{ABCR}\ket{xyz}^T,
\laeq{psourcestate}
}
where we denoted $X'Y'Z'$ simply by $T$.
Let $C$ be isomorphic to $C_1C_2C_3$ and $Z$ to $Z_LZ_R$. 
Fix an arbitrary $\epsilon>0$.
We apply \rLmm{tus2} under the following correspondense:
\alg{
S_1=\hat{A},
\quad
S_2=\hat{B},
\quad
S_3=RX'Y'.
\laeq{corresp}
}
Note that $\hat{R}=RX'Y'Z'$.
It follows that if the dimensions of $C_1$ and $C_2$ are sufficiently small (see the next subsection for the details), there exist $\sigma$ and $U$ that satisfy
\alg{
&
 \left\|
{\rm Tr}_{Z_RC_2C_3} \circ \ca{G}_\sigma^Z  \circ \ca{U}^{\hat{C}} ( \Psi^{\hat{C}\hat{B}\hat{R}} ) 
\right.
\nn \\
&
\left.\quad\quad\quad
-{\rm Tr}_{Z_RC_2C_3} \circ \ca{G}_\sigma^Z ( \Psi_{\rm av}^{\hat{C}\hat{B}\hat{R}} )
\right\|_1
\leq  
12\epsilon+6\delta,
\laeq{derby1}
\\
&
 \left\|
{\rm Tr}_{Z_RC_1C_3} \circ \ca{G}_\sigma^Z  \circ \ca{U}^{\hat{C}} ( \Psi^{\hat{C}\hat{A}\hat{R}} )
\right.
\nn \\
&
\left.\quad\quad\quad
 -{\rm Tr}_{Z_RC_1C_3} \circ \ca{G}_\sigma^Z ( \Psi_{\rm av}^{\hat{C}\hat{A}\hat{R}} )
\right\|_1
\leq  
12\epsilon+6\delta.
\laeq{derby2}
}
Let $\ket{\Psi_{\sigma,1}}^{C_1Z_L\hat{B}\hat{R}D_A}$ be a purification of ${\rm Tr}_{Z_RC_2C_3} \circ \ca{G}_\sigma^Z ( \Psi_{\rm av}^{\hat{C}\hat{B}\hat{R}} )$ with $D_A$ being the purifying system.
Similarly, let $\ket{\Psi_{\sigma,2}}^{C_2Z_L\hat{A}\hat{R}D_B}$ be a purification of ${\rm Tr}_{Z_RC_1C_3} \circ \ca{G}_\sigma^Z ( \Psi_{\rm av}^{\hat{C}\hat{A}\hat{R}} )$ with $D_B$ being the purifying system.
Due to Uhlmann's theorem (\cite{uhlmann1976transition}; see also e.g. Chapter 9 in \cite{wildetext}), there exist linear isometries
\alg{
V^{D_A\rightarrow Z_RC_2C_3\hat{A}},
\quad
W^{Z_RC_1C_3\hat{B}\rightarrow D_B}
}
such that
\begin{eqnarray}
\left\|
\ca{G}_\sigma^Z  \circ \ca{U}^{\hat{C}} ( \Psi^{\hat{C}\hat{A}\hat{B}\hat{R}} ) 
-
\ca{V}^{D_A\rightarrow Z_RC_2C_3\hat{A}}(\Psi_{\sigma,1})
\right\|_1
\nn \\
\leq
2\sqrt{12\epsilon+6\delta},
\quad\quad
\laeq{200a-11}
\end{eqnarray}
\begin{eqnarray}
\left\|
\ca{W}^{Z_RC_1C_3\hat{B}\rightarrow D_B}
\circ \ca{G}_\sigma^Z  \circ \ca{U}^{\hat{C}} ( \Psi^{\hat{C}\hat{A}\hat{B}\hat{R}} )
-
\Psi_{\sigma,2}
\right\|_1
\nn\\
\leq
2\sqrt{12\epsilon+6\delta}.
\quad\quad
\laeq{200a-22}
\end{eqnarray} 
We particularly choose $C_1$, $C_2$, $C_3$ and $Z_R$ so that they satisfy the isomorphism
\alg{
C_1\cong E_B, C_2\cong F_A, C_3\cong Q, Z_R\cong M. 
\laeq{yukue}
}
In addition, we introduce systems $C''$, $Z''$, $A_1$ and $B_2$ such that
\alg{
C''\cong C, Z''\cong Z,
A_1\cong E_A, B_2\cong F_B.
}
We consider the purifying systems to be $D_A\equiv Z_R\hat{C}''\hat{A}A_1$ and $D_B\equiv Z_R\hat{C}''\hat{B}B_2$, where $\hat{C}''=C''Z''$.

\subsubsection{Explicit Forms of The Purifications}

To obtain explicit forms of the purifications $\Psi_{\sigma,1}$ and $\Psi_{\sigma,2}$,
we define a state $\Psi_{\sigma}$ by
\alg{
\ket{\Psi_{\sigma}}^{\hat{A}\hat{B}\hat{C}''\hat{R}Z}
&:=\sum_{x,y,z}\sqrt{p_{xyz}}
\ket{x}^X\ket{y}^Y\ket{\sigma(z)}^Z\ket{z}^{Z''}
\nn \\
&\quad\quad\quad
\otm\ket{\psi_{xyz}}^{ABC''R}\ket{xyz}^T\!.
\!\!
\laeq{dfnPsisigma}
}
From the definition \req{sourcestate} of the source state $\Psi_s$, \req{psourcestate} of the purified source state $\Psi$ and \req{dfnPsisigma} of the state $\Psi_\sigma$,
it is straightforward to verify that the states are related simply by
\alg{
\ket{\Psi_{\sigma}}^{\hat{A}\hat{B}\hat{C}''\hat{R}Z}
=
G_\sigma^Z\circ P^{Z''\rightarrow Z''Z}
\ket{\Psi}^{\hat{A}\hat{B}\hat{C}''\hat{R}}
\!\!\!
\laeq{psisigmaGPpsi}
}
and
\alg{
{\rm Tr}_Z\otm\ca{C}^T(\Psi_{\sigma}^{\hat{A}\hat{B}\hat{C}''\hat{R}Z})
&
=
\Psi_s^{\hat{A}\hat{B}\hat{C}''\hat{R}}
\\
&
=
\ca{C}^T(\Psi^{\hat{A}\hat{B}\hat{C}''\hat{R}}).
\laeq{btf}
}
Here, 
Let $P^{Z''\rightarrow Z''Z}$ be a linear isometry defined by
\alg{
P^{Z''\rightarrow Z''Z}
:=
\sum_z\ket{z}^{Z''}\ket{z}^{Z}\bra{z}^{Z''},
}
and 
$\ca{C}$ be the completely dephasing operation on $T$ with respect to the basis $\{\ket{xyz}\}_{x,y,z}$.
The state $\Psi_{\sigma}$ is simply represented as
\alg{
\ket{\Psi_{\sigma}}^{\hat{A}\hat{B}\hat{C}''\hat{R}Z}
=
\sum_{z}\sqrt{p_{z}}
\ket{\sigma(z)}^Z
\ket{\psi_z}^{\hat{A}\hat{B}\hat{C}''RX'Y'}\ket{z}^{Z'}
\!\!.\!
}
where
\alg{
&
\ket{\psi_z}^{\hat{A}\hat{B}\hat{C}''RX'Y'}
\nn\\
&\quad:=\sum_{x,y}\sqrt{\frac{p_{xyz}}{p_z}}
\ket{x}^X\ket{y}^Y\ket{z}^{Z''}
\nn\\
&\quad\quad\quad\quad
\otm\ket{\psi_{xyz}}^{ABC''R}\ket{x}^{X'}\ket{y}^{Y'}.
}
It is convenient to note that
\alg{
\psi_z^{\hat{A}\hat{B}RX'Y'}=\sum_{x,y}\sqrt{\frac{p_{xyz}}{p_z}}
\:\psi_{xyz}^{ABR}\otm\proj{x}^{X'}\!\otm\proj{y}^{Y'}.
}

Due to \req{psourcestate} and \req{SEK2}, the averaged state in \req{derby1} is calculated to be
\begin{align}
\Psi_{\rm av}^{\hat{C}\hat{B}\hat{R}} 
\!=\!\sum_{z}p_z \proj{z}^{Z} \!\otimes\! \pi^{C} \!\otm\! \psi_{z}^{\hat{B}RX'Y'}\!\otm\!\proj{z}^{Z'},
\!
\end{align}
where $p_z=\sum_{x,y}p_{xyz}$.
It follows that
\begin{align}
&
{\rm Tr}_{Z_RC_2 C_3} \circ \ca{G}_\sigma^Z ( \Psi_{\rm av}^{\hat{C}\hat{B}\hat{R}} )
\nn\\
&\quad
=\sum_zp_z {\rm Tr}_{Z_R}[\proj{\sigma(z)}] \otm\pi^{C_1}
\nn\\
&\quad\quad\quad\quad\quad\quad
\otm\psi_{z}^{\hat{B}RX'Y'}\otm\proj{z}^{Z'}.
\end{align}
Thus, a purification $\Psi_{\sigma,1}$ of this state is given by
\alg{
\!\!
\ket{\Psi_{\sigma,1}}^{\hat{A}\hat{B}\hat{C}''\hat{R}A_1C_1Z}=\ket{\Psi_{\sigma}}^{\hat{A}\hat{B}\hat{C}''\hat{R}Z}\ket{\phi_1}^{A_1C_1},
\!\!
}
where $\phi_1$ is the maximally entangled state of Schmidt rank $d_{C_1}$.
In the same way, the purification $\Psi_{\sigma,2}$ is given by
\alg{
\!\!\!\!
\ket{\Psi_{\sigma,2}}^{\hat{A}\hat{B}\hat{C}''\hat{R}B_2C_2Z}
=\ket{\Psi_{\sigma}}^{\hat{A}\hat{B}\hat{C}''\hat{R}Z}\ket{\phi_2}^{B_2C_2}\!,\!\!
\laeq{gb}
}
with $\phi_2$ being the maximally entangled state of Schmidt rank $d_{C_2}$.
Substituting these to \req{200a-11} and \req{200a-22}, we arrive at
\alg{
&
\left\|
  \Psi^{\hat{C}\hat{A}\hat{B}\hat{R}} 
-
(\ca{G}_\sigma^Z  \circ \ca{U}^{\hat{C}})^\dagger\circ\ca{V}(\Psi_\sigma^{\hat{A}\hat{B}\hat{C}''\hat{R}}\otm\phi_1^{A_1C_1})
\right\|_1
\nn \\
&\quad\quad\quad\quad\quad\quad\quad\quad\quad\quad\quad
\leq
2\sqrt{12\epsilon+6\delta},
\laeq{200a-3-2-1} \\
&
\left\|
\ca{W}
\circ \ca{G}_\sigma^Z  \circ \ca{U}^{\hat{C}} ( \Psi^{\hat{C}\hat{A}\hat{B}\hat{R}} )
-
\Psi_{\sigma}^{\hat{A}\hat{B}\hat{C}''\hat{R}Z}\otm\phi_2^{B_2C_2}
\right\|_1
\nn\\
&\quad\quad\quad\quad\quad\quad\quad\quad\quad\quad\quad
\leq
2\sqrt{12\epsilon+6\delta}.
\laeq{200a-3-2-2} 
} 

Inequality \req{200a-3-2-1} implies that the operation $(\ca{G}_\sigma^Z  \circ \ca{U}^{\hat{C}})^\dagger\circ\ca{V}$ is a reverse Shannon protocol for the state $\Psi^{\hat{C}\hat{A}(\hat{B}\hat{R})} $, up to the action of a linear isometry $G_\sigma^Z\circ P^{Z''\rightarrow Z''Z}$ by which $\Psi_\sigma$ is obtained from $\Psi$ as \req{psisigmaGPpsi}.
Similarly, Inequality \req{200a-3-2-2} implies that the operation $\ca{W}\circ\ca{G}_\sigma^Z  \circ \ca{U}^{\hat{C}}$ is a Slepian-Wolf protocol for the state $\Psi^{\hat{C}\hat{B}(\hat{A}\hat{R})} $, up to the action of $G_\sigma^Z\circ P^{Z''\rightarrow Z''Z}$ (see Figure \ref{fig:stateredistributiondirect} in page \pageref{fig:stateredistributiondirect}).
We combine the two protocols to cancel out $(\ca{G}_\sigma^Z  \circ \ca{U}^{\hat{C}})^\dagger$ and $\ca{G}_\sigma^Z  \circ \ca{U}^{\hat{C}}$.
Due to the triangle inequality, it follows from \req{200a-3-2-1} and \req{200a-3-2-2} that
\begin{eqnarray}
\left\|
\ca{W}
\circ 
\ca{V}
 ( \Psi_\sigma^{\hat{A}\hat{B}\hat{C}''\hat{R}}\otm\phi_1^{A_1C_1} )
-
\Psi_{\sigma}^{\hat{A}\hat{B}\hat{C}''\hat{R}Z}\otm\phi_2^{B_2C_2}
\right\|_1
 \nn\\
\leq
4\sqrt{12\epsilon+6\delta}.
\quad\quad\quad
\laeq{200a-3} 
\end{eqnarray}

\subsubsection{Construction of The Encoding and Decoding Operations}

Define a partial isometry
\alg{
&
\!\!V_\sigma^{\hat{A}\hat{C}''A_1\rightarrow \hat{A}ZC_2C_3}
\nn\\
&\!\!\;
:=
V^{Z_RA_1\hat{A}\hat{C}''\rightarrow Z_RC_2C_3\hat{A}}
\circ
G_\sigma^Z\circ P^{Z''\rightarrow Z''Z}.
\laeq{dfnVsigma}
}
Applying the map ${\rm Tr}_Z\otm\ca{C}^T$ to Inequality \req{200a-3}, and using \req{psisigmaGPpsi} and \req{btf}, it follows that
\alg{
&
\left\|
{\rm Tr}_Z\circ
\ca{W}
\circ 
\ca{V}_\sigma
(
\Psi_s^{\hat{A}\hat{B}\hat{C}''\hat{R}}\otm\phi_1^{A_1C_1})
 \right.
 \nn\\
 &
 \quad\quad\quad\quad\quad
 \left.
-
\Psi_s^{\hat{A}\hat{B}\hat{C}''\hat{R}}\otm\phi_2^{B_2C_2}
\right\|_1
\leq
4\sqrt{12\epsilon+6\delta}.
\laeq{200a-4} 
}
We construct a protocol for state redistribution as follows:
In the first step, the sender performs the following encoding operation:
\alg{
&
\ca{E}^{\hat{A}\hat{C}''A_1\rightarrow \hat{A}Z_RC_2C_3}
\nn\\
&\quad
=
{\rm Tr}_{Z_L}\circ
\ca{V}_\sigma^{\hat{A}\hat{C}''A_1\rightarrow \hat{A}ZC_2C_3}
\circ
\ca{C}^{Z''},
\laeq{dfnVsignaenc}
}
where $\ca{C}^{Z''}$ is the completely dephasing operation on $Z''$ with respect to the basis $\{\ket{z_L}\ket{z_R}\}_{z_L,z_R}$.
The sender then sends the classical system $Z_R\cong M$ and the quantum system $C_3\cong Q$ to the receiver, who performs the decoding operation defined by
\alg{
&
\ca{D}^{Z_RC_1C_3\hat{B}\rightarrow B_2\hat{B}\hat{C}''}
\nn\\
&\quad
=
{\rm Tr}_{Z_R}\circ
\ca{W}^{Z_RC_1C_3\hat{B}\rightarrow Z_RB_2\hat{B}\hat{C}''}.
} 
Noting that ${\rm Tr}_Z={\rm Tr}_{Z_L}\otm{\rm Tr}_{Z_R}$, we obtain from \req{200a-4} that
\begin{eqnarray}
\left\|
\ca{D}
\circ 
\ca{E}
 (\Psi_s^{\hat{A}\hat{B}\hat{C}''\hat{R}}\otm\phi_1^{A_1C_1})
-
\Psi_s^{\hat{A}\hat{B}\hat{C}''\hat{R}}\otm\phi_2^{B_2C_2}
\right\|_1
\nn\\
\leq
4\sqrt{12\epsilon+6\delta}.
\quad\quad\quad
\laeq{200a-4-2} 
\end{eqnarray}
From \req{dfnVsigma} and \req{dfnVsignaenc}, it is straightforward to verify that $\ca{E}(\tau)$ is diagonal in $Z_R$ for any input state $\tau$.
Thus, the pair $(\ca{E},\ca{D})$ is a state redistribution protocol for the state $\Psi_s$ within the error $4\sqrt{12\epsilon+6\delta}$.

\subsubsection{Evaluation of Entropies}
\lsec{evalent}

We analyze conditions on the size of systems $C_1$ and $C_2$, in order that inequalities \req{derby1} and \req{derby2} are satisfied.
We use the partial bi-decoupling theorem (\rLmm{tus2}) under the correspondence \req{corresp}, which reads
\alg{
S_1=\hat{A},
\quad
S_2=\hat{B},
\quad
S_3=RX'Y'.
\laeq{corresp2}
}
It follows that inequalities \req{derby1} and \req{derby2} are satisfied if it holds that
\alg{
\log{\frac{d_{C_1}^2}{d_{Z_R}d_{C}}}
&\leq
H_{\rm min}^\epsilon(\hat{C}|\hat{B}\hat{R})_{\Psi}+\log{\frac{\delta^2}{2}},
\laeq{brd1}\\
\log{\frac{d_{C_1}^2}{d_{C}}}
&\leq
H_{\rm min}^\epsilon(\hat{C}|\hat{B}\hat{R})_{\ca{C}(\Psi)}+\log{\delta^2}, 
\laeq{brd2}\\
\log{\frac{d_{C_2}^2}{d_{Z_R}d_{C}}}
&\leq
H_{\rm min}^\epsilon(\hat{C}|\hat{A}\hat{R})_{\Psi}+\log{\frac{\delta^2}{2}}, 
\laeq{brd3}\\
\log{\frac{d_{C_2}^2}{d_{C}}}
&\leq
H_{\rm min}^\epsilon(\hat{C}|\hat{A}\hat{R})_{\ca{C}(\Psi)}+\log{\delta^2}. 
\laeq{brd4}
}
Using the duality of the smooth conditional entropy (\rLmm{duality}), and noting that $\Psi^{\hat{A}\hat{B}\hat{C}}=\Psi_s^{\hat{A}\hat{B}\hat{C}}$, the min entropies in the first and the third inequalities are calculated to be
\alg{
H_{\rm min}^\epsilon(\hat{C}|\hat{B}\hat{R})_{\Psi}
&=
-
H_{\rm max}^\epsilon(\hat{C}|\hat{A})_{\Psi}
\\
&
=
-
H_{\rm max}^\epsilon(CZ|AX)_{\Psi_s}, \\
H_{\rm min}^\epsilon(\hat{C}|\hat{A}\hat{R})_{\Psi}
&=
-
H_{\rm max}^\epsilon(\hat{C}|\hat{B})_{\Psi}
\\
&=
-
H_{\rm max}^\epsilon(CZ|BY)_{\Psi_s}.
}
Similarly, due to \rLmm{condminCQCQ} and \rLmm{SE11d} in \rApp{propSmEn}, and noting that $\ca{C}(\Psi)=\Psi_s$  because of \req{sourcestate} and \req{psourcestate}, we have
\alg{
&
H_{\rm min}^\epsilon(\hat{C}|\hat{B}\hat{R})_{\ca{C}(\Psi)}
\nn\\
&
=
H_{\rm min}^\epsilon(C|BRXYZ)_{\ca{C}(\Psi)}
\\
&
=
-H_{\rm max}^\epsilon(C|AXYZ)_{\Psi_s}
 }
 and
 \alg{
 &
H_{\rm min}^\epsilon(\hat{C}|\hat{A}\hat{R})_{\ca{C}(\Psi)}
\nn\\
&
=
H_{\rm min}^\epsilon(C|ARXYZ)_{\ca{C}(\Psi)}
\\&
=
-H_{\rm max}^\epsilon(C|BXYZ)_{\Psi_s}.
}
In addition, the isomorphism \req{yukue} implies
\alg{
&
\log{d_{C_1}}=e+e_0,
\;
\log{d_{C_2}}=e_0,
\\
&
\;
\log{d_{C_3}}=q,
\;
\log{d_{Z_R}}=c.
}
Substituting these relations to \req{brd1}-\req{brd4}, and noting that $d_C=d_{C_1}d_{C_2}d_{C_3}$, we arrive at
\alg{
c+q-e
&\geq
H_{\rm max}^\epsilon(CZ|AX)_{\Psi_s}-\log{\frac{\delta^2}{2}},
\laeq{direct1'}
\\
q-e
&\geq
H_{\rm max}^\epsilon(C|AXYZ)_{\Psi_s}-\log{\delta^2},
\\
c+q+e
&\geq
H_{\rm max}^\epsilon(CZ|BY)_{\Psi_s}-\log{\frac{\delta^2}{2}},
\\
q+e
&\geq
H_{\rm max}^\epsilon(C|BXYZ)_{\Psi_s}-\log{\delta^2}
}
and
$
q+e+2e_0
=\log{d_C}
$.
Combining these all together, we obtain the set of Ineqs.~\req{neon1}-\req{neon5} as a sufficient condition for the tuple $(c,q,e)$ to be achievable within the error $4\sqrt{12\epsilon+6\delta}$.
 \QED

\subsection{Proof of \rPrp{direct2} from \rPrp{direct}}
\lsec{prfthmdirect}

We prove \rPrp{direct2} based on  \rPrp{direct} by (i) modifying the first inequality \req{neon1},
and (ii) extending the rate region by incorporating teleportation and dense coding.

\subsubsection{Modification of Inequalities \req{neon1} and \req{matahi}}

We argue that the smooth conditional max entropy in the R.H.S. of Inequality \req{neon1} is modified to be $H_{\rm max}^\epsilon(C|AXZ)_{\Psi_s}$.
Consider a ``modified'' redistribution protocol as follows: 
In the beginning of the protocol, the sender prepares a copy of $Z$, which we denote by $\tilde{Z}$.
The sender then uses $X\tilde{Z}$ as the classical part of the side information, instead of $X$ alone, and apply the protocol presented in \rSec{consED}. 
The smooth max entropy corresponding to the first term in  \req{neon1} is then given by (see \rLmm{condmaxCQCQ})
\alg{
H_{\rm max}^\epsilon(CZ|AX\tilde{Z})_{\Psi_s}=H_{\rm max}^\epsilon(C|AXZ)_{\Psi_s}.
\laeq{dfnH1}
}
For the same reason, the term $H_{\rm max}^\epsilon(Z|AX)_{\Psi_s}$ in the condition \req{matahi} is modified to be $H_{\rm max}^\epsilon(Z|AX\tilde{Z})_{\Psi_s}$, which is no greater than zero (see \rLmm{onedimHminmax} and \rLmm{condmaxCQCQ}).
It should be noted that the entropies in the other three inequalities are unchanged by this modification.

\subsubsection{Extension of the rate region by Teleportation and Dense Coding}

To complete the proof of \rThm{direct}, we extend the achievable rate region given in \rPrp{direct} by incorporating teleportation and dense coding.
More precisely, we apply the following lemma that follows from teleportation and dense coding (see the next subsection for a proof):

\blmm{TPDCextension}
Suppose that a rate tuple $(\hat{c},\hat{q},\hat{e},\hat{e}_0)$ is achievable within the error $\delta$.
Then, for any $\lambda,\mu\geq 0$ and $e_0\geq0$ such that
\alg{
-\frac{\hat{c}}{2}\leq\lambda-\mu
\leq\hat{q},
\quad
\hat{e}_0
\leq
e_0,
}
the tuple $(c,q,e,e_0):=(\hat{c}+2\lambda-2\mu,\hat{q}-\lambda+\mu,\hat{e}+\lambda+\mu, e_0)$ is also achievable within the error $\delta$.
\elmm

\noindent
{\bf Proof of \rPrp{direct2}:}
Due to \rPrp{direct} and \rLmm{TPDCextension}, a tuple $(c,q,e,e_0)$ is achievable within the error $\delta$ if there exists $\lambda,\mu\geq 0$ and $\hat{e}_0\leq e_0$ such that the tuple 
\alg{
&
(\hat{c},\hat{q},\hat{e},\hat{e}_0)
:=
\nn\\
&\;
(c-2\lambda+2\mu,q+\lambda-\mu,e-\lambda-\mu,\hat{e}_0)
\!\!
\laeq{cqehat}
}
satisfies
\alg{
\hat{c}+\hat{q}-\hat{e}
&\geq
H_1,
\\
\hat{q}-\hat{e}
&\geq
H_2,
\\
\hat{c}+\hat{q}+\hat{e}
&\geq
H_3,
\\
\hat{q}+\hat{e}
&\geq
H_4,
\\
\hat{q}+\hat{e}+2\hat{e}_0
&=
\log{d_C}
}
and $\hat{c},\hat{q}\geq0$.
Here, we have denoted
the R.H.S.s of  Inequalities \req{neon2}-\req{neon4} by $H_2$, $H_3$ and $H_4$, respectively, and that of \req{dfnH1} by $H_1$.
Substituting \req{cqehat} to these inequalities yields
\alg{
c+q-e
&\geq
H_1-2\mu,
\laeq{neon1-4}
\\
q-e
&\geq
H_2-2\lambda,
\laeq{neon2-4}
\\
c+q+e
&\geq
H_3+2\lambda,
\laeq{neon3-4}
\\
q+e
&\geq
H_4+2\mu,
\laeq{neon4-4}
\\
q+e+2\hat{e}_0
&=
\log{d_C}+2\mu
\laeq{neon4-5}
}
and
\alg{
c-2\lambda+2\mu
&
\geq0,
\laeq{mois1}\\
q+\lambda-\mu
&\geq0.
\laeq{mois2}
}
Thus, it suffices to prove that, for any tuple $(c,q,e,e_0)$ satisfying Inequalities \req{neon00}-\req{neon05}, there exist $\hat{e}_0\leq e_0$ and $\lambda,\mu\geq 0$ such that the above inequalities hold. 
This is proved by noting that the inequality \req{neon00} is expressed as
\alg{
c+q+e-H_3
&
\geq
\max\{H_2,H_2'\}-q+e,
\laeq{dirprp1}\\
q+e-H_4
&
\geq
H_1-c-q+e,
\laeq{dirprp2}
}
where
\alg{
H_2':=H_{\rm min}^\epsilon(C|AXYZ)_{\Psi_s}-\log{\delta^2}.
}
The L.H.S. of \req{dirprp1} and \req{dirprp2} are nonnegative because of Inequalities \req{neon03} and \req{neon04}.
Thus, there exists $\lambda,\mu\geq0$ such that $2\lambda$ and $2\mu$ are in between both sides in \req{dirprp1} and \req{dirprp2}, respectively.
This implies \req{neon1-4}-\req{neon4-4}.
We particularly choose 
\alg{
\mu=\frac{1}{2}(q+e-H_4),
\quad
\hat{e}_0=
\frac{1}{2}(\log{d_C}-H_4).
}
A simple calculation leads to \req{neon4-5}.
Noting that $H_3\geq H_4$ by the data processing inequality, it follows from \req{neon3-4} that
\alg{
c-2\lambda\geq H_3-q-e \geq H_4-q-e=-2\mu,
}
which implies \req{mois1}.
Inequality \req{mois2} is obtained by combining \req{neon4-4} with
$
2\lambda\geq
\max\{H_2,H_2'\}-q+e
$.
Note that 
\alg{
&
H_2'+H_4
\nn\\
&
=
H_{\rm min}^\epsilon(C|AXYZ)_{\Psi_s}
\nn\\
&\quad\quad
+
H_{\rm max}^\epsilon(C|BXYZ)_{\Psi_s}
-2\log{\delta^2}
\\
&
=
H_{\rm min}^\epsilon(C|AXYZ)_{\Psi_s}
\nn\\
&\quad\quad
-
H_{\rm min}^\epsilon(C|ARXYZ)_{\Psi_s}
-2\log{\delta^2}
\\
&
\geq0,
}
where the third line follows from \rLmm{SE11d}.
This completes the proof of \rThm{direct}.
\QED

\subsubsection{Proof of \rLmm{TPDCextension} (see also Section IV in \cite{min08})}

We first consider the case where $\lambda-\mu\geq0$, and prove that the tuple $(c,q,e,e_0,\delta):=(\hat{c}+2\lambda-2\mu,\hat{q}-\lambda+\mu,\hat{e}+\lambda+\mu,\hat{e}_0,\delta)$ is achievable if a rate tuple $(\hat{c},\hat{q},\hat{e},\hat{e}_0,\delta)$ is achievable and $\hat{c},\hat{q}\geq0$.
Suppose that there exists a protocol $(\ca{E},\ca{D})$ with the classical communication cost $\hat{c}$, the quantum communication cost $\hat{q}$, the net entanglement cost  $\hat{e}$ and the catalytic entanglement cost $\hat{e}_0$ that achieves the state redistribution of the state $\Psi_s$ within the error $\delta$.
We construct a protocol $(\ca{E}',\ca{D}')$ such that the $\lambda-\mu$ qubits of quantum communication in the protocol $(\ca{E},\ca{D})$ is simulated by quantum teleportation, consuming $\lambda-\mu$ ebits of additional shared entanglement and $2\lambda-2\mu$ bits of classical communication.
The net costs of the resources are given by $\hat{c}+2\lambda-2\mu$, $\hat{q}-\lambda+\mu$, $\hat{e}+\lambda-\mu$ and the catalytic entanglement cost is $\hat{e}_0$, which implies achievability of the tuple $(\hat{c}+2\lambda-2\mu,\hat{q}-\lambda+\mu,\hat{e}+\lambda+\mu,\hat{e}_0,\delta)$.

Second, we consider the case where $\lambda-\mu\leq0$.
Suppose that there exists a protocol $(\ca{E},\ca{D})$ with the classical communication cost $\hat{c}$, the quantum communication cost $\hat{q}$ and the net entanglement cost  $\hat{e}$ that achieves the state redistribution of the state $\Psi_s$ within the error $\delta$.
We construct a protocol $(\ca{E}'',\ca{D}'')$ such that the $2\mu-2\lambda$ bits of classical communication in $(\ca{E},\ca{D})$ is simulated by dense coding, consuming $\mu-\lambda$ ebits of shared entanglement and $\mu-\lambda$ qubits of quantum communication.
The net costs of the resources are given by $\hat{c}-2\mu+2\lambda$, $\hat{q}+\mu-\lambda$, $\hat{e}+\mu-\lambda$ and the catalytic entanglement cost is $\hat{e}_0$, which implies achievability of the tuple $(\hat{c}-2\mu+2\lambda,\hat{q}+\mu-\lambda,\hat{e}+\mu+\lambda,\hat{e}_0,\delta)$.
\QED

\subsection{Proof of \rThm{direct} from \rPrp{direct2}}
\lsec{seru}

The achievability for the case of $d_C=1$ immediately follows from the condition \req{mongen} in \rPrp{direct2}. Thus, we only consider the case where $d_C\geq2$.

Let $\Pi$ be a projection onto a subspace $\ca{H}^{C_\Pi}\subseteq\ca{H}^C$ such that ${\rm dim}[\ca{H}^{C_\Pi}]=2^{H_{\rm max'}^{\epsilon^2/8}(C)_{\Psi_s}}$ and that ${\rm Tr}[\Pi\Psi_s^C]\geq1-\epsilon^2/8$.
Such a projection exists due to the definition of $H_{\rm max'}$ given by \req{dfnHrank}.
Consider the ``modified'' source state defined by
\alg{
\Psi_{s,\Pi}^{\hat{A}\hat{B}\hat{C}_\Pi\hat{R}}
:=
\Pi(\Psi_s^{\hat{A}\hat{B}\hat{C}\hat{R}})\Pi.
}
From the gentle measurement lemma (see \rLmm{gentlemeasurement} in \rApp{extUhlmann}), it holds that 
\alg{
P(\Psi_s,\Psi_{s,\Pi})
\leq
\frac{\epsilon}{2},
\quad
\|\Psi_s-\Psi_{s,\Pi}\|_1
\leq
\frac{\epsilon}{\sqrt{2}}.
\laeq{GMPsis}
}
Thus, due to the definitions of the smooth entropies \req{dfnmine} and \req{dfnmaxe}, we have
\alg{
H_{\rm  max}^{\epsilon/2}(CZ|BY)_{\Psi_s}
&
\geq
H_{\rm  max}^\epsilon(C_\Pi Z|BY)_{\Psi_{s,\Pi}}
\\
&
\geq
H_{\rm  max}^{3\epsilon/2}(CZ|BY)_{\Psi_s},
\\
H_{\rm  min}^{3\epsilon/2}(C|AXYZ)_{\Psi_s}
&
\geq
H_{\rm  min}^\epsilon(C_\Pi |AXYZ)_{\Psi_{s,\Pi}}
}
and so forth.

Suppose that the tuple $(c,q,e,e_0)$ satisfies Inequalities \req{neon00t}-\req{neon05t} in \rThm{direct}.
It follows that
\alg{
c+2q
&\geq
\max\{\tilde{H}_{I}^{(\epsilon,\epsilon)},\tilde{H}_{I\! I}^{(\epsilon)}\}_{\Psi_{s,\Pi}}-\log{(\delta^4/2)},
\laeq{neon00mod}
\\
c+q+e
&\geq
H_{\rm max}^\epsilon(C_\Pi Z|BY)_{\Psi_{s,\Pi}}-\log{(\delta^2/2)},
\laeq{neon03mod}
\\
q+e
&\geq
H_{\rm max}^\epsilon(C_\Pi |BXYZ)_{\Psi_{s,\Pi}}-\log{\delta^2},
\laeq{neon04mod}
\\
e_0
&\geq 
\frac{1}{2}(\log{d_{C_\Pi}}-H_{\rm max}^\epsilon(C_\Pi|BXYZ)_{\Psi_{s,\Pi}})
\nn\\
&\quad\quad\quad\quad\quad\quad\quad\quad\quad\quad+\log{\delta}.
\laeq{neon05mod}
}
Thus, due to \rPrp{direct2}, the tuple $(c,q,e,e_0)$ is achievable within an error $4\sqrt{12\epsilon+6\delta}$ for the state $\Psi_{s,\Pi}$.
That is, there exists 
a pair of an encoding CPTP map $\ca{E}_\Pi^{\hat{A}\hat{C}_\Pi E_A\rightarrow \hat{A}QMF_A}$ and a decoding CPTP map $\ca{D}_\Pi^{\hat{B}QME_B\rightarrow \hat{B}\hat{C}_\Pi F_B}$, such that
\alg{
&
\left\|
\ca{D}_\Pi\circ\ca{E}_\Pi(\Psi_{s,\Pi}^{\hat{A}\hat{B}\hat{C}_\Pi\hat{R}}\otm\Phi_{2^{e+e_0}}^{E_AE_B})
\right.
\nn\\
&\quad
\left.-\Psi_{s,\Pi}^{\hat{A}\hat{B}\hat{C}_\Pi\hat{R}}\otm\Phi_{2^{e_0}}^{F_AF_B}
\right\|_1
\leq
4\sqrt{12\epsilon+6\delta}.
\laeq{gcodeforpsipi}
}
Define an encoding map $\ca{E}^{\hat{A}\hat{C} E_A\rightarrow \hat{A}QMF_A}$ and a decoding map $\ca{D}^{\hat{B}QME_B\rightarrow \hat{B}\hat{C} F_B}$ for the state $\Psi_s$ by
\alg{
&
\ca{E}^{\hat{A}\hat{C} E_A\rightarrow \hat{A}QMF_A}
(\tau)
\nn\\
&\quad
=
\ca{E}_\Pi(\Pi^C\tau\Pi^C)
+{\rm Tr}[(I^C-\Pi^C)\tau]\xi_0,
}
where $\xi_0$ is an arbitrary fixed state on $\hat{A}QMF_A$,
and $\ca{D}=\ca{D}_\Pi$.
Note that the system $C_\Pi$ is naturally embedded into $C$.
By the triangle inequality, we have
\alg{
&
\left\|
\ca{D}\circ\ca{E}(\Psi_{s}^{\hat{A}\hat{B}\hat{C}\hat{R}}\otm\Phi_{2^{e+e_0}}^{E_AE_B})-\Psi_{s}^{\hat{A}\hat{B}\hat{C}\hat{R}}\otm\Phi_{2^{e_0}}^{F_AF_B}
\right\|_1
\nn\\
&
\leq
\left\|
\ca{D}\circ\ca{E}(\Psi_{s}^{\hat{A}\hat{B}\hat{C}\hat{R}}\otm\Phi_{2^{e+e_0}}^{E_AE_B})
\right.
\nn\\
&
\quad\quad\quad\quad
\left.-\ca{D}\circ\ca{E}(\Psi_{s,\Pi}^{\hat{A}\hat{B}\hat{C}\hat{R}}\otm\Phi_{2^{e+e_0}}^{E_AE_B})
\right\|_1
\nn\\
&\quad
+
\left\|
\ca{D}\circ\ca{E}(\Psi_{s,\Pi}^{\hat{A}\hat{B}\hat{C}\hat{R}}\otm\Phi_{2^{e+e_0}}^{E_AE_B})\right.
\nn\\
&\quad\quad\quad\quad\quad\quad\quad\quad\left.
-\Psi_{s,\Pi}^{\hat{A}\hat{B}\hat{C}\hat{R}}\otm\Phi_{2^{e_0}}^{F_AF_B}
\right\|_1
\nn\\
&
\quad
+
\left\|
\Psi_{s,\Pi}^{\hat{A}\hat{B}\hat{C}\hat{R}}\otm\Phi_{2^{e_0}}^{F_AF_B}-\Psi_{s}^{\hat{A}\hat{B}\hat{C}\hat{R}}\otm\Phi_{2^{e_0}}^{F_AF_B}
\right\|_1
\\
&
\leq
\left\|
\ca{D}_\Pi\circ\ca{E}_\Pi(\Psi_{s,\Pi}^{\hat{A}\hat{B}\hat{C}\hat{R}}\otm\Phi_{2^{e+e_0}}^{E_AE_B})\right.\nn\\
&
\quad\quad\quad\quad\quad\quad\quad\quad
\left.
-\Psi_{s,\Pi}^{\hat{A}\hat{B}\hat{C}\hat{R}}\otm\Phi_{2^{e_0}}^{F_AF_B}
\right\|_1
\nn\\
&\quad\quad
+
2
\left\|
\Psi_{s}^{\hat{A}\hat{B}\hat{C}\hat{R}}-\Psi_{s,\Pi}^{\hat{A}\hat{B}\hat{C}\hat{R}}
\right\|_1
\laeq{monn}\\
&\leq
4\sqrt{12\epsilon+6\delta}
+
\sqrt{2}
\epsilon,
}
Here, Inequality \req{monn} follows from $\ca{D}_\Pi\circ\ca{E}_\Pi(\Psi_{s,\Pi}^{\hat{A}\hat{B}\hat{C}\hat{R}}\otm\Phi_{2^{e+e_0}}^{E_AE_B})=\ca{D}\circ\ca{E}(\Psi_{s,\Pi}^{\hat{A}\hat{B}\hat{C}\hat{R}}\otm\Phi_{2^{e+e_0}}^{E_AE_B})$ and the monotonicity of the trace distance, and
the last line from \req{GMPsis} and \req{gcodeforpsipi}.
Hence, the tuple $(c,q,e,e_0)$ is achievable within an error $4\sqrt{12\epsilon+6\delta}+\sqrt{2}\epsilon$ for the state $\Psi_s$,
which completes the proof of \rThm{direct}.
\QED

\begin{figure*}[t]
\begin{center}
\includegraphics[bb={0 20 1033 506}, scale=0.32]{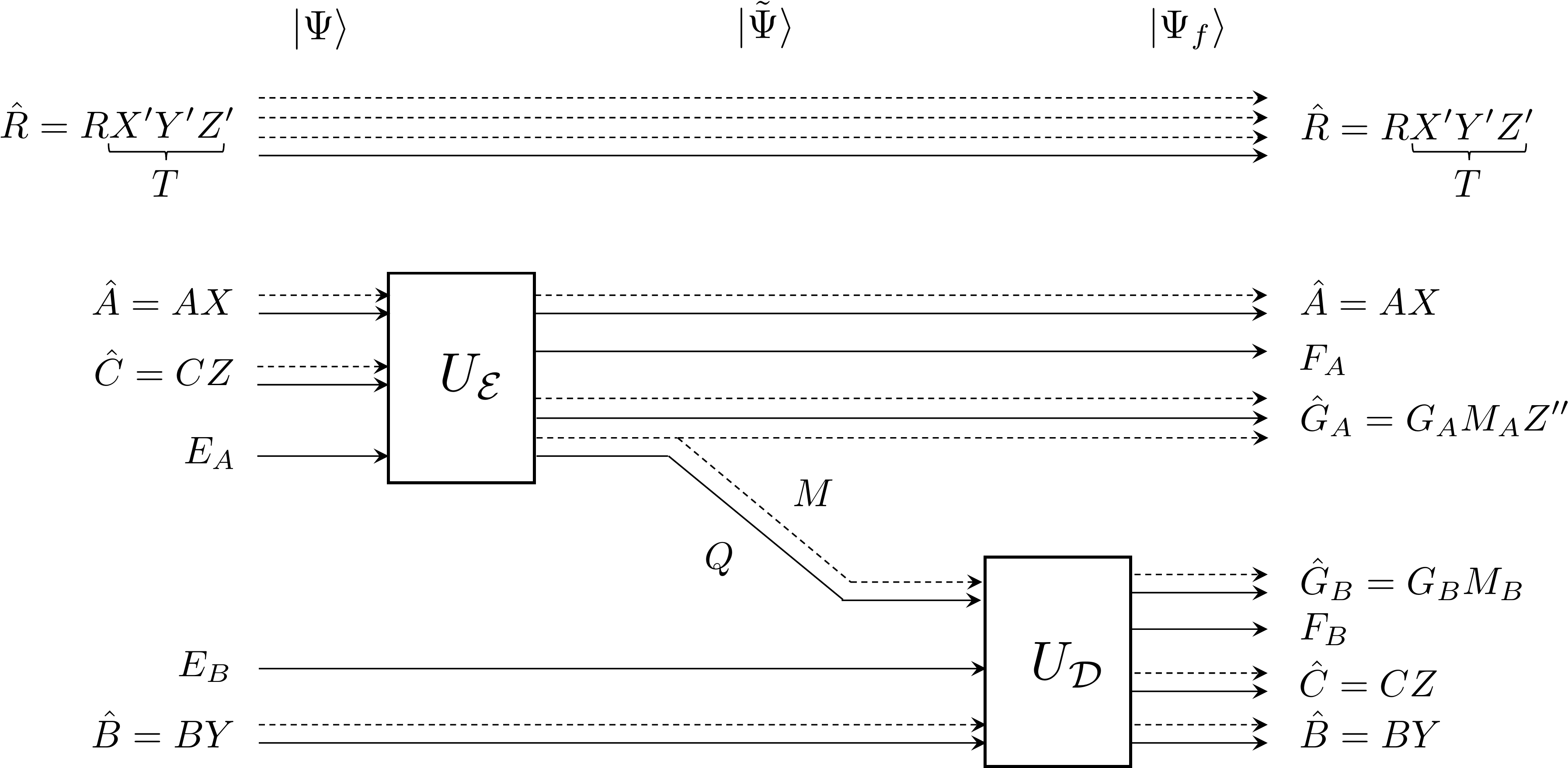}
\end{center}
\caption{
The purified picture of the task is depicted in the diagram. 
The black lines and the dashed lines represent classical and quantum systems, respectively.
}
\label{fig:C}
\end{figure*}

\section{Proof of The Converse Part\\ (\rThm{converse} and \rLmm{propDelta})}
\lsec{converse}

We prove the one-shot converse bound
(\rThm{converse}).
The proof proceeds as follows:
First, we construct quantum states that describe the state transformation in a redistribution protocol in a ``purified picture''.
Second, we prove four entropic inequalities that hold for those states.
Finally, we prove that the four inequalities imply the three inequalities in \rThm{converse}, thereby completing the proof of the converse bound.
We also analyze the properties of the function $\Delta^{(\epsilon,\delta)}$, and prove \rLmm{propDelta}.

\subsection{Construction of States}

Let $U_\ca{E}^{\hat{A}\hat{C}E_A\rightarrow \hat{A}QMF_A\hat{G}_A}$ and $U_\ca{D}^{\hat{B}QME_B\rightarrow \hat{B}\hat{C}F_B\hat{G}_B}$ be the Stinespring dilations of the encoding operation $\ca{E}$ and the decoding operation $\ca{D}$, respectively, i.e., 
\alg{
\ca{E}={\rm Tr}_{\hat{G}_A}\circ\ca{U}_\ca{E},
\quad
\ca{D}={\rm Tr}_{\hat{G}_B}\circ\ca{U}_\ca{D}.
}
We define the ``purified'' source state  $\ket{\Psi}$ by
\alg{
&
\ket{\Psi}^{ABCRXYZT}:=
 \nn\\
&
\quad\quad
\sum_{x,y,z}\sqrt{p_{xyz}}
\ket{x}^X\ket{y}^Y\ket{z}^Z
\ket{\psi_{xyz}}^{ABCR}\ket{xyz}^T,
\laeq{psourcestateeee}
}
and consider the states
\alg{
&
\ket{\tilde{\Psi}}^{\hat{A}QMF_A\hat{G}_A\hat{B}\hat{R}E_B}:=U_\ca{E}\ket{\Psi}^{\hat{A}\hat{B}\hat{C}\hat{R}}\ket{\Phi_{2^{e+e_0}}}^{E_AE_B},
\laeq{dfnOmega} \\
&
\ket{\Psi_f}^{\hat{A}\hat{B}\hat{C}\hat{R}F_AF_B\hat{G}_A\hat{G}_B}:=U_\ca{D}\ket{\tilde{\Psi}}.
\laeq{dfnPsif}
}
The state $\tilde{\Psi}$ is a purification of the state after the encoding operation,  and $\Psi_f$ is the one after the decoding operation.
See Figure \ref{fig:C} for the diagram.

\begin{figure*}[t]
\begin{center}
\includegraphics[bb={0 20 1417 256}, scale=0.3]{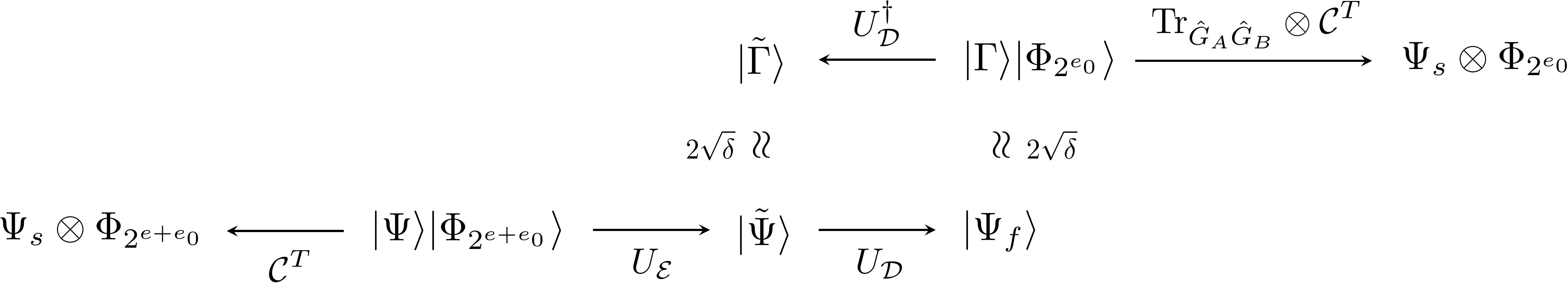}
\end{center}
\caption{
Relations among the states $\tilde{\Psi}$, $\tilde{\Gamma}$, $\Psi$, $\Gamma$ and $\Psi_s$ are depicted. 
}
\label{fig:D}
\end{figure*}

Due to the relation \req{relTDPD} between the trace distance and the purified distance,
the condition \req{qeec} implies that
\alg{
P
\left(
\ca{C}^T(\Psi_f)^{\hat{A}\hat{B}\hat{C}\hat{R}F_AF_B},\Psi_s^{\hat{A}\hat{B}\hat{C}\hat{R}}\otm\Phi_{2^{e_0}}^{F_AF_B}
\right)
\leq
2\sqrt{\delta},
\laeq{qeed}
}
with $\ca{C}^T$ being the completely dephasing operation on $T$ with respect to the basis $\{\ket{xyz}\}$.
Due to an extension of Uhlmann's theorem (see \rLmm{extUhlmann1} in \rApp{extUhlmann}), there exists a pure state $|\Gamma\rangle^{\hat{A}\hat{B}\hat{C}\hat{G}_A\hat{G}_B\hat{R}}$, which is represented in the form of 
\alg{
|\Gamma\rangle
&=\sum_{x,y,z}\sqrt{p_{xyz}}\ket{x}^X\ket{y}^Y\ket{z}^Z
\nn\\
&\quad\quad
\ket{\psi_{xyz}}^{ABCR}
\ket{\phi_{xyz}}^{\hat{G}_A\hat{G}_B}\ket{xyz}^T, 
\laeq{dfntipsi}
} 
such that 
\begin{eqnarray}
P\left(
\Psi_f^{\hat{A}\hat{B}\hat{C}\hat{R}F_AF_B\hat{G}_A\hat{G}_B},
\Gamma^{\hat{A}\hat{B}\hat{C}\hat{G}_A\hat{G}_B\hat{R}}\otm\Phi_{2^{e_0}}^{F_AF_B}
\right)
\nn\\
\leq
2\sqrt{\delta}.
\quad\quad
\laeq{qeef}
\end{eqnarray}
Using this state, we define
\alg{
&
\ket{\tilde{\Gamma}}^{\hat{A}QMF_A\hat{G}_A\hat{B}\hat{R}E_B}
\nn\\
&\quad
:=
U_{\ca{D}}^\dagger
\ket{\Gamma}^{\hat{A}\hat{B}\hat{C}\hat{G}_A\hat{G}_B\hat{R}}\ket{\Phi_{2^{e_0}}}^{F_AF_B}.
\laeq{dfntildePsi}
}
Due to the isometric invariance of the purified distance, it follows from \req{qeef} and \req{dfnPsif} that
\alg{
P\left(
\tilde{\Psi}^{\hat{A}QMF_A\hat{G}_A\hat{B}\hat{R}E_B},
\tilde{\Gamma}^{\hat{A}QMF_A\hat{G}_A\hat{B}\hat{R}E_B}
\right)
\leq
2\sqrt{\delta}.
\laeq{qeef2}
}
Relations among the states defined as above are depicted in Figure \ref{fig:D}.
Some useful properties of these states are presented in the following, and will be used in the proof of the converse part.

\subsubsection{Decomposition of $U_{\ca{E}}$ and $U_{\ca{D}}$}

Since $M$ is a classical system, we may, without loss of generality, assume that $U_\ca{E}$ and $U_\ca{D}$ are decomposed as
\alg{
&
U_\ca{E}
=
\sum_m
\ket{m}^M\ket{m}^{M_A}
\!\otm
v_{m}^{\hat{A}\hat{C}E_A\rightarrow \hat{A}QF_A\tilde{G}_A},
\laeq{dfnUE}\\
&
U_{\ca{D}}
=
\sum_m\ket{m}^{M_B}\bra{m}^M\!\otm u_m^{\hat{B}QE_B\rightarrow \hat{B}\hat{C}F_BG_B}
\!.\!
\laeq{dfnUD}
}
Here, $M_A$ and $M_B$ are quantum systems isomorphic to $M$ with the fixed orthonormal basis $\{\ket{m}\}_m$, the operators $u_m$ are linear isometries, and 
$\hat{G}_A$ and $\hat{G}_B$ are such that $\hat{G}_A\equiv \tilde{G}_AM_A$ and $\hat{G}_B\equiv G_BM_B$.
It follows that
\alg{
U_{\ca{D}}\circ U_{\ca{E}}
=
\sum_m
\ket{m}^{M_A}\ket{m}^{M_B}\otm(u_m\circ v_m).
\laeq{dfnUDE}
}
Since $Z$ is a classical system, we may further assume that $v_m$ are decomposed as
\alg{
v_{m}
:=
\sum_z
\ket{z}^{Z''}\bra{z}^{Z}
\otm
v_{m,z}^{\hat{A}CE_A\rightarrow \hat{A}QF_AG_A},
\laeq{dfnUDE2}
}
where $Z''$ is a system isomorphic to $Z$ with the fixed orthonormal basis $\{\ket{z}\}_z$ and $\tilde{G}_A\equiv G_AZ''$. 
The operators $v_{m,z}$ are linear operators such that $\sum_m v_{m,z}^\dagger v_{m,z}=I$ for all $z$.
It should be noted that $\hat{G}_A=G_AM_AZ''$.

\subsubsection{Properties of $\tilde{\Psi}$ and $\Psi_f$}

Since $\ket{\tilde{\Psi}}$ is defined as \req{dfnOmega} by $U_{\ca{E}}$ that is in the form of \req{dfnUE}, it is decomposed into
\alg{
\ket{\tilde{\Psi}}
=
\sum_m\sqrt{q_m}\ket{m}^{M}\ket{m}^{M_A}\ket{\tilde{\Psi}_{m}},
}
with some probability distribution $\{q_m\}_m$ and pure states $\{\ket{\tilde{\Psi}_{m}}\}_m$.
Thus, we have
\alg{
\ca{C}^M(\tilde{\Psi})
=
\sum_{m}q_{m}\proj{m}^M\!\otm\!\proj{m}^{M_A}\!\otm\!\proj{\tilde{\Psi}_{m}},
\laeq{tildemsimmz2}
}
where
$\ca{C}^M$ is the completely dephasing operation on $M$ with respect to the basis $\{\ket{m}\}_m$.
Similarly, due to \req{dfnUDE}, \req{dfnOmega} and \req{dfnPsif}, the state  $\ket{\Psi_f}$ is decomposed into
\alg{
\ket{\Psi_f}
=
\sum_{m}\sqrt{q_m}\ket{m}^{M_A}\ket{m}^{M_B}\ket{\Psi_{f,m}}.
\laeq{tildemsimmz}
}
From \req{dfnUDE2}, it holds that $\bra{z_1}^{Z'}\bra{z_2}^{Z''}\ket{\Psi_f}\propto\delta_{z_1,z_2}$.
Thus, the states $\ket{\Psi_{f,m}}$ are further decomposed into
\alg{
\ket{\Psi_{f,m}}
=
\sum_{z}\sqrt{q_{z|m}}\ket{z}^{Z''}\ket{\Psi_{f,m,z}}\ket{z}^{Z'}.
\laeq{tildemsimmz3}
}

\subsubsection{Properties of $\Gamma$}

From the definition \req{dfntipsi}, it follows that
\alg{
&
\ca{C}^T(\Gamma)
=
\nn\\
&\quad
\sum_{x,y,z}p_{xyz}\proj{xyz}^{XYZ}
\otm
\proj{\psi_{xyz}}^{ABCR}
\nn\\
&\quad\quad\quad
\otm
\proj{\phi_{xyz}}^{\hat{G}_A\hat{G}_B}
\otm
\proj{xyz}^T
\laeq{dfntipsiii}
}
and that
\alg{
{\rm Tr}_T(\Gamma)
&
=
\sum_{x,y,z}p_{xyz}\proj{xyz}^{XYZ}
\nn\\
&\quad
\otm
\proj{\psi_{xyz}}^{ABCR}
\otm
\proj{\phi_{xyz}}^{\hat{G}_A\hat{G}_B}.
\laeq{dfntipsiii2}
}
Both states are ensembles of pure states on $ABCR\hat{G}_A\hat{G}_B$, classically labelled by $xyz$ on $XYZ$ or $T$, that are decoupled between $ABCR$ and $\hat{G}_A\hat{G}_B$.
It follows from \req{dfntipsiii} that
\alg{
{\rm Tr}_{\hat{G}_A\hat{G}_B}\otm\ca{C}^T(\Gamma)
=
\Psi_s^{\hat{A}\hat{B}\hat{C}\hat{R}}.
\laeq{PTgammaf}
}

Due to \req{tildemsimmz}, \req{tildemsimmz3} and \rLmm{extUhlmann2} in \rApp{extUhlmann}, we may, without loss of generality, assume that $\ket{\phi_{xyz}}$ is in the form of
\alg{
\ket{\phi_{xyz}}^{\hat{G}_A\hat{G}_B}
=
\ket{\phi_{xyz}'}^{G_AM_A\hat{G}_B}\ket{z}^{Z''}
\laeq{mardock}
}
and
\alg{
&
\ket{\phi_{xyz}'}^{G_AM_A\hat{G}_B}
\nn\\
&
\quad
:=
\sum_m
\sqrt{p_{m|xyz}}\ket{m}^{M_A}\ket{m}^{M_B}\ket{\phi_{m,xyz}}^{G_AG_B}.
}
Substituting this to \req{dfntipsiii}, we have
\alg{
&
\ca{C}^T(\Gamma)^{AG_AM_AXYZT}
\nn\\
&
=\sum_{x,y,z}p_{xyz}\proj{xyz}^{XYZ}
\otm
\proj{z}^{Z''}
\nn\\
&
\quad\quad
\otm
\psi_{xyz}^{A}
\otm
\phi_{xyz}^{G_AM_A}
\otm
\proj{xyz}^T
\!.\!
\laeq{fstar}
}
Thus, the state $\ca{C}^T(\Gamma)$ given by  is classically coherent in $ZZ''$.
Denoting $p_{xyz}p_{m|xyz}$ by $p_{m,xyz}$, it follows from \req{dfntipsiii} that
\alg{
&
\ca{C}^T\circ\ca{C}^{M_A}(\Gamma^{AG_AM_A\hat{G}_BT})
\nn\\
&
=\sum_{x,y,z}p_{m,xyz}
\:
\psi_{xyz}^{A}
\otm
\proj{m}^{M_A}\otm\proj{m}^{M_B}
\nn\\
&\quad\quad\quad\quad
\otm
\proj{\phi_{m,xyz}}^{G_AG_B}
\otm
\proj{xyz}^T,
\laeq{dfntipsiii55}
}
with $\ca{C}^{M_A}$ being the completely dephasing operation on $M_A$ with respect to the basis $\{\ket{m}\}_m$.
It should also be noted that
\alg{
&
\Gamma^{AG_AM_AXYZ}
\nn\\
&
\quad
=\sum_{x,y,z}p_{m,xyz}
\proj{m}^{M_A}
\!\otm\!
\psi_{xyz}^{A}
\nn\\
&
\quad\quad\quad\quad
\otm
\phi_{m,xyz}^{G_A}
\!\otm\!
\proj{xyz}^{XYZ}.
\laeq{sstar}
}

\subsection{Inequalities for Proving  \rThm{converse}}

As an intermediate goal for the proof of \rThm{converse}, we prove that the following four inequalities hold for the states $\Psi_s$ and $\Gamma$ defined by \req{sourcestate} and \req{dfntipsi}, respectively:
\alg{
&
c+q-e
\geq
H_{\rm min}^{\epsilon}(AXCZ)_{\Psi_s}
-
H_{\rm max}^{\epsilon}(AXZ)_{\Psi_s}
\nn\\
&\quad\quad\quad\quad
-H_{\rm min}^{7\epsilon+2\sqrt{\delta}}(G_A|M_AAXZ)_{\Gamma}-4f(\epsilon),
\laeq{convv1}
\\
&
q-e
\geq
H_{\rm min}^{\epsilon}(AC|XYZ)_{\Psi_s}
-
H_{\rm max}^{\epsilon}(A|XYZ)_{\Psi_s}
\nn\\
&\quad\quad\quad\quad
-
H_{\rm min}^{5\epsilon+2\sqrt{\delta}}(G_AM_A|XYZ)_{\Gamma}
-3f(\epsilon),
\laeq{convv000}
\\
&
c+q+e
\geq
H_{\rm min}^{\epsilon}(BYCZ)_{\Psi_s}
-
H_{\rm min}^{12\epsilon+6\sqrt{\delta}}(BY)_{\Psi_s}
\nn\\
&\quad\quad\quad\quad
+
H_{\rm min}^{5\epsilon+2\sqrt{\delta}}(G_AM_A|XYZ)_{\Gamma}
-f(\epsilon),
\laeq{convv3}
\\
&
q+e
\geq
H_{\rm min}^{\epsilon}(BC|XYZ)_{\Psi_s}
-H_{\rm min}^{11\epsilon+8\sqrt{\delta}}(B|XYZ)_{\Psi_s}
\nn\\
&\quad\quad\quad\quad
+H_{\rm min}^{7\epsilon+6\sqrt{\delta}}(G_A|M_AAXYZ)_{\Gamma}
-2f(\epsilon),
\laeq{convv4}
}
where
$
f(x):=-\log{(1-\sqrt{1-x^2})}
$.
The proof of these inequalities will be given in the following subsections.
We will extensively use the properties of the smooth conditional entropies, which are summarized in \rApp{propSmEn}.

\subsubsection{Proof of Inequality \req{convv1}}

We start with
\alg{
&
e+e_0+H_{\rm min}^{\epsilon}(AXCZ)_{\Psi_s}
\nn\\
&
=e+e_0+H_{\rm min}^{\epsilon}(AXCZ)_{\Psi}
\laeq{convmod4-0}\\
&\leq
H_{\rm min}^{\epsilon}(AXCZE_A)_{\Psi_s\otm\Phi_{2^{e+e_0}}}
\laeq{convmod4-1}\\
&=
H_{\rm min}^{\epsilon}(AXF_A\hat{G}_AQM)_{\tilde{\Psi}}
\laeq{convmod4-2}\\
&\leq
H_{\rm max}^{\epsilon}(QM)
+
H_{\rm min}^{4\epsilon}(AXF_A\hat{G}_A|QM)_{\tilde{\Psi}}
\nn\\
&\quad\quad\quad\quad\quad\quad\quad\quad\quad\quad\quad\quad\quad\quad
+2f(\epsilon)
\laeq{convmod4-3}\\
&\leq
c+q
+
H_{\rm min}^{4\epsilon}(AXF_A\hat{G}_A|M)_{\tilde{\Psi}}+2f(\epsilon),
\laeq{convmod4-3-2}
}
where \req{convmod4-0} follows from $\Psi_s^{\hat{A}\hat{C}}=\Psi^{\hat{A}\hat{C}}$;
\req{convmod4-1} from the superadditivity of the smooth conditional min entropy for product state (\rLmm{SE3});
\req{convmod4-2} from the fact that $\ket{\tilde{\Psi}}$ is obtained  from $\ket{\Psi}\ket{\Phi_{2^{e+e_0}}}$ by an isometry $U_{\ca{E}}$ as \req{dfnOmega}, under which the smooth conditional entropy is invariant (\rLmm{invCEiso});
\req{convmod4-3} from the chain rule \req{CRminmaxmin};
and \req{convmod4-3-2} from the dimension bound (\rLmm{SE4}).

The third term in \req{convmod4-3-2} is further calculated as
\alg{
&
H_{\rm min}^{4\epsilon}(AXF_A\hat{G}_A|M)_{\tilde{\Psi}}
\laeq{convmod4-4}\\
&\leq
H_{\rm min}^{4\epsilon}(AXF_A\hat{G}_A|M)_{\ca{C}^M(\tilde{\Psi})}
\laeq{convmod4-5}\\
&=
H_{\rm min}^{4\epsilon}(AXZ''F_AG_AM_A|M)_{\ca{C}^M(\tilde{\Psi})}
\laeq{convmod4-5-2}\\
&=
H_{\rm min}^{4\epsilon}(AXZ''F_AG_A|M_A)_{\tilde{\Psi}}
\laeq{convmod4-6}\\
&\leq
H_{\rm min}^{4\epsilon+2\sqrt{\delta}}(AXZ''F_AG_A|M_A)_{\tilde{\Gamma}}
\laeq{convmod4-7}\\
&=
H_{\rm min}^{4\epsilon+2\sqrt{\delta}}(AXZ''F_AG_A|M_A)_{\Gamma\otm\Phi_{2^{e_0}}}
\laeq{convmod4-8}\\
&\leq
H_{\rm min}^{4\epsilon+2\sqrt{\delta}}(AXZ''G_A|M_A)_{\Gamma}+e_0
\laeq{convmod4-9}\\
&=
H_{\rm min}^{4\epsilon+2\sqrt{\delta}}(AXZG_A|M_A)_{\Gamma}+e_0
\laeq{convmod4-9-2}\\
&\leq
H_{\rm max}^{\epsilon}(AXZ|M_A)_{\Gamma}
\nn\\
&\quad\quad
+H_{\rm min}^{7\epsilon+2\sqrt{\delta}}(G_A|M_AAXZ)_{\Gamma}+e_0+2f(\epsilon)
\laeq{convmod4-10}\\
&\leq
H_{\rm max}^{\epsilon}(AXZ)_{\Gamma}
\nn\\
&\quad\quad
+H_{\rm min}^{7\epsilon+2\sqrt{\delta}}(G_A|M_AAXZ)_{\Gamma}+e_0+2f(\epsilon)
\laeq{convmod4-11}\\
&=
H_{\rm max}^{\epsilon}(AXZ)_{\Psi_s}
\nn\\
&\quad\quad
+H_{\rm min}^{7\epsilon+2\sqrt{\delta}}(G_A|M_AAXZ)_{\Gamma}+e_0+2f(\epsilon).
\laeq{convmod4-12}
}
Here, \req{convmod4-5} follows from the monotonicity of the smooth conditional entropy (\rLmm{Hminmonotonicity});
\req{convmod4-5-2} from $\hat{G}_A\equiv G_AM_AZ''$;
\req{convmod4-6} from \rLmm{condminCQCQ} and the fact that $M_A$ is a classical copy of $M$ as \req{tildemsimmz2};
\req{convmod4-7} from the continuity of the smooth conditional entropy (\rLmm{SE12}) and the fact that $\tilde{\Gamma}$ and  $\tilde{\Psi}$ are $2\sqrt{\delta}$-close with each other as \req{qeef2};
\req{convmod4-8} from the fact that $\tilde{\Gamma}$ is converted to $\Gamma$ by $U_{\ca{D}}$ as \req{dfntildePsi}, which does not change the reduced state on $AXZ''F_AG_AM_A$;
\req{convmod4-9} from the dimension bound (\rLmm{SE4});
\req{convmod4-9-2} from the fact that $Z''$ is a classical copy of $Z$, due to \req{fstar};
\req{convmod4-10} from the chain rule \req{CRminmaxmin};
\req{convmod4-11} from the fact that conditioning reduces the entropy due to the monotonicity of the smooth conditional entropy (\rLmm{Hminmonotonicity});
and \req{convmod4-12} from the fact that $\Gamma^{AXZ}=\Psi_s^{AXZ}$.

Combining these inequalities, we obtain
\alg{
&
e+e_0+H_{\rm min}^{\epsilon}(AXCZ)_{\Psi_s}
\nn\\
&
\leq
c+q+2f(\epsilon)
+
H_{\rm max}^{\epsilon}(AXZ)_{\Psi_s}
\nn\\
&\quad
+H_{\rm min}^{7\epsilon+2\sqrt{\delta}}(G_A|M_AAXZ)_{\Gamma}+e_0+2f(\epsilon),
}
which implies \req{convv1}.

\subsubsection{Proof of Inequality \req{convv000}}

We have
\alg{
&
e_0+H_{\rm min}^{2\epsilon+2\sqrt{\delta}}(\hat{A}\hat{G}_A|T)_{\ca{C}^T(\Gamma)}
\laeq{convmod1-0}\\
&
=
e_0+H_{\rm min}^{2\epsilon+2\sqrt{\delta}}(\hat{B}\hat{C}R\hat{G}_B|T)_{\ca{C}^T(\Gamma)}
\laeq{convmod1-2}\\
&
\geq
H_{\rm min}^{2\epsilon+2\sqrt{\delta}}(\hat{B}\hat{C}RF_B\hat{G}_B|T)_{\ca{C}^T(\Gamma)\otm\Phi_{2^{e_0}}}
\laeq{convmod1-3}\\
&
=
H_{\rm min}^{2\epsilon+2\sqrt{\delta}}(\hat{B}RE_BQM|T)_{\ca{C}^T(\tilde{\Gamma})}
\laeq{convmod1-4}\\
&
\geq
H_{\rm min}^{\epsilon+2\sqrt{\delta}}(\hat{B}RE_BM|T)_{\ca{C}^T(\tilde{\Gamma})}
\nn\\
&
\quad\quad
+
H_{\rm min}(Q|\hat{B}RE_BMT)_{\ca{C}^T(\tilde{\Gamma})}
-f(\epsilon)
\laeq{convmod1-5}\\
&
\geq
H_{\rm min}^{\epsilon+2\sqrt{\delta}}(\hat{B}RE_BM|T)_{\ca{C}^T(\tilde{\Gamma})}
-
q
-f(\epsilon).
\laeq{convmod1-6}
}
Here,
\req{convmod1-2} is from the fact that $\Gamma$ is a pure state on $\hat{A}\hat{B}\hat{C}\hat{R}\hat{G}_A\hat{G}_B$ as \req{dfntipsi}, which is transformed by $\ca{C}^T$ to an ensemble of classically-labelled pure states, to which \rLmm{SE11} is applicable;
\req{convmod1-3} from the dimension bound (\rLmm{SE4});
\req{convmod1-4} from the fact that $\tilde{\Gamma}$ is obtained from $\Gamma\otm\Phi_{2^{e_0}}$ by an isometry as \req{dfntildePsi} under which the smooth conditional entropy is invariant (\rLmm{invCEiso});
\req{convmod1-5} from the chain rule \req{CRminminmin};
and \req{convmod1-6} from the dimension bound (\rLmm{SE2}).

The first term in \req{convmod1-6} is further calculated to be
\alg{
&
H_{\rm min}^{\epsilon+2\sqrt{\delta}}(\hat{B}RE_BM|T)_{\ca{C}^T(\tilde{\Gamma})}
\\
&\geq
H_{\rm min}^{\epsilon}(\hat{B}RE_BM|T)_{\ca{C}^T(\tilde{\Psi})}
\laeq{convmod2-2}\\
&=
H_{\rm min}^{\epsilon}(\hat{A}F_A\hat{G}_AQ|T)_{\ca{C}^T(\tilde{\Psi})}
\laeq{convmod2-3}\\
&=
H_{\rm min}^{\epsilon}(\hat{A}F_A\hat{G}_AQM|T)_{\ca{C}^T\otm\ca{C}^M(\tilde{\Psi})}
\laeq{convmod2-4}\\
&\geq
H_{\rm min}^{\epsilon}(\hat{A}F_A\hat{G}_AQM|T)_{\ca{C}^T(\tilde{\Psi})}
\laeq{convmod2-5}\\
&=
H_{\rm min}^{\epsilon}(\hat{A}\hat{C}E_A|T)_{\ca{C}^T(\Psi)\otm\Phi_{e+e_0}}
\laeq{convmod2-6}\\
&\geq
H_{\rm min}^{\epsilon}(\hat{A}\hat{C}|T)_{\ca{C}^T(\Psi)}+e+e_0
\laeq{convmod2-7}\\
&=
H_{\rm min}^{\epsilon}(AC|XYZ)_{\Psi_s}+e+e_0.
\laeq{convmod2-8}
}
Inequality \req{convmod2-2} is from the continuity of the smooth conditional entropy (\rLmm{SE12}) and the fact that $\tilde{\Gamma}$ and  $\tilde{\Psi}$ are $2\sqrt{\delta}$-close with each other as \req{qeef2};
\req{convmod2-3} from \rLmm{SE11} and the fact that $\tilde{\Psi}$ is a pure state on $\hat{A}\hat{B}\hat{R}QMF_A\hat{G}_AE_B$ as \req{dfnOmega}, which is transformed by $\ca{C}^T$ to an ensemble of classically-labelled pure states;
\req{convmod2-4} from $\hat{G}_A=G_AM_AZ''$ and the fact that $M$ is a classical copy of $M_A$ as \req{tildemsimmz2};
\req{convmod2-5} from the monotonicity of the smooth conditional min entropy under unital maps (\rLmm{Hminmonotonicity});
\req{convmod2-6} from the isometric invariance of the smooth conditional entropy (\rLmm{invCEiso}) and the fact that $\tilde{\Psi}$ is obtained by an isometry $U_{\ca{E}}$ from $\Psi$ as \req{dfnOmega};
\req{convmod2-7} from the superadditivity of the smooth conditional entropy (\rLmm{SE3});
and \req{convmod2-8} from $\ca{C}^T(\Psi)=\Psi_s$ and the property of the smooth conditional entropy for CQ states (\rLmm{condminCQCQ}).

The second term in \req{convmod1-0} is bounded as
\alg{
&
H_{\rm min}^{2\epsilon+2\sqrt{\delta}}(\hat{A}\hat{G}_A|T)_{\ca{C}^T(\Gamma)}
\nn\\
&=
H_{\rm min}^{2\epsilon+2\sqrt{\delta}}(AG_AM_A|XYZ)_{\Gamma}
\laeq{convmod3-1}\\
&\leq
H_{\rm max}^{\epsilon}(A|XYZ)_{\Gamma}
\nn\\
&\quad
+
H_{\rm min}^{5\epsilon+2\sqrt{\delta}}(G_AM_A|AXYZ)_{\Gamma}
+2f(\epsilon)
\laeq{convmod3-2}\\
&=
H_{\rm max}^{\epsilon}(A|XYZ)_{\Psi_s}
\nn\\
&\quad\quad
+
H_{\rm min}^{5\epsilon+2\sqrt{\delta}}(G_AM_A|XYZ)_{\Gamma}
+2f(\epsilon).
\laeq{convmod3-3}
}
Here, \req{convmod3-1} follows from $\hat{G}_A\equiv G_AM_AZ''$ and the fact that $\ca{C}^T(\Gamma)$ is classically coherent in $XX'$ and in $ZZ''$ because of \req{fstar};
\req{convmod3-2} from the chain rule \req{CRminmaxmin};
and \req{convmod3-3} from $\Gamma^{AXYZ}=\Psi_s^{AXYZ}$ and the fact that the system $A$ in the conditioning part is decoupled from $G_AM_A$ when conditioned by $XYZ$ as \req{dfntipsiii2} in addition to \rLmm{SE1}.

Combining these all together, we arrive at
\alg{
&
\!
e_0\!+\!H_{\rm max}^{\epsilon}(A|XYZ)_{\Psi_s}\!
\nn\\
&\quad+\!H_{\rm min}^{5\epsilon+2\sqrt{\delta}}(G_AM_A|XYZ)_{\Gamma}
\!+\!2f(\epsilon)
\!
\nn\\
&
\geq
H_{\rm min}^{\epsilon}(AC|XYZ)_{\Psi_s}
+e+e_0-q-f(\epsilon).
}
This completes the proof of Ineq.~\req{convv000}.

\subsubsection{Proof of Inequality \req{convv3}}

We first calculate
\alg{
&
H_{\rm min}^{\epsilon}(BYCZ)_{\Psi_s}
\nn\\
&=
H_{\rm min}^{\epsilon}(BYCZ)_{\Gamma}
\laeq{convmod5-2}\\
&\leq
H_{\rm min}^{12\epsilon+4\sqrt{\delta}}(BYCZF_B\hat{G}_B)_{\Gamma\otm\Phi_{2^{e_0}}}
\nn\\
&\quad\quad
-
H_{\rm min}^{5\epsilon+2\sqrt{\delta}}(F_B\hat{G}_B|BYCZ)_{\Gamma\otm\Phi_{2^{e_0}}}
+f(\epsilon)
\laeq{convmod5-3}\\
&\leq
H_{\rm min}^{12\epsilon+4\sqrt{\delta}}(BYCZF_B\hat{G}_B)_{\Gamma\otm\Phi_{2^{e_0}}}
\nn\\
&\quad\quad
-
e_0-H_{\rm min}^{5\epsilon+2\sqrt{\delta}}(\hat{G}_B|BYCZ)_{\Gamma}
+f(\epsilon)
\laeq{convmod5-4}
\\
&=
H_{\rm min}^{12\epsilon+4\sqrt{\delta}}(BYE_BQM)_{\tilde{\Gamma}}
\nn\\
&\quad\quad
-
e_0-H_{\rm min}^{5\epsilon+2\sqrt{\delta}}(\hat{G}_B|BYCZ)_{\Gamma}
+f(\epsilon).
\laeq{convmod5-4-2}
}
Here, \req{convmod5-2} follows from $\Psi_s^{BYCZ}=\Gamma^{BYCZ}$;
\req{convmod5-3} from the chain rule \req{CRminminmin};
\req{convmod5-4} from the superadditivity of the smooth conditional entropy for product states (\rLmm{SE3}); and
\req{convmod5-4-2} from the fact that $\tilde{\Gamma}$ is obtained by an isometry $U_{\ca{D}}^\dagger$ from $\Gamma\otm\Phi_{2^{e_0}}$ as \req{dfntildePsi}. 

The first term in \req{convmod5-4-2} is further calculated to be
\alg{
&
H_{\rm min}^{12\epsilon+4\sqrt{\delta}}(BYE_BQM)_{\tilde{\Gamma}}
\laeq{convmod5-5}\\
&\leq
H_{\rm min}^{12\epsilon+6\sqrt{\delta}}(BYE_BQM)_{\tilde{\Psi}}
\laeq{convmod5-6}\\
&\leq
H_{\rm min}^{12\epsilon+6\sqrt{\delta}}(BYE_B)_{\tilde{\Psi}}+c+q
\laeq{convmod5-7}\\
&=
H_{\rm min}^{12\epsilon+6\sqrt{\delta}}(BYE_B)_{\Psi\otm\Phi_{e+e_0}}+c+q
\laeq{convmod5-8}\\
&\leq
H_{\rm min}^{12\epsilon+6\sqrt{\delta}}(BY)_{\Psi_s}+e+e_0+c+q,
\laeq{convmod5-9}
}
where \req{convmod5-6} follows from the continuity of the smooth conditional entropy (\rLmm{SE12}) and the fact that $\tilde{\Gamma}$ and  $\tilde{\Psi}$ are $2\sqrt{\delta}$-close with each other as \req{qeef2};
\req{convmod5-7} from the dimension bound (\rLmm{SE4});
\req{convmod5-8} from the fact that $\tilde{\Psi}$ is converted to $\Psi\otm\Phi_{e+e_0}$ by an operation $U_{\ca{E}}$ by Alice as \req{dfnOmega}, which does not change the reduced state on $BYE_B$;
and \req{convmod5-9} from the dimension bound (\rLmm{SE4}) and $\Psi_s^{BY}=\Psi^{BY}$.

For the third term in \req{convmod5-4-2}, we have
\alg{
&
H_{\rm min}^{5\epsilon+2\sqrt{\delta}}(\hat{G}_B|BYCZ)_{\Gamma}
\laeq{convmod5-10}\\
&\geq
H_{\rm min}^{5\epsilon+2\sqrt{\delta}}(\hat{G}_B|BCXYZ)_{\Gamma}
\laeq{convmod5-11}\\
&=
H_{\rm min}^{5\epsilon+2\sqrt{\delta}}(\hat{G}_B|XYZ)_{\Gamma}
\laeq{convmod5-12}\\
&=
H_{\rm min}^{5\epsilon+2\sqrt{\delta}}(\hat{G}_A|XYZ)_{\Gamma}
\laeq{convmod5-13}\\
&=
H_{\rm min}^{5\epsilon+2\sqrt{\delta}}(G_AM_A|XYZ)_{\Gamma}
\laeq{convmod5-14}
}
Here, \req{convmod5-11} is from the monotonicity of the smooth conditional entropy (\rLmm{Hminmonotonicity});
\req{convmod5-12} from the fact that $\Gamma$ is decoupled between $BC$ and $\hat{G}_B$ when conditioned by $XYZ$ as \req{dfntipsiii2}, and the property of the smooth conditional entropy (\rLmm{SE1});
\req{convmod5-13} from \rLmm{SE11} and the fact that $\Gamma^{\hat{G}_A\hat{G}_BXYZ}$ is an ensemble of classically-labelled pure states on $\hat{G}_A\hat{G}_B$ as \req{dfntipsiii2};
and \req{convmod5-14} from $\hat{G}_A\equiv G_AM_AZ''$, \rLmm{condminCQCQ} and the fact that $Z''$ is a classical copy of $Z$ due to \req{fstar}.

Combining these all together, we arrive at
\alg{
&
H_{\rm min}^{\epsilon}(BYCZ)_{\Psi_s}
\nn\\
&
\leq
H_{\rm min}^{12\epsilon+6\sqrt{\delta}}(BY)_{\Psi_s}
+
e+c+q+f(\epsilon)
\nn\\
&\quad\quad
-H_{\rm min}^{5\epsilon+2\sqrt{\delta}}(G_AM_A|XYZ)_{\Gamma}.
}

\subsubsection{Proof of Inequality \req{convv4}}

We have
\alg{
&
e+e_0+H_{\rm min}^{11\epsilon+8\sqrt{\delta}}(B|XYZ)_{\Psi_s}
\\
&=
e+e_0+H_{\rm min}^{11\epsilon+8\sqrt{\delta}}(ACR|XYZ)_{\Psi_s}
\laeq{Vconv1-0}\\
&=
e+e_0+H_{\rm min}^{11\epsilon+8\sqrt{\delta}}(\hat{A}\hat{C}R|T)_{\Psi_s}
\laeq{Vconv1-0-2}\\
&
=e+e_0+H_{\rm min}^{11\epsilon+8\sqrt{\delta}}(\hat{A}\hat{C}R|T)_{\ca{C}^T(\Psi)}
\laeq{Vconv1-1} \\
&
\geq
H_{\rm min}^{11\epsilon+8\sqrt{\delta}}(\hat{A}\hat{C}E_AR|T)_{\ca{C}^T(\Psi)\otm\Phi_{2^{e+e_0}}}
\laeq{dorodoro}\\
&
=
H_{\rm min}^{11\epsilon+8\sqrt{\delta}}(\hat{A}QMF_A\hat{G}_AR|T)_{\ca{C}^T(\tilde{\Psi})}
\laeq{Vconv2-2}\\
&
\geq
H_{\rm min}^{10\epsilon+8\sqrt{\delta}}(\hat{A}MF_A\hat{G}_AR|T)_{\ca{C}^T(\tilde{\Psi})}
\nn\\
&\quad
+
H_{\rm min}(Q|\hat{A}MF_A\hat{G}_ART)_{\ca{C}^T(\tilde{\Psi})}
-f(\epsilon)
\laeq{Vconv3-2}\\
&
\geq
H_{\rm min}^{10\epsilon+8\sqrt{\delta}}(\hat{A}MF_A\hat{G}_AR|T)_{\ca{C}^T(\tilde{\Psi})}
-q
-f(\epsilon)
\laeq{Vconv3-2-2}
\\
&
=
H_{\rm min}^{10\epsilon+8\sqrt{\delta}}(\hat{B}E_BQ|T)_{\ca{C}^T(\tilde{\Psi})}
-q
-f(\epsilon),
\laeq{Vconv3-2-3}
}
where 
\req{Vconv1-0} follows from \rLmm{SE11};
\req{Vconv1-0-2} from \rLmm{condminCQCQ} and the fact that $T=X'Y'Z'$ is a classical copy of $XYZ$;
\req{Vconv1-1} from $\Psi_s=\ca{C}^T(\Psi)$,
\req{dorodoro} from the dimension bound (\rLmm{SE4}),
\req{Vconv2-2} from the fact that $\tilde{\Psi}$ is obtained from $\Psi\otm\Phi_{2^{e+e_0}}$ by applying the isometry $U_{\ca{E}}$ as \req{dfnOmega},  under which the smooth conditional entropy is invariant (\rLmm{invCEiso}),
\req{Vconv3-2} from the chain rule \req{CRminminmin},
\req{Vconv3-2-2} from the dimension bound (\rLmm{SE2}),
and \req{Vconv3-2-3} from \rLmm{SE11} and the fact that $\tilde{\Psi}$ is a pure state on $\hat{A}\hat{B}\hat{R}F_A\hat{G}_AQME_B$ as \req{dfnOmega}, which is converted by $\ca{C}^T$ to an ensemble of classically-labelled pure states.

The first term in \req{Vconv3-2-3} is further calculated to be
\alg{
&
H_{\rm min}^{10\epsilon+8\sqrt{\delta}}(\hat{B}E_BQ|T)_{\ca{C}^T(\tilde{\Psi})}
\\
&
=
H_{\rm min}^{10\epsilon+8\sqrt{\delta}}(\hat{B}E_BQ|T)_{\ca{C}^T\otm\ca{C}^M(\tilde{\Psi})}
\laeq{ttt00}\\
&
\geq
H_{\rm min}^{10\epsilon+8\sqrt{\delta}}(\hat{B}E_BQ|TM)_{\ca{C}^T\otm\ca{C}^M(\tilde{\Psi})}
\laeq{ttt1}\\
&
=
H_{\rm min}^{10\epsilon+8\sqrt{\delta}}(\hat{B}E_BQM|TM_A)_{\ca{C}^T\otm\ca{C}^{M_A}(\tilde{\Psi})}
\laeq{ttt2}\\
&
=
H_{\rm min}^{10\epsilon+8\sqrt{\delta}}(\hat{B}\hat{C}F_B\hat{G}_B|TM_A)_{\ca{C}^T\otm\ca{C}^{M_A}(\Psi_f)}
\laeq{ttt0}\\
&
\geq
H_{\rm min}^{10\epsilon+6\sqrt{\delta}}(\hat{B}\hat{C}F_B\hat{G}_B|TM_A)_{\ca{C}^T\otm\ca{C}^{M_A}(\Gamma)\otm\Phi_{2^{e_0}}}
\laeq{ttt0-2}\\
&
\geq
H_{\rm min}^{7\epsilon+6\sqrt{\delta}}(\hat{G}_B|TM_A)_{\ca{C}^T\otm\ca{C}^{M_A}(\Gamma)\otm\Phi_{2^{e_0}}}
\nn\\
&\quad\quad
+
H_{\rm min}^{\epsilon}(\hat{B}\hat{C}F_B|T\hat{G}_BM_A)_{\ca{C}^T\otm\ca{C}^{M_A}(\Gamma)\otm\Phi_{2^{e_0}}}
\nn\\
&\quad\quad\quad\quad\quad\quad\quad\quad\quad\quad\quad\quad\quad
-f(\epsilon).
\laeq{ttt}
}
Inequality \req{ttt00} is due to the fact that $\ca{C}^M$ does not change the reduced state on $\hat{B}E_BQT$;
\req{ttt1} from the monotonicity of the conditional entropy (\rLmm{Hminmonotonicity});
\req{ttt2} from the property of the conditional entropy for classical-quantum states (\rLmm{condminCQCQ}) and the fact that $M_A$ is a classical copy of $M$ as \req{tildemsimmz2};
\req{ttt0} from the fact that $\Psi_f$ is obtained from $\tilde{\Psi}$ by the isometry $U_{\ca{D}}$ as \req{dfnPsif}, under which the smooth conditional entropy is invariant; 
\req{ttt0-2} from the continuity (\rLmm{SE12}) and the fact that $\Gamma\otm\Phi_{2^{e_0}}$ is $2\sqrt{\delta}$-close to $\Psi_f$ as \req{qeef};
and \req{ttt} from the chain rule \req{CRminminmin}.

The second term in \req{ttt} is further calculated as
\alg{
&
H_{\rm min}^{\epsilon}(\hat{B}\hat{C}F_B|T\hat{G}_BM_A)_{\ca{C}^T\otm\ca{C}^{M_A}(\Gamma)\otm\Phi_{2^{e_0}}}
\laeq{rashi1}\\
&\geq
H_{\rm min}^{\epsilon}(\hat{B}\hat{C}|T\hat{G}_BM_A)_{\ca{C}^T\otm\ca{C}^{M_A}(\Gamma)}+e_0
\laeq{rashi2-0}\\
&\geq
H_{\rm min}^{\epsilon}(\hat{B}\hat{C}|T\hat{G}_BM_A)_{\ca{C}^T(\Gamma)}+e_0
\laeq{rashi2-0-2}\\
&=
H_{\rm min}^{\epsilon}(\hat{B}\hat{C}|T)_{\ca{C}^T(\Gamma)}+e_0
\laeq{rashi2}\\
&=
H_{\rm min}^{\epsilon}(\hat{B}\hat{C}|T)_{\Psi_s}+e_0
\laeq{rashi3-2}\\
&=
H_{\rm min}^{\epsilon}(BC|XYZ)_{\Psi_s}+e_0,
\laeq{rashi4}
}
where 
\req{rashi2-0} follows from the superadditivity of the smooth conditional entropy (\rLmm{SE3});
\req{rashi2-0-2} from the monotonicity of the smooth conditional entropy (\rLmm{Hminmonotonicity});
\req{rashi2} from \rLmm{SE1} and the fact that the state $\ca{C}^T(\Gamma)$ is decoupled between $\hat{B}\hat{C}$ and $\hat{G}_BM_A$ when conditioned by $T$ as \req{dfntipsiii};
\req{rashi3-2} from Equality \req{PTgammaf};
and \req{rashi4} from \rLmm{condminCQCQ}.

 The first term in \req{ttt} is calculated as
\alg{
&
H_{\rm min}^{7\epsilon+6\sqrt{\delta}}(\hat{G}_B|TM_A)_{\ca{C}^T\otm\ca{C}^{M_A}(\Gamma)}
\laeq{kota1}\\
&
=
H_{\rm min}^{7\epsilon+6\sqrt{\delta}}(G_A|TM_A)_{\Gamma}
\laeq{kota2}\\
&
=
H_{\rm min}^{7\epsilon+6\sqrt{\delta}}(G_A|M_AXYZ)_{\Gamma}
\laeq{kota3}\\
&
=
H_{\rm min}^{7\epsilon+6\sqrt{\delta}}(G_A|M_AAXYZ)_{\Gamma},
\laeq{kota4}
}
where
\req{kota2} is from $\hat{G}_B=G_BM_B$, Equality \req{dfntipsiii55} and \rLmm{SE11};
\req{kota3} from \rLmm{condminCQCQ} and the fact that $T=X'Y'Z'$ is a copy of $XYZ$ as \req{dfntipsi};
and 
\req{kota4} from \rLmm{SE1} and the fact that the state $\Gamma$ is decoupled between $A$ and $G_A$ when conditioned by $M_AXYZ$ as \req{sstar}.

Combining these all together, we arrive at
\alg{
&
e+e_0+H_{\rm min}^{11\epsilon+8\sqrt{\delta}}(B|XYZ)_{\Psi_s}
\nn\\
&\geq
-q
+H_{\rm min}^{7\epsilon+6\sqrt{\delta}}(G_A|M_AAXYZ)_{\Gamma}
\nn\\
&\quad
+H_{\rm min}^{\epsilon}(BC|XYZ)_{\Psi_s}+e_0
-2f(\epsilon).
}
This completes the proof of Inequality \req{convv4}.
\QED

\subsection{Proof of \rThm{converse} from Inequalities \req{convv1}-\req{convv4}}

Since $\Gamma$ is diagonal in $M_AXYZ$ as \req{sstar}, and due to the properties of the smooth conditional entropies for classical-quantum states (\rLmm{SE1}), we have
\alg{
&
H_{\rm min}^{5\epsilon+2\sqrt{\delta}}(G_AM_A|XYZ)_{\Gamma}
\geq0,
\\
&
H_{\rm min}^{7\epsilon+6\sqrt{\delta}}(G_A|M_AAXYZ)_{\Gamma}
\geq
0.
}
Thus, Inequalities \req{convv3} and \req{convv4} implies Inequalities \req{convv03} and \req{convv04} in \rThm{converse}, respectively.
Summing up both sides in \req{convv000} and \req{convv3} yields
\alg{
c+2q
&
\geq
H_{\rm min}^{\epsilon}(AC|XYZ)_{\Psi_s}
-
H_{\rm max}^{\epsilon}(A|XYZ)_{\Psi_s}
\nn\\
&\quad
+H_{\rm min}^{\epsilon}(BYCZ)_{\Psi_s}
\nn\\
&\quad
-
H_{\rm min}^{12\epsilon+6\sqrt{\delta}}(BY)_{\Psi_s}
-4f(\epsilon)
\nn\\
&=
\tilde{H}_{I}'^{(\epsilon,\delta)}-4f(\epsilon).
\laeq{doso1}
}
Similarly, combining Inequalities \req{convv1} and \req{convv4}, we obtain
\alg{
c+2q
&\geq
H_{\rm min}^{\epsilon}(AXCZ)_{\Psi_s}
-
H_{\rm max}^{\epsilon}(AXZ)_{\Psi_s}
\nn\\
&\quad\quad
+
H_{\rm min}^{\epsilon}(BC|XYZ)_{\Psi_s}
\nn\\
&\quad\quad
-H_{\rm min}^{11\epsilon+8\sqrt{\delta}}(B|XYZ)_{\Psi_s}
\nn\\
&\quad\quad
-H_{\rm min}^{7\epsilon+2\sqrt{\delta}}(G_A|M_AAXZ)_{\Gamma}
\nn\\
&\quad\quad
+H_{\rm min}^{7\epsilon+6\sqrt{\delta}}(G_A|M_AAXYZ)_{\Gamma}
-6f(\epsilon)
\\
&=
\tilde{H}_{I\!I}'^{(\epsilon,\delta)}
-
\Delta_\Gamma'^{(\epsilon,\delta)}
-6f(\epsilon),
\laeq{doso2}
}
where we have defined
\alg{
\Delta_\Gamma'^{(\epsilon,\delta)}
:=
&
H_{\rm min}^{7\epsilon+2\sqrt{\delta}}(G_A|M_AAXZ)_{\Gamma}
\nn\\
&
-
H_{\rm min}^{7\epsilon+6\sqrt{\delta}}(G_A|M_AAXYZ)_{\Gamma}.
}
In the following, we prove that
\alg{
\Delta_\Gamma'^{(\epsilon,\delta)}
\leq
\Delta^{(\epsilon,\delta)}.
}
Combining this with \req{doso2} in addition to \req{doso1},
we arrive at Inequality \req{convv00} in \rThm{converse}.

We start by noting that
\alg{
\Delta_\Gamma'^{(\epsilon,\delta)}
&
=
H_{\rm min}^{7\epsilon+2\sqrt{\delta}}(G_A|M_AAX'Z')_{\Gamma}
\nn\\
&
\quad\quad
-
H_{\rm min}^{7\epsilon+6\sqrt{\delta}}(G_A|M_AAX'Y'Z')_{\Gamma}
\\
&
\leq
\tilde{I}_{\rm min}^{7\epsilon+4\sqrt{\delta}}(G_A:Y'|M_AAX'Z')_{\tilde{\Psi}}
}
The first line follows from \rLmm{condminCQCQ} and the fact that $XYZ$ is a copy of $X'Y'Z'$ as \req{dfntipsi}, and
 the second line from the continuity bounds for the smooth conditional entropy (\rLmm{SE12}) and the definition of the smooth conditional min mutual information \req{dfnmimMI}.
Hence, it suffices to prove that there exists an operation $\ca{F}:\hat{A}\hat{C}\rightarrow AG_AM_A$ satisfying
\alg{
\ca{F}(\Psi_s^{\hat{A}\hat{C}\hat{R}})=\tilde{\Psi}^{AG_AM_A\hat{R}},
\quad
\ca{C}^{M_A}\circ\ca{F}=\ca{F}
\laeq{fpsiseqtildepsi}
}
and that $\tilde{\Psi}$ satisfies the condition 
\begin{eqnarray}
\!
\inf_{\{\omega_{xyz}\}}
\!
P
\!
\left(
\!
\tilde{\Psi}^{AG_AM_A\hat{R}},
\sum_{x,y,z}p_{xyz}
\psi_{xyz}^{A\hat{R}}
\otm
\omega_{xyz}^{G_AM_A}
\!
\right)
\nn\\
\leq
2\sqrt{\delta}.
\quad\quad
\laeq{conditionaldecoupling2}
\end{eqnarray}

Recall that the state $\ket{\tilde{\Psi}}$ is obtained by an encoding isometry $U_\ca{E}^{\hat{A}\hat{C}E_A\rightarrow \hat{A}QMF_A\hat{G}_A}$ from $\ket{\Psi}\ket{\Phi_{2^{e+e_0}}}$ as \req{dfnOmega}, where $\hat{G}_A=G_AM_AZ''$.
We define an operation $\ca{F}:\hat{A}\hat{C}\rightarrow AG_AM_A$ by
\alg{
\ca{F}(\tau)
:=
{\rm Tr}_{QMF_AXZ''}\circ\ca{U}_{\ca{E}}(\tau\otm\pi_{2^{e+e_0}}^{E_A}).
}
Noting that $U_\ca{E}$ is in the form of \req{dfnUE},
this implies \req{fpsiseqtildepsi}.
To obtain the decoupling condition \req{conditionaldecoupling2}, note that, since $\tilde{\Psi}$ is converted by an operation by Bob to $\Psi_f$ as \req{dfnPsif}, it holds that $\tilde{\Psi}^{AG_AM_A\hat{R}}=\Psi_f^{AG_AM_A\hat{R}}$.
Thus, tracing out $\hat{B}\hat{C}F_AF_B\hat{G}_BXZ''$ in \req{qeef}, we obtain
\alg{
P\left(
\tilde{\Psi}^{A\hat{R}G_AM_A},
\Gamma^{A\hat{R}G_AM_A}
\right)
\leq
2\sqrt{\delta}.
}
Due to \req{dfntipsi}, 
the state $\Gamma$ is in the form of
\alg{
\!
\Gamma^{A\hat{R}G_AM_A}
=
\!\sum_{x,y,z}p_{xyz}
\psi_{xyz}^{AR}
\otm
\phi_{xyz}^{G_AM_A}
\!\otm
\proj{xyz}^T.
\!
} 
This implies \req{conditionaldecoupling2} and completes the proof of Inequality \req{convv00}.
\QED

\subsection{Property of $\Delta^{(\epsilon,\delta)}$ (Proof of \rLmm{propDelta})}
\lsec{propDelta}

Due to the definition of the smooth conditional min mutual information \req{dfnmimMI} and \req{dfnDeltaed}, it is straightforward to verify that $\Delta^{\epsilon,\delta}\geq0$.
The equality holds if $Y'\cong Y$ is a one-dimensional system, that is, if there is no classical side information at the decoder.
In the case where there is neither quantum message nor quantum side information at the encoder, i.e.
\alg{
d_A=d_C=1,
\quad
\hat{A}=X, 
\quad
\hat{C}=Z,
\laeq{ssff}
} 
the source state $\Psi_s$ is represented as
\alg{
\!
\Psi_s^{XZ\hat{B}\hat{R}}
\!=\!
\sum_{x,y,z}
p_{xyz}
\proj{x}^X
\!\!\otm\!
\proj{y}^Y
\!\!\otm\!
\proj{z}^Z
\!\!\otm
\psi_{xyz}^{B\hat{R}}.
\!
}
Thus, for any CPTP map $\ca{F}:XZ\rightarrow G_AM_A$, we have
\alg{
\ca{F}(\Psi_s)^{G_AM_A\hat{R}}
=
\sum_{x,y,z}
p_{xyz}
\omega_{xz}^{G_AM_A}
\otm
\psi_{xyz}^{\hat{R}},
}
where $\omega_{xz}:=\ca{F}(\proj{x}^X
\otm\proj{z}^Z)$.
It follows that
\alg{
&
\ca{F}(\Psi_s)^{G_AM_AX'Y'Z'}
\nn\\
&\;
=
\sum_{x,z}
p_{xz}
\omega_{xz}^{G_AM_A}
\otm
\proj{xz}^{X'Z'}
\nn\\
&\quad\quad
\otm
\left(
\sum_y
p_{y|xz}
\proj{y}^{Y'}
\right),
}
and consequently, $\tilde{I}_{\rm min}^{7\epsilon+4\sqrt{\delta}}(G_A:Y'|M_AX'Z')=0$.
This implies $\Delta^{\epsilon,\delta}=0$, and completes the proof of \rLmm{propDelta}.
\QED

\section{Conclusion}
\lsec{conclusion}

In this paper, we investigated the state redistribution of classical and quantum hybrid sources in the one-shot scenario.
We analyzed the costs of classical communication, quantum communication and entanglement.
We obtained the direct bound and the converse bound for those costs in terms of smooth conditional entropies.
In most of the cases that have been analyzed in the previous literatures, the two bounds coincide in the asymptotic limit of infinitely many copies and vanishingly small error.
Various coding theorems for two-party source coding tasks are systematically obtained by reduction from our results,
including the ones that have not been analyzed in the previous literatures.

To investigate the protocol that are covered by our result, but have not been addressed in the previous literature, in detail is left as a future work.
Another direction is to explore the family of quantum communication protocols in the presence of classical side information only at the decoder.
It would also be beneficial to analyze the relation between our results and the one-shot bounds for entanglement-assisted communication of classical and quantum messages via a noisy quantum channel \cite{wakakuwa2020randomized}.

\section*{Acknowledgement}

This work was supported by JSPS KAKENHI (Grant No.~18J01329), and by JST, PRESTO Grant Number JPMJPR1865, Japan.

\bibliographystyle{plain}

\onecolumn\newpage
\appendix

\section{Definitions and Properties of Smooth Entropies}
\lapp{propSmEn}

In this appendix, we summarize the properties of the smooth conditional entropies that are used in the main text.
For the properties of the purified distance used in some of the proofs, see \rApp{extUhlmann}.

\subsection{Basic Properties}

\blmm{duality}
{\bf (duality: see e.g. \cite{tomamichel2010duality})}
For any subnormalized pure state $|\psi\rangle$ on system $ABC$, and for any $\epsilon>0$, 
$H_{\rm max}^\epsilon(A|B)_\psi=
-
H_{\rm min}^\epsilon(A|C)_\psi$.
\end{lmm}

\begin{lmm}\label{lmm:Hminmonotonicity}
{\bf(monotonicity: Theorem 18 in  \cite{tomamichel2010duality} and Theorem 6.2 in \cite{T16})}
For any $\rho^{AB} \in \ca{S}_\leq(\ca{H}^{AB})$, $0\leq\epsilon\leq\sqrt{{\rm Tr}[\rho]}$,  any unital CPTP map $\ca{E}:A\rightarrow C$ and any CPTP map $\ca{F}:B\rightarrow D$, it holds that $H_{\rm min}^\epsilon(A|B)_\rho\leq H_{\rm min}^\epsilon(C|D)_{\ca{E}\otm\ca{F}(\rho)}$.
\elmm

\begin{lmm}[isometric invariance: Lemma 13 in  \cite{tomamichel2010duality}]\label{lmm:invCEiso}
For any $\epsilon\geq0$, $\rho^{AB} \in \ca{S}_\leq(\ca{H}^{AB})$ and any linear isometries $U:A\rightarrow C$ and $V:B\rightarrow D$, $H_{\rm min}^\epsilon(A|B)_\rho=H_{\rm min}^\epsilon(C|D)_{\ca{U}\otm\ca{V}(\rho)}$.
\elmm

\begin{lmm}[additivity: see Section I C  in \cite{konig2009operational}]\label{lmm:addcondmax}
For any $\rho\in\ca{S}(\ca{H}^{AB})$ and $\sigma\in\ca{S}(\ca{H}^{CD})$, it holds that
\alg{
\!\!
H_{\rm max}(AC|BD)_{\rho\otm\sigma}
=
H_{\rm max}(A|B)_{\rho}
+
H_{\rm max}(C|D)_{\sigma}
.
\!\!
}
\elmm

\begin{lmm}\label{lmm:SE3}
{\bf (superadditivity: Lemma A.2 in \cite{DBWR2010})}
For any states $\rho^{AB}$, $\sigma^{CD}$ and any $\epsilon,\epsilon'\geq0$, it holds that
\alg{
\!\!
H_{\rm min}^{\epsilon+\epsilon'}(AC|BD)_{\rho\otm\sigma}
\geq
H_{\rm min}^{\epsilon}(A|B)_{\rho}
\!+\!
H_{\rm min}^{\epsilon'}(C|D)_{\sigma}.
\!\!\!
}
\elmm

\begin{lmm}[chain rule: see \cite{vitanov2013chain}]\label{lmm:chainrule}
For any $\epsilon>0$, $\epsilon',\epsilon''\geq0$ and $\rho\in\ca{S}_\leq(\ca{H}^{ABC})$, 
it holds that
\alg{
H_{\rm min}^{\epsilon+\epsilon'+2\epsilon''}(AB|C)_\rho
&
\geq
H_{\rm min}^{\epsilon'}(B|C)_\rho
+
H_{\rm min}^{\epsilon''}(A|BC)_\rho
-f(\epsilon),
\laeq{CRminminmin}\\
H_{\rm min}^{\epsilon'}(AB|C)_\rho
&
\leq
H_{\rm max}^{\epsilon''}(B|C)_\rho
+
H_{\rm min}^{\epsilon+\epsilon'+2\epsilon''}(A|BC)_\rho
+2f(\epsilon),
\laeq{CRminmaxmin}
}
where 
\alg{
f(\epsilon):=-\log{(1-\sqrt{1-\delta^2})}.
}
\elmm

\begin{lmm}\label{lmm:SE2}
{\bf(dimension bounds: Corollary of Lemma 20 in \cite{tomamichel2010duality})}
For any state $\rho^{AB}$ and $\epsilon\geq0$, it holds that
\alg{
H_{\rm min}^\epsilon(A|B)_\rho
&\geq
-\log{d_A},
\\
H_{\rm max}^\epsilon(A|B)_\rho
&\leq
\log{d_A}.
}
\elmm

\begin{lmm}[dimension bound: Lemma 21 in \cite{datta2011apex}]\label{lmm:SE4}
For any state $\rho^{ABC}$ and $\epsilon>0$, it holds that
\alg{
H_{\rm min}^{\epsilon}(AB|C)_{\rho}
\leq
H_{\rm min}^{\epsilon}(A|C)_{\rho}
+
\log{d_B}.
}
\elmm

\blmm{SE12}{\bf (continuity)}
For any $\epsilon,\delta\geq0$, any $\rho^{AB}$ and $\sigma^{AB}\in\ca{B}^\delta(\rho)$, it holds that
\alg{
&
H_{\rm min}^{\epsilon+\delta}(A|B)_\rho
\geq
H_{\rm min}^{\epsilon}(A|B)_\sigma.
\laeq{mincont}
}
\elmm

\bprf
Let $\hat{\sigma}^{AB}\in\ca{B}^{\epsilon}(\sigma)$ be such that $H_{\rm min}^{\epsilon}(A|B)_\sigma=H_{\rm min}(A|B)_{\hat{\sigma}}$.
Due to the triangle inequality for the purified distance, it holds that
\alg{
P(\rho,\hat{\sigma})
\leq
P(\rho,\sigma)
+
P(\sigma,\hat{\sigma})
\leq
\epsilon+\delta,
}
which implies $\hat{\sigma}\in\ca{B}^{\epsilon+\delta}(\rho)$.
Thus, we obtain Inequality \req{mincont} as
\alg{
H_{\rm min}^{\epsilon}(A|B)_\sigma
=H_{\rm min}(A|B)_{\hat{\sigma}}
\leq
\sup_{\hat{\rho}\in\ca{B}^{\epsilon+\delta}(\rho)}H_{\rm min}(A|B)_{\hat{\rho}}
=
H_{\rm min}^{\epsilon+\delta}(A|B)_\rho.
}
\QED
\eprf

\blmm{onedimHminmax}{\bf (one-dimensional system.)}
Suppose that $d_A=1$. For any $\epsilon\geq0$ and $\rho\in\ca{S}(\ca{H}^{AB})$, it holds that
\alg{
&
0
\leq
H_{\rm min}^\epsilon(A|B)_\rho
\leq
-\log{(1-2\epsilon)},
\laeq{katt1}\\
&
0
\geq
H_{\rm max}^\epsilon(A|B)_\rho
\geq
\log{(1-2\epsilon)}.
\laeq{katt2}
}
\elmm

\bprf
Since $d_A=1$, there exists a fixed vector $\ket{e}\in\ca{H}^A$ such that $I^A=\proj{e}$ and that any $\tilde{\rho}\in\ca{S}_\leq(\ca{H}^{AB})$ is represented as $\proj{e}^A\otm\tilde{\rho}^B$.
Due to the definition of the smooth conditional min entropy, we have
\alg{
H_{\rm min}^\epsilon(A|B)_\rho
&
\geq
H_{\rm min}(A|B)_\rho
\\
&
=
\sup_{\sigma^B \in \ca{S}_=(\ca{H}^B)}H_{\rm min}(A|B)_{\rho|\sigma}
\\
&
\geq
H_{\rm min}(A|B)_{\rho^{AB}|\rho^B}
\\
&=
\sup \{ \lambda \in \mathbb{R}| 2^{-\lambda} I^A \otimes \rho^B \geq \rho^{AB} \}
\\
&=
\sup \{ \lambda \in \mathbb{R}| 2^{-\lambda} \proj{e}^A \otimes \rho^B \geq \proj{e}^A\otm\rho^{B} \}
\\
&=0.
}
This implies the first inequality in \req{katt1}.
To prove the second inequality in \req{katt1}, let $\hat{\rho}\in\ca{B}^\epsilon(\rho)$ and $\sigma^B \in \ca{S}_=(\ca{H}^B)$ be such that 
\alg{
H_{\rm min}^\epsilon(A|B)_\rho
=
H_{\rm min}(A|B)_{\hat{\rho}}
=
H_{\rm min}(A|B)_{\hat{\rho}|\sigma}.
}
By definition, it holds that
\alg{
2^{-H_{\rm min}^\epsilon(A|B)_\rho} I^A \otimes \sigma^B \geq \hat{\rho}^{AB},
}
which is equivalent to
\alg{
2^{-H_{\rm min}^\epsilon(A|B)_\rho} \proj{e}^A \otimes \sigma^B \geq \proj{e}^A\otm\hat{\rho}^{B}.
}
By taking the trace in both sides, we obtain
\alg{
2^{-H_{\rm min}^\epsilon(A|B)_\rho} \geq {\rm Tr}[\hat{\rho}].
}
The R.H.S. of the above inequality is evaluated as
\alg{
{\rm Tr}[\hat{\rho}]
=
\|\hat{\rho}\|_1
\geq
\|\rho\|_1-\|\rho-\hat{\rho}\|_1
\geq
1-2\epsilon,
}
where the last line follows from \req{relTDPD} and the condition $\hat{\rho}\in\ca{B}^\epsilon(\rho)$.
This implies the second inequality in \req{katt1}.
Inequality \req{katt2} follows due to the duality relation (\rLmm{duality}).
\QED
\eprf

\subsection{Classical-Quantum States}

\begin{lmm}[Lemma A.5 in  \cite{DBWR2010}]\label{lmm:condminCQ}
For any state $\rho^{ABK} \in \ca{S}_=(\ca{H}^{ABK})$ in the form of
\alg{
\rho^{ABK}=\sum_kp_k\rho_k^{AB}\otm\proj{k}^K,
\laeq{cqsmoothstate}
}
where $\rho_k \in \ca{S}_=(\ca{H}^{AB})$, $\inpro{k}{k'}=\delta_{k,k'}$ and $\{p_k\}_k$ is a normalized probability distribution, it holds that 
\alg{
H_{\rm min}(A|BK)_\rho=-\log\left(\sum_kp_k\cdot2^{-H_{\rm min}(A|B)_{\rho_k}}\right).
}
\elmm

\begin{lmm}[Lemma A.7 in  \cite{DBWR2010}]\label{lmm:condminCQCQ}
For any state $\rho^{ABK_1K_2} \in \ca{S}_\leq(\ca{H}^{ABK_1K_2})$ in the form of
\alg{
\rho^{ABK_1K_2}=\sum_kp_k\rho_k^{AB}\otm\proj{k}^{K_1}\otm\proj{k}^{K_2},
\laeq{rhoclk1k2}
}
where $\inpro{k}{k'}=\delta_{k,k'}$, and for any $\epsilon\geq0$, it holds that 
\alg{
H_{\rm min}^\epsilon(AK_1|BK_2)_\rho=H_{\rm min}^\epsilon(A|BK_2)_\rho=H_{\rm min}^\epsilon(A|BK_1)_\rho.
}
\elmm

\begin{lmm}[Lemma 29 in \cite{wakakuwa2021one}]\label{lmm:condmaxCQCQ}
In the same setting as in \rLmm{condminCQCQ}, it holds that 
\alg{
H_{\rm max}^\epsilon(AK_1|BK_2)_\rho=H_{\rm max}^\epsilon(A|BK_2)_\rho=H_{\rm max}^\epsilon(A|BK_1)_\rho.
}
\elmm

\blmm{SE1}
Consider a state in the form of
\alg{
\rho^{ACK}=\sum_kp_k\rho_k^{A}\otm\sigma_k^C\otm\proj{k}^K.
}
For any $\epsilon>0$, it holds that
\alg{
H_{\rm min}^{\epsilon}(A|CK)_{\rho}
=
H_{\rm min}^{\epsilon}(A|K)_{\rho}
\geq
0.
}
\elmm

\bprf
It is straightforward to verify that there exists a quantum operation $\ca{E}:K\rightarrow CK$ such that $\rho^{ACK}=\ca{E}(\rho^{AK})$.
Due to the monotonicity of the smooth conditional min entropy under operations on the conditioning system, we have
\alg{
&
H_{\rm min}^{\epsilon}(A|K)_{\rho^{AK}}
\leq
H_{\rm min}^{\epsilon}(A|CK)_{\rho^{ACK}}
=
H_{\rm min}^{\epsilon}(A|CK)_{\ca{E}(\rho^{AK})}
\leq
H_{\rm min}^{\epsilon}(A|K)_{\rho^{AK}},
}
which implies $H_{\rm min}^{\epsilon}(A|CK)_{\rho}=H_{\rm min}^{\epsilon}(A|K)_{\rho}$.
The non-negativity follows due to \rLmm{condminCQ} as
\alg{
H_{\rm min}^{\epsilon}(A|K)_{\rho}
\geq
H_{\rm min}(A|K)_{\rho}
=
-\log\left(\sum_kp_k\cdot2^{-H_{\rm min}(A)_{\rho_k}}\right)
\geq
-\log\left(\sum_kp_k\right)
=0,
}
which completes the proof.
\QED
\eprf

\subsection{Classically-labelled Pure States}

\blmm{SE11d}
Consider a state in the form of
\alg{
\rho^{ABCK}=\sum_kp_k\proj{\psi_k}^{ABC}\otm\proj{k}^K.
\laeq{CQpurestate}
}
For any $\epsilon>0$, it holds that
\alg{
&
H_{\rm max}^{\epsilon}(A|BK)_{\rho}
=
-H_{\rm min}^{\epsilon}(A|CK)_{\rho}.
\laeq{minsymd}
}
\elmm

\bprf
It is straightforward to verify that a purification of the state $\rho$, defined by \req{CQpurestate}, is given by
\alg{
\ket{\psi_\rho}^{ABCKK'}=\sum_k\sqrt{p_k}\ket{\psi_k}^{ABC}\ket{k}^K\ket{k}^{K'}.
\laeq{CQpurestate2}
}
Due to the duality of the smooth conditional entropies (\rLmm{duality}), we have
\alg{
H_{\rm max}^{\epsilon}(A|BK)_{\rho}
=
H_{\rm max}^{\epsilon}(A|BK)_{\psi_\rho}
=
-H_{\rm min}^{\epsilon}(A|CK')_{\psi_\rho}
=
-H_{\rm min}^{\epsilon}(A|CK)_{\rho},
}
which completes the proof.
\QED
\eprf

\blmm{SE11}
Consider the same setting as in \rLmm{SE11d}.
For any $\epsilon>0$, it holds that
\alg{
&
H_{\rm min}^{\epsilon}(A|K)_{\rho}
=
H_{\rm min}^{\epsilon}(B|K)_{\rho}.
\laeq{minsym}
}
\elmm

\bprf
To prove \req{minsym},
let $\hat{\rho}^{AK}\in\ca{B}^\epsilon(\rho)$ and $\varsigma\in\ca{S}_=(\ca{H}^K)$ be such that
\alg{
H_{\rm min}^{\epsilon}(A|K)_{\rho}
=
H_{\rm min}(A|K)_{\hat{\rho}}
=
H_{\rm min}(A|K)_{\hat{\rho}|\varsigma}.
}
With $\ca{C}$ being the completely dephasing operation on $K$ with respect to the basis $\{\ket{k}\}_k$, it holds that
\alg{
P(\ca{C}(\hat{\rho}),\ca{C}(\rho))
\leq
P(\hat{\rho}),\rho)
\leq
\epsilon.
}
In addition, if
\alg{
2^{-\lambda} I^A \otimes \varsigma^K \geq \hat{\rho}^{AK},
\laeq{mincond1}
}
then
\alg{
2^{-\lambda} I^A \otimes \ca{C}(\varsigma)^K \geq {\rm id}^A\otm\ca{C}^K(\hat{\rho}^{AK}).
}
Thus, without loss of generality, we may assume that both $\hat{\rho}^{AK}$ and $\varsigma$ are diagonal in $\{\ket{k}\}_k$.
That is, we may assume that $\hat{\rho}^{AK}$ and $\varsigma$ are in the form of
\alg{
\hat{\rho}^{AK}
=
\sum_k
\hat{p}_k
\hat{\rho}_k^A\otm\proj{k}^K,
\quad
\varsigma
=
\sum_kq_k\proj{k}.
}
Suppose that the Schmidt decomposition of $\ket{\psi_k}$ is given by
\alg{
\ket{\psi_k}
=
\sum_j\sqrt{\mu_{j|k}}\ket{e_{j|k}}^A\ket{f_{j|k}}^B,
}
Define linear operators $v_k:\ca{H}^A\rightarrow\ca{H}^B$ and $V:\ca{H}^A\otm\ca{H}^K\rightarrow\ca{H}^B\otm\ca{H}^K$ by
\alg{
v_k:=\sum_j\ket{f_{j|k}}^B\bra{e_{j|k}}^A
\quad
(\forall k)
}
and $V:=\sum_kv_k\otm\proj{k}^K$.
It is straightforward to verify that $\rho^{BK}=V\rho^{AK}V^\dagger$.
Thus, due to the monotonicity of the purified distance under trace non-increasing CP maps (Lemma 7 in \cite{tomamichel2010duality}), it holds that
\alg{
P(\rho^{BK},V\hat{\rho}^{AK}V^\dagger)
\leq
P(\rho^{AK},\hat{\rho}^{AK})
\leq
\epsilon.
}
Applying $V$ to the both sides in condition \req{mincond1}, it follows that
\alg{
2^{-\lambda} V(I^A\otm\varsigma^K)V^\dagger
 \geq V\hat{\rho}^{AK}V^\dagger.
}
Noting that $I^B\geq (v_k^\dagger v_k)^B$,
this implies that
\alg{
2^{-\lambda} I^B\otm\varsigma^K
 \geq V\hat{\rho}^{AK}V^\dagger.
}
Thus, we arrive at
\alg{
H_{\rm min}^{\epsilon}(A|K)_{\rho}
\leq
H_{\rm min}^{\epsilon}(B|K)_{\rho}.
}
By exchanging the roles of $A$ and $B$, we also obtain the converse inequality.
This completes the proof of Equality \req{minsym}.
\eprf

\section{Properties of The Purified Distance}
\lapp{extUhlmann}

We summarize the properties of the purified distance, used in \rApp{propSmEn} to prove the properties of the smooth conditional entropies. 
%
%

\blmm{purifiedDmonot}
{\bf (monotonicity: Lemma 7 in  \cite{tomamichel2010duality})}
For any subnormalized states $\rho,\sigma\in\ca{S}_\leq(\ca{H})$ and for any completely positive trace non-increasing map $\ca{E}$, it holds that $P(\rho,\sigma)\geq P(\ca{E}(\rho),\ca{E}(\sigma))$.
 Consequently, for any linear isometry $\ca{U}$, it holds that $P(\rho,\sigma)= P(\ca{U}(\rho),\ca{U}(\sigma))$
\elmm

\blmm{extUhlmann0}
For any normalized state $\rho$ on system $A$ and any normalized pure state $\ket{\phi}$ on system $AB$, the purified distance satisfies
\alg{
P(\rho^A,\phi^A)
=
\min_{|\psi\rangle^{AB}}
P(\proj{\psi},\proj{\phi})
=
\sqrt{1-\max_{|\psi\rangle^{AB}}|\langle\psi|\phi\rangle|^2},
}
where the minimum and the maximum are taken over all purifications $\ket{\psi}$ of $\rho$.
\elmm

\bprf
Follows from Definition 4 and Lemma 8 in \cite{tomamichel2010duality}.
\QED
\eprf

\blmm{extUhlmann1}
Consider a state $\Gamma$ on $KAB$ and a pure state $\ket{\Psi}$ on $KABC$ in the form of
\alg{
\ket{\Psi}=\sum_k\sqrt{p_k}\ket{k}^K\ket{\psi_k}^{ABCD},
\quad\quad
\Gamma=\sum_kp_k\proj{k}^K\otm\proj{\gamma_k}^{AB}.
}
There exists a set of pure states $\{\ket{\phi_k}\}_k$ on $CD$ such that, for the state
\alg{
\ket{\tilde{\Gamma}}&=\sum_k\sqrt{p_k}\ket{k}^K\ket{\gamma_k}^{AB}\ket{\phi_k}^{CD},
\laeq{kilkil}
} 
it holds that
\alg{
\!\!\!
P\left(
\proj{\tilde{\Gamma}}\!,\proj{\Psi}
\right)
=
P\left(
\Gamma^{KAB}
,\:
\ca{C}^K\!\circ\!{\rm Tr}_{CD}(\proj{\Psi})
\right),
\!\!
}
where $\ca{C}$ is the completely dephasing operation on $K$ with respect to the basis $\{\ket{k}\}_k$.
\elmm

\bprf
It is straightforward to verify that 
a purification of the state $\ca{C}^K\!\circ\!{\rm Tr}_{CD}(\proj{\Psi})$ is given by
\alg{
\ket{\Psi_p}=\sum_k\sqrt{p_k}\ket{k}^K\ket{\psi_k}^{ABCD}\ket{k}^{K'},
}
and that any purification of the state $\Gamma^{KAB}$ to the system $KABCDK'$ is in the form of
\alg{
\ket{\Gamma_p}=\sum_k\sqrt{p_k}\ket{k}^K\ket{\gamma_k}^{AB}\ket{\xi_k}^{CDK'},
}
with $\{\ket{\xi_k}\}_k$ being a set of orthogonal states.
A simple calculation yields
\alg{
|\langle\Psi_p|\Gamma_p\rangle|
=
\sum_kp_k
|(\langle\psi_k|^{ABCD}\langle k|^{K'})(|\gamma_k\rangle^{AB}|\xi_k\rangle^{CDK'})|.
}
The maximum of the above quantity over all orthogonal $\{\ket{\xi_k}\}_k$ is achieved by $\{\ket{\xi_k}\}_k$ that is decomposed into $|\xi_k\rangle^{CDK'}=|\phi_k\rangle^{CD}|k\rangle^{K'}$.
Using this $\{|\phi_k\rangle\}_k$, we define a state $|\tilde{\Gamma}\rangle$ by
\alg{
\ket{\tilde{\Gamma}}:=\sum_k\sqrt{p_k}\ket{k}^K\ket{\gamma_k}^{AB}\ket{\phi_k}^{CD}
\laeq{kilkilff}
} 
and a purification of $\Gamma^{KAB}$ by
\alg{
\ket{\Gamma_p^*}:=\sum_k\sqrt{p_k}\ket{k}^K\ket{\gamma_k}^{AB}\ket{\phi_k}^{CD}\ket{k}^{K'}.
\laeq{kilkili}
} 
It follows that
\alg{
&
\max_{\{\xi_k\}_k}
|\langle\Psi_p|\Gamma_p\rangle|
=
|\langle\Psi_p|\Gamma_p^*\rangle|
\\
&
=
\sum_kp_k
|\langle\psi_k|^{ABCD}|\gamma_k\rangle^{AB}|\phi_k\rangle^{CD}|
\laeq{bilkili}
\\
&=|\langle\Psi|\tilde{\Gamma}\rangle|.
}
In addition, the states $\ket{\Psi_p}$ and $\ket{\Gamma_p^*}$ are obtained by a linear isometry $P^{K\rightarrow KK'}:=\sum_k\ket{k}^K\ket{k}^{K'}\bra{k}$ from $\ket{\Psi}$ and $\ket{\tilde{\Gamma}}$as
\alg{
\ket{\Psi_p}=P^{K\rightarrow KK'}\ket{\Psi},
\quad
\ket{\Gamma_p^*}=P^{K\rightarrow KK'}\ket{\tilde{\Gamma}}
}
Thus, due to the property of the purified distance (\rLmm{extUhlmann0} and \rLmm{purifiedDmonot}),
it follows that
\alg{
P\left(
\proj{\tilde{\Gamma}},\proj{\Psi}
\right)
=
P\left(
\proj{\Gamma_p^*},\proj{\Psi_p},
\right)
=
P\left(
\Gamma^{KAB}
,\:
\ca{C}^K\!\circ\!{\rm Tr}_{CD}(\proj{\Psi})
\right),
}
which completes the proof.
\QED
\eprf

\blmm{extUhlmann2}
Consider the same setting as in \rLmm{extUhlmann1}, and assume that $C$ and $D$ are composite systems $C_0M_C$ and $D_0M_D$, respectively, where $M_C$ and $M_D$ are isomorphic quantum systems with an orthonormal basis $\{\ket{m}\}_m$.
In addition, suppose that the state $\Psi$ is classically coherent in $M_CM_D$, i.e., that
\alg{
\|
\langle m|^{M_C}\langle m'|^{M_D}\ket{\Psi}\|
\propto
\delta_{m,m'}.
}
Then, without loss of generality, we may assume that the states $|\phi_k\rangle$ are classically coherent in $M_CM_D$.
\elmm

\bprf
It is straightforward to verify that the state $\Psi$ is classically coherent in $M_CM_D$ if and only if all $\psi_k$ are classically coherent in $M_CM_D$.
Consequently, the maximum of each term in \req{bilkili} is achieved by $\phi_k$ that is classically coherent in $M_CM_D$, which completes the proof.
\QED
\eprf

\blmm{gentlemeasurement}
{\bf (gentle measurement: Lemma 5 in \cite{ogawa2002new} and Corollary of Lemma 7 in \cite{berta2010uncertainty})}
Let $\epsilon\in(0,1]$, $\rho\in\ca{S}(\ca{H})$ and $\Lambda\in\ca{L}(\ca{H})$ be such that $0\leq\Lambda\leq I$ and ${\rm Tr}[\Lambda\rho]\geq1-\epsilon$.
It holds that
\alg{
&
\|\rho-\sqrt{\Lambda}\rho\sqrt{\Lambda}\|_1
\leq
2\sqrt{\epsilon},
\quad
P(
\rho,
\sqrt{\Lambda}\rho\sqrt{\Lambda}
)
\leq
\sqrt{2\epsilon}.
}
\elmm

\begin{figure*}[t]
\begin{center}
\includegraphics[bb={0 30 1658 2408}, scale=0.26]{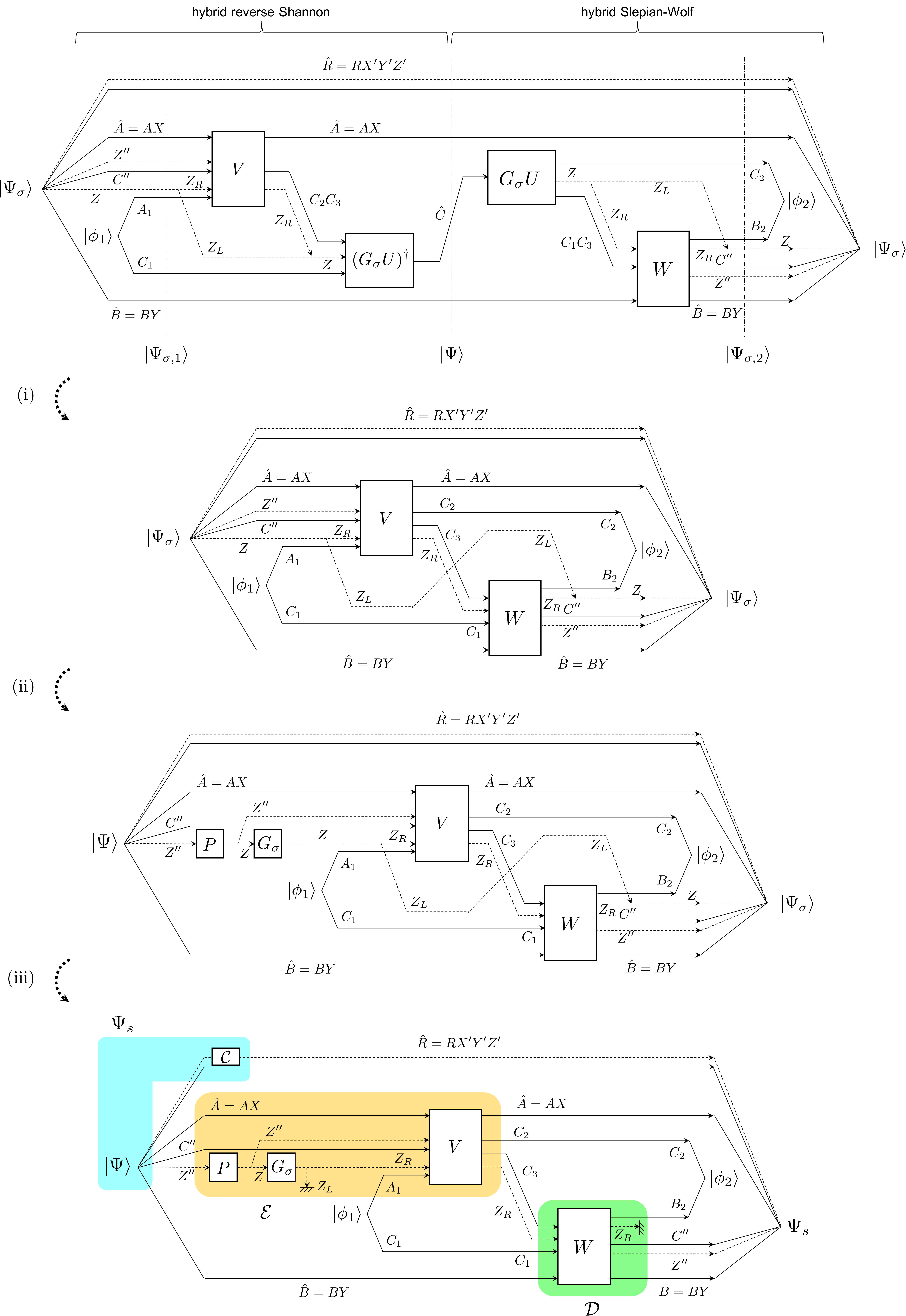}
\end{center}
\caption{
The construction of encoding and decoding operations in the proof of the direct part is depicted.
(i) is obtained by cancelling out $G_\sigma U$ and $(G_\sigma U)^\dagger$, corresponding to Inequality \req{200a-3} obtained from \req{200a-11} and \req{200a-22}. 
(ii) follows from the fact that the state $\ket{\Psi_\sigma}$ is obtained from $\ket{\Psi}$ by applying $P$ and $G_\sigma$, due to \req{psisigmaGPpsi}.
In (iii), we trace out $Z\equiv Z_LZ_R$ and apply the completely dephasing operation $\ca{C}$ to $X'Y'Z'$. See Inequalities \req{200a-4} and \req{200a-4-2} that are obtained from \req{200a-3}.
Note that the source state $\Psi_s$ is obtained from $\ket{\Psi}$ and $\ket{\Psi_\sigma}$ as \req{btf}.
}
\label{fig:stateredistributiondirect}
\end{figure*}

\end{document}

%% file: preferences.tex
{\theorembodyfont{\normalfont\it}
\theoremheaderfont{\normalfont\bf}
\newtheorem{thm}{ Theorem}
\newtheorem{dfn}[thm]{ Definition}
\newtheorem{lmm}[thm]{ Lemma}

\newtheorem{crl}[thm]{ Corollary}
\newtheorem{asm}[thm]{ Assumption}
\newtheorem{prp}[thm]{ Proposition}
\newtheorem{cjt}[thm]{ Conjecture}}

{\theorembodyfont{\normalfont}
\theoremheaderfont{\normalfont\it}
\newtheorem{prf}{ Proof:}}

{\theorembodyfont{\normalfont}
\theoremheaderfont{\normalfont\it}
}

{\theorembodyfont{\normalfont}
\theoremheaderfont{\normalfont\it}
}

{\theorembodyfont{\normalfont}
\theoremheaderfont{\normalfont\it}
}

{\theorembodyfont{\normalfont}
\theoremheaderfont{\normalfont\it}
}

{\theorembodyfont{\normalfont}
\theoremheaderfont{\normalfont\it}
}

{\theorembodyfont{\normalfont}
\theoremheaderfont{\normalfont\it}
\newtheorem{rmk}{ Remark.}}

\newcommand{\bra}[1]{\mbox{$\langle#1|$}}

\newcommand{\ket}[1]{\mbox{$|#1\rangle$}}

\newcommand{\inpro}[2]{\mbox{$\left\langle#1|#2\right\rangle$}}
\newcommand{\outpro}[2]{\mbox{$\ket{#1}\!\bra{#2}$}}

\newcommand{\proj}[1]{\mbox{$\ket{#1}\!\bra{#1}$}}

\newcommand{\alg}[1]{\begin{align}#1\end{align}}

\newcommand{\nn}{\nonumber}
\newcommand{\ca}[1]{{\mathcal #1}}
\newcommand{\mbb}[1]{{\mathbb #1}}

\newcommand{\bthm}[1]{\begin{thm}\label{thm:#1}}
\newcommand{\ethm}{\end{thm}}

\newcommand{\rThm}[1]{Theorem \ref{thm:#1}}
\newcommand{\blmm}[1]{\begin{lmm}\label{lmm:#1}}
\newcommand{\elmm}{\end{lmm}}
\newcommand{\rLmm}[1]{Lemma \ref{lmm:#1}}

\newcommand{\bdfn}[1]{\begin{dfn}\label{dfn:#1}}
\newcommand{\edfn}{\end{dfn}}

\newcommand{\basm}[1]{\begin{asm}\label{asm:#1}}
\newcommand{\easm}{\end{asm}}
\newcommand{\bprp}[1]{\begin{prp}\label{prp:#1}}
\newcommand{\eprp}{\end{prp}}

\newcommand{\rPrp}[1]{Proposition \ref{prp:#1}}
\newcommand{\bcrl}[1]{\begin{crl}\label{crl:#1}}
\newcommand{\ecrl}{\end{crl}}

\newcommand{\bcjt}[1]{\begin{cjt}\label{cjt:#1}}
\newcommand{\ecjt}{\end{cjt}}

\newcommand{\bprf}{\begin{prf}}
\newcommand{\eprf}{\end{prf}}
\newcommand{\brmk}{\begin{rmk}}
\newcommand{\ermk}{\end{rmk}}
\newcommand{\laeq}[1]{\label{eq:#1}}
\newcommand{\req}[1]{(\ref{eq:#1})}

\newcommand{\QED}{\hfill$\blacksquare$}
\newcommand{\lsec}[1]{\label{sec:#1}}
\newcommand{\rsec}[1]{\ref{sec:#1}}
\newcommand{\rSec}[1]{Section \ref{sec:#1}}
\newcommand{\lapp}[1]{\label{app:#1}}

\newcommand{\rApp}[1]{Appendix \ref{app:#1}}

\newcommand{\bitem}{\begin{itemize}}
\newcommand{\entem}{\end{itemize}}
\newcommand{\benum}{\begin{enumerate}}
\newcommand{\ennum}{\end{enumerate}}

\newcommand{\otm}{\otimes}